\documentclass[12pt]{article}
\usepackage{epsf}

\def\hybrid{\topmargin -20pt    \oddsidemargin 0pt
        \headheight 0pt \headsep 0pt
        \textwidth 6.25in       % A4 paper
        \textheight 9.5in       % A4 paper
        \marginparwidth .875in
        \parskip 5pt plus 1pt   \jot = 1.5ex}
\def\cQ{{\cal Q}}
\def\cG{{\cal G}}
\def\cL{{\cal L}}

\def\cH{{\cal H}}
\def\ket#1{|{#1}\rangle}
\def\noi{\noindent}
\def\half{{1\over2}}
\def\baselinestretch{1.2}

\catcode`\@=11

\def\marginnote#1{}
\def\draftlabel#1{{\@bsphack\if@filesw {\let\thepage\relax
   \xdef\@gtempa{\write\@auxout{\string
      \newlabel{#1}{{\@currentlabel}{\thepage}}}}}\@gtempa
   \if@nobreak \ifvmode\nobreak\fi\fi\fi\@esphack}
        \gdef\@eqnlabel{#1}}
\def\@eqnlabel{}
\def\@vacuum{}
\def\draftmarginnote#1{\marginpar{\raggedright\scriptsize\tt#1}}

\def\draft{\oddsidemargin -.2truein
        \def\@oddfoot{\sl preliminary draft \hfil
        \rm\thepage\hfil\sl\today\quad\militarytime}
        \let\@evenfoot\@oddfoot \overfullrule 3pt
        \let\label=\draftlabel
        \let\marginnote=\draftmarginnote
   \def\@eqnnum{(\theequation)\rlap{\kern\marginparsep\tt\@eqnlabel}%
\global\let\@eqnlabel\@vacuum}  }

\def\preprint{\twocolumn\sloppy\flushbottom\parindent 2em
        \leftmargini 2em\leftmarginv .5em\leftmarginvi .5em
        \oddsidemargin -.5in    \evensidemargin -.5in
        \columnsep .4in \footheight 0pt
        \textwidth 10.in        \topmargin  -.4in
        \headheight 12pt \topskip .4in
        \textheight 6.9in \footskip 0pt
        \def\@oddhead{\thepage\hfil\addtocounter{page}{1}\thepage}
        \let\@evenhead\@oddhead \def\@oddfoot{} \def\@evenfoot{} }

\def\numberbysection{\@addtoreset{equation}{section}
        \def\theequation{\thesection.\arabic{equation}}}

\def\underline#1{\relax\ifmmode\@@underline#1\else
        $\@@underline{\hbox{#1}}$\relax\fi}

\def\titlepage{\@restonecolfalse\if@twocolumn\@restonecoltrue
\onecolumn
     \else \newpage \fi \thispagestyle{empty}\c@page\z@
        \def\thefootnote{\fnsymbol{footnote}} }

\def\endtitlepage{\if@restonecol\twocolumn \else \newpage \fi
        \def\thefootnote{\arabic{footnote}}
        \setcounter{footnote}{0}}  %\c@footnote\z@ }

\catcode`@=12
\relax

\def\figcap{\section*{Figure Captions\markboth
        {FIGURECAPTIONS}{FIGURECAPTIONS}}\list
        {Figure \arabic{enumi}:\hfill}{\settowidth\labelwidth{Figure
999:}
        \leftmargin\labelwidth
        \advance\leftmargin\labelsep\usecounter{enumi}}}
 \relax
\def\tablecap{\section*{Table Captions\markboth
        {TABLECAPTIONS}{TABLECAPTIONS}}\list
        {Table \arabic{enumi}:\hfill}{\settowidth\labelwidth{Table
999:}
        \leftmargin\labelwidth
        \advance\leftmargin\labelsep\usecounter{enumi}}}
 \relax
\def\reflist{\section*{References\markboth
        {REFLIST}{REFLIST}}\list
        {[\arabic{enumi}]\hfill}{\settowidth\labelwidth{[999]}
        \leftmargin\labelwidth
        \advance\leftmargin\labelsep\usecounter{enumi}}}
 \relax
\makeatletter
\newcounter{pubctr}
\def\publist{\@ifnextchar[{\@publist}{\@@publist}}
\def\@publist[#1]{\list
        {[\arabic{pubctr}]\hfill}{\settowidth\labelwidth{[999]}
        \leftmargin\labelwidth
        \advance\leftmargin\labelsep
        \@nmbrlisttrue\def\@listctr{pubctr}
        \setcounter{pubctr}{#1}\addtocounter{pubctr}{-1}}}
\def\@@publist{\list
        {[\arabic{pubctr}]\hfill}{\settowidth\labelwidth{[999]}
        \leftmargin\labelwidth
        \advance\leftmargin\labelsep
        \@nmbrlisttrue\def\@listctr{pubctr}}}
 \relax
\makeatother
\newskip\humongous \humongous=0pt plus 1000pt minus 1000pt

\newif\ifdtup

\font\Scbig=cmss10 scaled\magstep1
\font\Scscr=cmss8 scaled\magstep1
\font\Scscrscr=cmss8
\newfam\Scfam
\textfont\Scfam=\Scbig
\scriptfont\Scfam=\Scscr
\scriptscriptfont\Scfam=\Scscrscr
\def\Sc{\fam\Scfam}

\relax
\hybrid
\def\lvm{\leavevmode\hbox to\parindent{\hfill}}

\def\thefootnote{\fnsymbol{footnote}}
\def\BE{\begin{equation}}
\def\EE{\end{equation}}
\def\BA{\begin{eqnarray}}
\def\EA{\end{eqnarray}}

\def\D{\Delta}

\def\th{\theta}

\def\P{\Phi}

\def\tt{\bar\tau}

\def\lvm{\leavevmode\hbox to\parindent{\hfill}}
\def\bar{\overline}
\def\req#1{(\ref{#1})}

\def\L{\left}
\def\R{\right}

\def\BE{\begin{equation}}
\def\EE{\end{equation} \vskip 0.30\baselineskip}
\def\BA{\begin{array}}
\def\EA{\end{array}}

\def\noi{\noindent}

\def\frac#1#2{{\textstyle{{#1}\over{#2}}}}
\def\half{{1\over2}}

\def\Kr#1{\delta_{{#1},0}}
\def\ket#1{|{#1}\rangle}

\def\ccases#1#2{\L\{\!\new\BA{l}{#1}\\ {#2}\EA\R.}

\def\cA{{\cal A}}
\def\cG{{\cal G}}
\def\cH{{\cal H}}

\def\cL{{\cal L}}

\def\cQ{{\cal Q}}

\def\cU{{\cal U}}

\def\open#1{\mbox{{\bf{#1}}}}
\def\oI{{\open I}}

\def\oZ{{\open Z}}

\def\ctop{{\Sc c}}
\def\htop{{\Sc h}}

\def\kp{\ket\phi}

\def\ie{{\it i.e.}}

\def\Qz{\cQ_0}
\def\Gz{\cG_0}
\def\Qn{$\Qz$}
\def\Gn{$\Gz$}
\def\kc{{\ket{\chi}}}
\def\kcc#1#2#3{{\kc_{#1}^{({#2}){#3}}}}

\newif\ifold \oldtrue \def\new{\oldfalse}
\let\ssection=\section
\def\section{\setcounter{equation}{0}\ssection}

\begin{document}
\renewcommand{\theequation}{\thesection.\arabic{equation}}
\newcommand{\beq}{\begin{equation}}
\newcommand{\eeq}[1]{\label{#1}\end{equation}}
\newcommand{\ber}{\begin{eqnarray}}
\newcommand{\eer}[1]{\label{#1}\end{eqnarray}}
\begin{titlepage}
\begin{center}

\hfill IMAFF-96/40,  NIKHEF-96-008\\
\hfill hep-th/9701041
\vskip .4in

{\large \bf  Families of Singular and Subsingular Vectors of the 
Topological N=2 Superconformal Algebra}
\vskip .4in

{\bf Beatriz Gato-Rivera}$^{a,b}$ {\bf and Jose Ignacio Rosado}$^a$\\
\vskip .3in

${\ }^a$ {\em Instituto de Matem\'aticas y F\'\i sica Fundamental, CSIC,\\
 Serrano 123, Madrid 28006, Spain} \footnote{e-mail address:
bgato@pinar1.csic.es}\\
\vskip .2in

${\ }^b${\em NIKHEF-H, Kruislaan 409, NL-1098 SJ Amsterdam, The Netherlands}\\

\vskip .5in

\end{center}

\begin{center} {\bf ABSTRACT } \end{center}
\begin{quotation}
We analyze several issues concerning the singular vectors of the Topological
N=2 Superconformal algebra. First we investigate  
which types of singular vectors exist, regarding the relative U(1) 
charge and the BRST-invariance properties, finding four different types
in chiral Verma modules and twenty-nine different types in 
complete Verma modules. Then we study the family structure of the 
singular vectors, every member of a family being mapped to any other member
by a chain of simple transformations involving the spectral flows. The 
families of singular vectors in chiral Verma modules follow a unique pattern 
(four vectors) and contain subsingular vectors. We write down these families 
until level 3, identifying the subsingular vectors. The families of singular 
vectors in complete Verma modules follow infinitely many different patterns, 
grouped roughly in five main kinds. We present a particularly interesting 
thirty-eight-member family at levels 3, 4, 5, and 6, 
as well as the complete set of singular
vectors at level 1 (twenty-eight different types). Finally we analyze the 
D\"orrzapf conditions leading to two linearly independent singular vectors of
the same type, at the same level in the same Verma module, and we write down
four examples of those pairs of singular vectors, which belong to the same 
thirty-eight-member family.

\end{quotation}
\vskip 0.3cm

October 1997
\end{titlepage}

\def\baselinestretch{1.2}
\baselineskip 17 pt
\section{Introduction and Notation}\lvm

In the last few years, singular vectors of infinite
dimensional algebras corresponding to conformal field theories
are drawing much attention.
Far from being empty objects that one simply would like to get rid
of, they rather contain an amazing amount of useful information.
For example, as a general feature,
their decoupling from all other states in the
corresponding Verma module gives rise to
differential equations which can be solved for correlators of
conformal fields.
Also, their possible vanishing in
the Fock space of the theories is directly connected with the existence
of extra states in the Hilbert space that are not primary and not secondary
(not included in any Verma modules) \cite{MM}. 
In some specific theories the corresponding singular vectors are, in
 addition, directly related to Lian-Zuckermann states 
 \cite{LZ} \cite{MV}.   

Regarding the construction of singular vectors, 
using either the ``fusion" method or the ``analytic continuation"
method, explicit general expressions have been obtained for the singular
vectors of the Virasoro algebra \cite{VirSV}, the Sl(2) 
Kac-Moody algebra \cite{MFF},
 the Affine algebra $A_1^{(1)}$ \cite{BaSo}, the N=1
Superconformal algebra \cite{N1S}, the Antiperiodic 
N=2 Superconformal algebra \cite{Doerr1}, \cite{Doerr2},
 and some W algebras \cite{Walg}.
There is also the method of construction
of singular vertex operators, which produce singular vectors
when acting on the vacuum \cite{KaMa1}.
 In some cases it is possible
to transform singular vectors of an algebra into singular vectors
of the same or a different algebra, simplifying notably the computation
of the latter ones. For example, 
Kac-Moody singular vectors have been transformed into Virasoro ones,
 by using the Knizhnik-Zamolodchikov equation \cite{GP},
and singular vectors of W algebras have been obtained
out of $A_2^{(1)}$ singular vectors via a quantum version of the
highest weight Drinfeld-Sokolov gauge transformations \cite{FGP}.

The singular vectors of the Topological N=2 Superconformal 
algebra have been considered mostly in chiral Verma modules. 
Some interesting features of such topological
singular vectors have appeared in a series of papers, starting
in the early nineties. For example, in \cite{BeSe2} and \cite{BeSe3} 
 it was shown that the uncharged BRST-invariant
singular vectors, in the ``mirror bosonic string" realization (KM)
of the Topological algebra, are related to Virasoro constraints
on the KP $\tau$ -function. In \cite{Sem}
an isomorphism was uncovered between the uncharged BRST-invariant
singular vectors and the singular vectors of the Sl(2) Kac-
Moody algebra (this was proved until level four).
In \cite{MV} the singular vectors were related to
Lian-Zuckermann states.
Some properties of the singular vectors in the
DDK and KM realizations were analyzed in \cite{BJI2}. In 
\cite{BJI3} the complete set of singular vectors at level 2
(four types in chiral Verma modules) was written down, 
together with the ``universal" spectral flow automorphism $\cA$
which transforms all types of 
topological singular vectors back into singular vectors.

In this paper we investigate several issues related to the singular 
vectors of the Topological N=2 Superconformal algebra, 
denoted as topological singular vectors, considering
chiral as well as complete Verma modules. The results are presented 
as follows. In section 2 we discuss the possible types of topological 
singular vectors which may exist, taking into account the relative
U(1) charge and the BRST-invariance properties of the vector and
of the primary state on which it is built. 
We use an algebraic mechanism, the ``cascade effect",
which provides a necessary (although not sufficient) condition for 
the existence of singular vectors of a given type,
finding four different types in chiral Verma
modules and twenty-nine different types in complete Verma modules.
All these types of topological singular vectors exist 
already at level 1, except one type which only exists at level zero. 

In section 3
we analyze a set of mappings which transform topological singular 
vectors into each other (of a different or of the same Verma module).
These mappings give rise to family structures which depend on the types 
of singular vectors and Verma modules involved.

In section 4 we derive the family structure corresponding to 
singular vectors in chiral Verma modules. We find a unique
structure consisting of four singular vectors, one of each type
at the same level, involving generically two different chiral Verma 
modules. We write down the complete families until level 3. 
These families contain subsingular vectors; we identify them in the
given families and we conjecture an infinite tower 
of them for higher levels.  

In section 5 we derive the family structure corresponding to
singular vectors in complete Verma modules. We find an infinite
number of different patterns which can be roughly grouped in five
main kinds. Then we derive the spectra
of conformal weights $\D$ and U(1) charges $\htop$ corresponding to the
complete Verma modules which contain generic and chiral singular vectors.
 
In section 6 we analyze some conditions under which the chains of
mappings act inside a Verma module, transforming some types of singular
vectors into singular vectors of exactly the same types at the same level.
We then analyze the D\"orrzapf equations, originally written for the
Antiperiodic N=2 Superconformal algebra, leading to the
existence of two linearly
independent singular vectors of the same type, at the same level in the 
same Verma module. We present examples which prove that some (at least) 
of those singular vectors are transformed into each other by the 
mappings described here, \ie\  they belong to the same families,
and we conjecture that the same is true for all of them; that is, that
the two partners in each D\"orrzapf pair belong to the same family. 

Section 7 is devoted to conclusions and final remarks. In 
Appendix A we describe the ``cascade effect",
in Appendix B we write down the whole set of singular vectors at 
level 1 in complete Verma modules, and in 
Appendix C we present a particularly interesting thirty-eight-member 
family of singular vectors at levels 3, 4, 5, and 6. 

\vskip .3in
\noi
{\bf Notation}
\vskip .17in
\noi
{\it Highest weight (h.w.) states} denote states annihilated by all
the positive modes of the generators of the algebra, \ie\
${\ } \cL_{n \geq 1} \kc =  \cH_{n \geq 1} \kc =  {\cG}_{n \geq 1} \kc
=  {\cQ}_{n \geq 1} \kc = 0 {\ }$.

\noi
{\it Primary states} denote non-singular h.w. states. 

\noi
{\it Secondary or descendant states} denote states obtained by acting on 
the h.w. states with the negative modes of the generators of the algebra
and with the fermionic zero modes \Qn\ and \Gn\ . The fermionic zero 
modes can also interpolate between two h.w. states at the same footing
(two primary states or two singular vectors).

\noi
{\it Chiral topological states} $\kc^{G,Q}$ are states
annihilated by both $\cG_0$ and $\cQ_0$, \ie\ 
$\cG_0\kc^{G,Q}=\cQ_0\kc^{G,Q}=0$.

\noi
{\it $\cG_0$-closed topological states} $\kc^G$ denote  
non-chiral states annihilated by $\cG_0$, \ie\  $\cG_0\kc^G=0$.

\noi
{\it $\cQ_0$-closed topological states} $\kc^Q$ denote  
non-chiral states annihilated by $\cQ_0$, \ie\ 
$\cQ_0\kc^Q=0$ (they are BRST-invariant since $\cQ_0$ is the
BRST charge).

\noi
{\it \Gn-exact topological states} are $\cG_0$-closed 
or chiral states that can
be expressed as the action of $\cG_0$ on another state:
$\ket{\gamma}=\cG_0\kc$.

\noi
{\it \Qn-exact topological states} are $\cQ_0$-closed 
or chiral states that can
be expressed as the action of $\cQ_0$ on another state:
$\ket{\gamma}=\cQ_0\kc$.

\noi
{\it No-label topological states} $\kc$ denote states  
that cannot be expressed as linear combinations of $\cG_0$-closed
and $\cQ_0$-closed states.

\noi
{\it The Verma module} associated to a h.w. state consists of the
h.w. state plus the set of secondary states built on it. For some
Verma modules the h.w. state is degenerate, the fermionic zero
modes interpolating between the two h.w. states.

\noi
{\it Null vectors} are zero-norm states. 

\noi
{\it Singular vectors} are h.w. null vectors.

\noi
{\it Generic singular vectors} are \Gn-closed and \Qn-closed singular
vectors built on \Gn-closed or \Qn-closed primary states.

\noi
{\it Subsingular vectors} are non-h.w. null vectors not descendants of 
any singular vectors, which become singular (\ie\ h.w.) in the quotient 
of the Verma module by a submodule generated by singular vectors.

\noi
{\it Secondary singular vectors} are singular vectors built on singular
vectors. The level-zero secondary singular vectors cannot ``come back" to 
the singular vectors on which they are built by acting with \Gn\ or \Qn\ .

\noi
The Topological N=2 superconformal algebra will be denoted as 
{\it the Topological algebra}.

\noi
The Antiperiodic N=2 superconformal algebra will be denoted as 
{\it the NS algebra}.

\noi
The singular vectors of the Topological algebra will be denoted as
{\it topological singular vectors}.

\noi
The singular vectors of the NS algebra will be denoted as
{\it NS singular vectors}.

\section{Singular Vectors of the Topological Algebra}\lvm

\subsection{Basic Concepts}

\vskip .17in
\noi
{\it The Topological algebra}

The algebra obtained by applying the topological twists on the
NS algebra reads \cite{DVV}

\BE\new\BA{lclclcl}
\L[\cL_m,\cL_n\R]&=&(m-n)\cL_{m+n}\,,&\qquad&[\cH_m,\cH_n]&=
&{\ctop\over3}m\Kr{m+n}\,,\\
\L[\cL_m,\cG_n\R]&=&(m-n)\cG_{m+n}\,,&\qquad&[\cH_m,\cG_n]&=&\cG_{m+n}\,,
\\
\L[\cL_m,\cQ_n\R]&=&-n\cQ_{m+n}\,,&\qquad&[\cH_m,\cQ_n]&=&-\cQ_{m+n}\,,\\
\L[\cL_m,\cH_n\R]&=&\multicolumn{5}{l}{-n\cH_{m+n}+{\ctop\over6}(m^2+m)
\Kr{m+n}\,,}\\
\L\{\cG_m,\cQ_n\R\}&=&\multicolumn{5}{l}{2\cL_{m+n}-2n\cH_{m+n}+
{\ctop\over3}(m^2+m)\Kr{m+n}\,,}\EA\qquad m,~n\in\oZ\,.\label{topalgebra}
\EE

\noi
where $\cL_m$ and $\cH_m$ are the bosonic generators corresponding
to the energy momentum tensor (Virasoro generators)
 and the topological $U(1)$ current respectively, while
$\cQ_m$ and $\cG_m$ are the fermionic generators corresponding
to the BRST current and the spin-2 fermionic current
respectively. The eigenvalues of $\cL_0$ and $\cH_0$ correspond to
the conformal weight $\D$ and the U(1) charge $\htop$ of the states.
The ``topological" central charge $\ctop$ is the central charge 
corresponding to the NS algebra. This algebra is 
topological because the Virasoro generators can be expressed as
$\cL_m={1\over2}\{\cG_m,\cQ_0\}$, where $\cQ_0$ is the BRST charge. This
implies, as is well known, that the correlators of the fields do not
depend on the metric.

\vskip .17in
\noi
{\it Topological twists}

The two possible topological twists of the NS 
superconformal generators are:

\BE\new\BA{rclcrcl}
\cL^{(1)}_m&=&\multicolumn{5}{l}{L_m+\half(m+1)H_m\,,}\\
\cH^{(1)}_m&=&H_m\,,&{}&{}&{}&{}\\
\cG^{(1)}_m&=&G_{m+\half}^+\,,&\qquad &\cQ_m^{(1)}&=&G^-_{m-\half}
\,,\label{twa}\EA\EE

\noi
and

\BE\new\BA{rclcrcl}
\cL^{(2)}_m&=&\multicolumn{5}{l}{L_m-\half(m+1)H_m\,,}\\
\cH^{(2)}_m&=&-H_m\,,&{}&{}&{}&{}\\
\cG^{(2)}_m&=&G_{m+\half}^-\,,&\qquad &
\cQ_m^{(2)}&=&G^+_{m-\half}\,.\label{twb}\EA\EE

\noi
These twists, which we denote as $T_{W1}$ and $T_{W2}$,  
are mirrored under the interchange $H_m \leftrightarrow -H_m$, 
${\ } G^{+}_r \leftrightarrow G^{-}_r$. Observe that the h.w. conditions
$G^{\pm}_{1/2}\, \ket{\chi_{NS}} = 0$ of the NS algebra 
read $\Gz \kc = 0$ after the corresponding twists.
Therefore, any h.w. state of the NS algebra results
in a \Gn-closed or chiral state of the Topological algebra, which is also
h.w. as the reader can easily verify by inspecting the twists
\req{twa} and \req{twb}. Conversely, any \Gn-closed or chiral h.w.
topological state (and only these) transforms into a h.w. state
of the NS algebra.

\vskip .17in
\noi
{\it Topological states}

From the anticommutator $\{ \cQ_0, \cG_0\} = 2 \cL_0 $ one deduces that
a topological state (primary or secondary)
 with non-zero conformal weight can be either
 \Gn-closed, or \Qn-closed, or a linear combination of 
 both types. One deduces also that \Qn-closed (\Gn-closed) 
topological states with non-zero conformal weight 
 are \Qn-exact (\Gn-exact) as well. The topological
states with zero conformal weight, however, can be \Qn-closed (satisfying 
$\cQ_0\cG_0\kc^Q=0$), or \Gn-closed (satisfying $\cG_0\cQ_0\kc^G=0$), 
or chiral, or no-label (satisfying $\cQ_0\cG_0\kc=-\cG_0\cQ_0\kc$).
Hence, since physical states in string theory are BRST cohomology 
classes\footnote{The reader not familiar with this issue may benefit from
reading section 3.2 of ref. \cite{GSW}.}, 
only the topological states which are chiral, or 
\Qn-closed with zero conformal weight, may be physical (\Qn-closed but 
not \Qn-exact).

\vskip .17in
\noi
{\it Topological primaries}

Of special interest are the topological chiral primaries
$\ket{0,\htop}^{G,Q}$, annihilated by 
both \Gn\ and \Qn . Since
the conformal weight $\Delta$ of these
primaries is zero the only quantum number
carried by them is the $U(1)$ charge $\htop$:
${\cal H}_0 \ket{0,\htop}^{G,Q} = \htop \ket{0,\htop}^{G,Q}$.
Regarding the twists \req{twa} and \req{twb}, a key observation is
that $(G^{+}_{1/2}, G^{-}_{-1/2})$ results in
$(\cG^{(1)}_0, \cQ^{(1)}_0)$ and $(G^{-}_{1/2}, G^{+}_{-1/2})$ gives
$(\cG^{(2)}_0, \cQ^{(2)}_0)$. Therefore the topological chiral primaries
$\ket{\P^{(1)}}$ and $\ket{\P^{(2)}}$ 
 correspond to the antichiral primaries
(\ie\ $G^{-}_{-1/2}\ket{\P^{(1)}}=0$) and to the chiral
primaries (\ie\ $G^{+}_{-1/2}\ket{\P^{(2)}}=0$)
of the NS algebra, respectively.
In our analysis we will consider also the topological primaries
without additional constraints, 
\ie\ either \Gn-closed or \Qn-closed primaries,
denoted as $\ket{\D,\,\htop}^G$ and $\ket{\D,\,\htop}^Q$,
or no-label primaries, denoted as $\ket{0,\,\htop}$. We will not consider
primaries $\ket{\D,\,\htop}$ which can be expressed as linear 
combinations of \Gn-closed and \Qn-closed primaries.

\newpage
\noi
{\it Topological secondaries}

As a first classification of the topological secondary states one considers
their level $l$, their {\it relative} U(1) charge $q$
and their transformation properties under
\Qn\ and \Gn\ (BRST-invariance properties). The level $l$ 
and the relative charge $q$ are defined as the difference between
the conformal weight and U(1)
charge of the secondary state and the conformal weight and U(1) charge
of the primary state on which it is built.
Hence the topological secondary states will be denoted as $\kc_l^{(q)G}$
($\cG_0$-closed), $\kc_l^{(q)Q}$ ($\cQ_0$-closed), 
$\kc_l^{(q)G,Q}$ (chiral), and $\kc_l^{(q)}$ (no-label).
For convenience we will also indicate the conformal weight $\D$,
the U(1) charge $\htop$, and the BRST-invariance properties
of the primary state on which the secondary is built. Observe that
the conformal weight and the total U(1) charge of the secondary
states are given by $\D + l$ and $\htop+q$, respectively. Thus the
chiral and no-label secondary states satisfy $\D + l=0$.

\vskip .17in
\noi
{\it Chiral Verma modules}

The Verma modules $V(\ket{0,\htop}^{G,Q})$, built on chiral primaries,
will be denoted as chiral Verma modules. They are not complete because
the chirality constraint is an additional constraint on the primary state
not required by the algebra.

\vskip .17in
\noi
{\it Generic Verma modules}

The Verma modules $V(\ket{\D,\htop}^G)$ and $V(\ket{\D,\htop}^Q)$, built on
$\cG_0$-closed and $\cQ_0$-closed primaries
without additional constraints, will be denoted as
generic Verma modules. They are complete in the sense that the constraints
of being annihilated either by $\cG_0$ or by $\cQ_0$ are required by
the algebra. Namely, any state
$\ket{\D,\htop}$ with $\D\neq0$ is either $\cG_0$-closed or $\cQ_0$-closed
or a linear combination of both types. 
Furthermore for $\D\neq0$ the h.w. state of
any generic Verma module is degenerate, \ie\ there are two primary
states. The reason is that the action of
\Qn\ on $\ket{\D,\htop}^G$ produces another primary state:
$\cQ_0\ket{\D,\htop}^G=\ket{\D,\htop-1}^Q$, and similarly the action of
\Gn\ on $\ket{\D,\htop}^Q$ produces another primary state:
$\cG_0\ket{\D,\htop}^Q=\ket{\D,\htop+1}^G$. Therefore, for $\D\neq0$ the
Verma modules $V(\ket{\D,\htop}^G)$ and $V(\ket{\D,\htop}^Q)$, 
built on the primaries $\ket{\D,\htop}^G$ and $\ket{\D,\htop}^Q$, are
equivalent to the Verma modules $V(\ket{\D,\htop-1}^Q)$ and
$V(\ket{\D,\htop+1}^G)$, built on the primaries $\ket{\D,\htop-1}^Q$
and $\ket{\D,\htop+1}^G$, respectively. For $\D=0$, however,
$\, \cQ_0\ket{0,\htop}^G$ and $\, \cG_0\ket{0,\htop}^Q$ are not primary 
states but level-zero chiral singular vectors instead, denoted as
$\kc_{0,\ket{0,\htop}^G}^{(-1)G,Q}$ and
$\kc_{0,\ket{0,\htop}^Q}^{(1)G,Q}$ respectively, so
that the h.w. states of the Verma modules $V(\ket{0,\htop}^G)$ and
$V(\ket{0,\htop}^Q)$ are not degenerate.

\vskip .17in
\noi
{\it No-label Verma modules}

The Verma modules $V(\ket{0,\htop})$, built on no-label primaries, will
be denoted as no-label Verma modules. They are complete, obviously,
since the no-label primaries are annihilated only by the positive modes
of the generators of the algebra.

\subsection{Types of topological singular vectors}\lvm

Let us discuss which types of topological secondary states can be
singular, a given type being defined by the relative charge $q$, the
BRST-invariance properties of the state itself, and the
BRST-invariance properties of the primary on which it is built.

First of all, we will consider only four different
types of topological primaries:
 $\ket{\D,\htop}^G$, $\ket{\D,\htop}^Q$ and $\ket{0,\htop}$, without
additional constraints, and $\ket{0,\htop}^{G,Q}$ with the chirality
constraint. Therefore we will not consider primaries which are linear
combinations of two or more of these types, neither primaries of these
types with additional constraints (other than the chirality constraint)
giving rise to incomplete Verma modules.

As to the secondary states $\kc_l^{(q)G}$, $\kc_l^{(q)Q}$, $\kc_l^{(q)G,Q}$ 
and $\kc_l^{(q)}$ (the latter two 
with zero conformal weight, \ie\ satisfying $\D+l=0$), taking
into account that there are four kinds of topological primaries to
be considered, a naive estimate would give sixteen different types
of them for every allowed value of the relative charge $q$, which in turn
is determined by the level $l$.
This is incorrect, however, because
the chiral primaries $\ket{0,\htop}^{G,Q}$ have no secondaries of
types $\kc_l^{(q)G,Q}$ and $\kc_l^{(q)}$, and the no-label 
primaries $\ket{0,\htop}$ have no secondaries of type $\kc_l^{(q)}$, and
have a secondary of type $\kc_l^{(q)G,Q}$ only for $l=0, {\ } q=0$.
Observe that there are no chiral secondary states in chiral Verma modules.

Thus there are twelve different types of secondary states, 
with well defined BRST-invariance properties, for
every non-zero allowed value of $q$, and thirteen types for $q=0$
(although the extra type only exists for level zero).
 However, only a few of these types admit
singular vectors. The reason is that, when one imposes the h.w. conditions,
 the allowed values of $q$ reduce drastically and, in addition, not all 
the twelve types do exist for a given non-zero allowed value of $q$, as
we will see.

\vskip .14in

The question now arises as whether any singular vector with non-zero
conformal weight can be decomposed into a \Gn-closed singular vector
plus a \Qn-closed singular vector.  From the anticommutator
$\{\Qz , \Gz \}=2 \cL_0$ one obtains the decomposition

\BE \kc_l= {1\over 2(\D+l)} {\ } \Gz \Qz \kc_l + 
   {1\over 2(\D+l)} {\ } \Qz \Gz \kc_l = \kc_l^G + \kc_l^Q \,. \EE

\noi 
If $\kc_l$ is a singular vector, \ie\ satisfies the h.w. conditions
${\ } \cL_{n \geq 1} \kc =  \cH_{n \geq 1} \kc =  {\cG}_{n \geq 1} \kc
=  {\cQ}_{n \geq 1} \kc = 0 {\ }$, then $\, \Gz \Qz \kc_l$  
and $\, \Qz \Gz \kc_l$ satisfy the h.w. conditions 
too, as one deduces straightforwardly using the Topological algebra
\req{topalgebra}. Therefore,  regarding singular vectors with
non-zero conformal weight, we can restrict ourselves to \Gn-closed and
to \Qn-closed singular vectors. 

We have not found a rigorous method so far to deduce which types of
singular vectors do exist. We have identified an algebraic mechanism,
however, which is a key fact underlying whether or not a
given type of topological secondary state admits singular vectors. This
mechanism, which we denote ``the cascade effect", consists  of the
vanishing in cascade of the coefficients of the ``would-be" singular
vector when the h.w. conditions are imposed, alone or in combination with
the BRST and/or anti-BRST-invariance conditions $\cQ_0\kc=0$ 
and/or $\cG_0\kc=0$.
As is explained in Appendix A, the starting of the cascade effect, which
occurs in most of the possible types, is very easy to determine.
Once it starts, the cascade effect goes on until the end, getting rid
of all the coefficients of the would-be singular vectors. As a result the 
types of secondary states for which the cascade effect takes place
do not admit singular vectors. The rigorous proofs of these 
statements will be presented in a forthcoming paper \cite{DB1}.

The cascade effect takes place in all cases for 
$|q|>2$, and also for $|q|=2$ in chiral Verma
modules. The types of singular vectors allowed by the cascade effect are:

\vskip .17in

- Four types built on chiral
 primaries $\ket{0,\htop}^{G,Q}$:

\BE
\begin{tabular}{r|l l l}
{\ }& $q=-1$ & $q=0$ & $q=1$\\
\hline\\
$\cG_0$-closed & $-$ & $\kc_l^{(0)G}$ & $\kc_l^{(1)G}$ \\
$\cQ_0$-closed & $\kc_l^{(-1)Q}$ & $\kc_l^{(0)Q}$ & $-$ \\
\end{tabular}
\label{tabl1}
\EE

- Ten types built on $\cG_0$-closed primaries
$\ket{\D,\htop}^G$:

\BE
\begin{tabular}{r|l l l l}
{\ }& $q=-2$ & $q=-1$ & $q=0$ & $q=1$\\
\hline\\
\Gn-closed & $-$ & $\kc_l^{(-1)G}$ & $\kc_l^{(0)G}$ & $\kc_l^{(1)G}$\\
\Qn-closed & $\kc_l^{(-2)Q}$ & $\kc_l^{(-1)Q}$ & $\kc_l^{(0)Q}$ & $-$ \\
chiral & $-$ & $\kc_l^{(-1)G,Q}$ & 
$\kc_l^{(0)G,Q}$ & $-$ \\
no-label & $-$ & $\kc_l^{(-1)}$ &
$\kc_l^{(0)}$ & $-$\\
\end{tabular}
\label{tabl2}
\EE

- Ten types built on $\cQ_0$-closed primaries
$\ket{\D,\htop}^Q$:

\BE
\begin{tabular}{r|l l l l}
{\ }& $q=-1$ & $q=0$ & $q=1$ & $q=2$\\
\hline\\
$\cG_0$-closed & $-$ & $\kc_l^{(0)G}$ & $\kc_l^{(1)G}$ & $\kc_l^{(2)G}$\\
$\cQ_0$-closed & $\kc_l^{(-1)Q}$ & $\kc_l^{(0)Q}$ & $\kc_l^{(1)Q}$ & $-$ \\
chiral & $-$ & $\kc_l^{(0)G,Q}$ & $\kc_l^{(1)G,Q}$ & 
$-$ \\
no-label & $-$ & $\kc_l^{(0)}$ &
$\kc_l^{(1)}$ & $-$\\
\end{tabular}
\label{tabl3}
\EE

- Nine types built on no-label primaries
$\ket{0,\htop}$:

\BE
\begin{tabular}{r|l l l l l}
{\ }& $q=-2$ & $q=-1$ & $q=0$ & $q=1$ & $q=2$\\
\hline\\
$\cG_0$-closed & $-$ & $\kc_l^{(-1)G}$ & $\kc_l^{(0)G}$ & $\kc_l^{(1)G}$ &
$\kc_l^{(2)G}$\\
$\cQ_0$-closed & $\kc_l^{(-2)Q}$ &$\kc_l^{(-1)Q}$ & 
$\kc_l^{(0)Q}$ & $\kc_l^{(1)Q}$ & $-$ \\
chiral & $-$ & $-$ & $\kc_0^{(0)G,Q}$ & $-$ & $-$ \\
\end{tabular}
\label{tabl4}
\EE

\vskip .17in
\noi
The chiral and no-label singular vectors satisfy $\D+l=0$. Observe that 
the chiral type on no-label primaries only exists for level zero. It 
is given by $\kc_{0,\ket{0,\htop}}^{(0)G,Q}=\cG_0\cQ_0\ket{0,\htop}$. 

The cascade effect allows therefore four different types 
of singular vectors in chiral
Verma modules and twenty-nine different types
in complete Verma modules. It turns out that all these types of
singular vectors exist at low levels, as we will see, in spite of 
the fact that the cascade effect provides a necessary, but
not sufficent, condition for their existence. 
In fact all these types can be constructed at level 1, except the 
type that only exists at level zero. In addition, the generic and the 
chiral singular vectors must necessarily exist, as we will explain.  

An important observation is that some types of topological singular
vectors admit two-dimensional spaces (see section 6). That is, in some
Verma modules one finds two linearly independent singular vectors of
the same type at the same level. In those cases our notation does not
distinguish betwen the two singular vectors and an additional label is
necessary. The rigorous analysis of this issue is beyond the scope
of this paper and will be performed in a next publication \cite{DB1}.

\vskip  0.17in
\noi
{\it Generic singular vectors}

There are twelve types of generic singular vectors, \ie\
$\cG_0$-closed and
$\cQ_0$-closed singular vectors in generic Verma modules,
as can be seen in tables \req{tabl2} and \req{tabl3}. These singular
vectors are directly related to the singular vectors of the NS algebra,
via the topological twists and the mappings that we will analyze
in next section and in section 5.
Namely, four of the generic
singular vectors ($\kc_{l,\ket{\D,\htop}^G}^{(0)G}$,
$\kc_{l,\ket{\D,\htop}^Q}^{(1)G}$, $\kc_{l,\ket{\D,\htop}^G}^{(-1)Q}$,
$\kc_{l,\ket{\D,\htop}^Q}^{(0)Q}$)
can be mapped
to the uncharged NS singular vectors, whereas the remaining eight types
($\kc_{l,\ket{\D,\htop}^G}^{(-2)G}$, $\kc_{l,\ket{\D,\htop}^Q}^{(-1)G}$,
$\kc_{l,\ket{\D,\htop}^G}^{(0)Q}$, $\kc_{l,\ket{\D,\htop}^Q}^{(1)Q}$,
$\kc_{l,\ket{\D,\htop}^G}^{(-1)G}$, $\kc_{l,\ket{\D,\htop}^Q}^{(0)G}$,
$\kc_{l,\ket{\D,\htop}^G}^{(1)G}$, $\kc_{l,\ket{\D,\htop}^Q}^{(2)G}$)
can be mapped to the charged NS
singular vectors. This argument shows that the
generic types of topological singular vectors must necessarily exist.

\vskip 0.17in
\noi
{\it Chiral singular vectors}

An important observation is that chiral singular vectors
$\kc_{l}^{(q)G,Q}\,$ can be regarded as particular cases of \Gn-closed
singular vectors $\kc_{l}^{(q)G}\,$ and  also as particular cases
of \Qn-closed singular vectors $\kc_{l}^{(q)Q}\,$. That is, 
some \Gn-closed and \Qn-closed singular vectors may ``become"
chiral (although not necessarily) when the conformal weight of the
singular vector turns out to be zero, \ie\ $\D+l=0$.
Notice, however, that there are singular vectors of several types, 
for example $\kc_{l,\ket{\D,\htop}^G}^{(1)G}\,$, that 
cannot ``become" chiral since the cascade effect prevents the existence
of chiral singular vectors of the corresponding types, for example
$\kc_{l,\ket{\D,\htop}^G}^{(1)G,Q}$. Observe
also that there are no chiral singular vectors in chiral Verma modules.
The chiral singular vectors are related 
to the singular vectors of the NS algebra, via the untwistings and 
the mappings, like the generic singular vectors.

\vskip 0.17in
\noi 
{\it Equivalent types of singular vectors}

It happens that not all the types of singular vectors shown
in tables \req{tabl2} and \req{tabl3} are inequivalent since,
for $\D\neq0$, the primaries of types $\ket{\D,\htop}^G$ and
$\ket{\D,\htop}^Q$ can be mapped into each other
inside the same Verma module, as we explained, producing
a modification of $\pm1$ in the U(1) charges $\htop$ and $q$ of the
singular vectors, so that
the total U(1) charge remains the same. For example, the singular vectors
 $\,\kc_{l,\ket{\D,\htop}^G}^{(q)G}\,$ are equivalent (for
$\D\neq0$) to the singular vectors 
$\,\kc_{l,\ket{\D,\htop-1}^Q}^{(q+1)G}$ if we
express $\ket{\D,\htop}^G=\cG_0\ket{\D,\htop-1}^Q$.
Furthermore it turns out that some of these types
of singular vectors only exist
for $\D\neq0$. Namely all four types of no-label singular vectors
in generic Verma modules, and the uncharged chiral singular vectors
$\kc_{l,\ket{\D,\htop}^G}^{(0)G,Q}$ and
$\kc_{l,\ket{\D,\htop}^Q}^{(0)G,Q}$.
As a result, the four types of no-label singular vectors reduce to 
two inequivalent types, and the uncharged chiral singular vectors
can be expressed as charged
chiral singular vectors $\kc_{l,\ket{\D,\htop-1}^Q}^{(1)G,Q}$
and $\kc_{l,\ket{\D,\htop+1}^G}^{(-1)G,Q}$ respectively. The latter
do exist for $\D=0$, however.

An important observation is that certain spectral flow mappings distinguish
between some types of singular vectors and the equivalent types. Therefore,
for practical purposes the pairs of equivalent singular vectors
must be taken into account separately 
(this issue will be discussed in next section and in section 5).

\section{Mappings between Topological Singular Vectors}\lvm

Inside a given Verma module $V(\Delta,\htop)$ and for a given level $l$
the topological singular vectors are connected
by the action of \Qn\ and \Gn\ in the following way:

\begin{eqnarray}  \Qz \kcc{l}{q}{G} \to \kcc{l}{q-1}{Q} &,& 
 \qquad\Gz \kcc{l}{q}{Q} \to \kcc{l}{q+1}{G}\\ 
  \Qz \kcc{l}{q}{ } \to \kcc{l}{q-1}{Q} &,&  
 \qquad\Gz \kcc{l}{q}{ } \to \kcc{l}{q+1}{G} \label {QGh} 
\end{eqnarray}

\noi
These arrows can be reversed (up to constants), using \Gn\ and
\Qn\ respectively, only if the conformal weight of the singular vectors
is different from zero, \ie\ $\D+l\neq0$. Otherwise, on the right-hand side
of the arrows one obtains secondary singular vectors
which cannot ``come back" to
the singular vectors on the left-hand side and are at level 
zero with respect to these. In particular
this always happens to no-label singular vectors 
$\kc_l^{(q)}$, in the second row, since they always satisfy $\D+l=0$,
while this never happens to singular vectors in chiral Verma modules,
for which $0+l>0$.

Hence \Gn\ and \Qn\ interpolate between two singular vectors
with non-zero conformal weight, in both directions, whereas they 
produce secondary singular vectors when acting on singular vectors
with zero conformal weight.

In what follows we will discuss the action of the spectral
flow transformations which map topological singular vectors back to 
topological singular vectors, interpolating between different Verma
modules. A very detailed analysis of the spectral flows has been
made recently in ref. \cite{B1}, from which we borrow the notation
and some of the results.

\vskip 0.17in
\noi 
{\it The universal odd spectral flow $\cA$}

The universal odd spectral flow automorphism $\cA_1$, 
 denoted simply as $\cA$, transforms all kinds of
primary states and singular vectors back into primary states and
singular vectors, mapping chiral states to chiral states.
It is given by \cite{BJI3} \cite{B1}

\BE\new\BA{rclcrcl}
\cA \, \cL_m \, \cA^{-1}&=& \cL_m - m\cH_m\,,\\
\cA \, \cH_m \, \cA^{-1}&=&-\cH_m - {\ctop\over3} \delta_{m,0}\,,\\
\cA \, \cQ_m \, \cA^{-1}&=&\cG_m\,,\\
\cA {\ } \cG_m \, \cA^{-1}&=&\cQ_m\,.\
\label{autom} \EA\EE

\noi
with  $\cA^{-1} = \cA$. 
It transforms the $(\cL_0,\cH_0)$ eigenvalues $(\D,\htop)$
of the states as $(\D,-\htop-{\ctop\over3})$, 
reversing the relative charge of the secondary states and 
leaving the level 
invariant, as a consequence. In addition, $\cA$ also reverses 
the BRST-invariance
properties of the states (primary as well as secondary) mapping 
\Gn-closed (\Qn-closed)
states into \Qn-closed (\Gn-closed)
states, and chiral states into chiral states.

For chiral Verma modules the action of $\cA$ results
therefore in the mappings \cite{BJI3}

\BE {\cal A}\, \kcc{l,\, \htop}{q}{Q} \to \kcc{l,\,  
-\htop-{\ctop\over3}}{-q}{G}\ , \ \ \
\ \ {\cal A} \,\kcc{l,\, \htop}{q}{G} \to \kcc{l,
\, -\htop-{\ctop\over3}}{-q}{Q}\,\, , 
\label{AAh} \EE

\noi
where $\htop$ and $-\htop-{\ctop\over3}$ denote the chiral primaries
$\ket{0,\htop}^{G,Q}$ and
$\ket{0,-\htop-{\ctop\over3}}^{G,Q}$, respectively. Observe
that these two mappings are each other's inverse.

For complete Verma modules the action of $\cA$ results in the mappings:

\begin{eqnarray}
{\cal A}\, \kcc{l,\, \ket{\D,\,\htop}^G}{q}{G} \to \kcc{l,\,
\ket{\D,-\htop-{\ctop\over3}}^Q}{-q}{Q}\ , \ \ \
\ \ {\cal A} \,\kcc{l,\, \ket{\D,\,\htop}^G}{q}{Q} \to \kcc{l,
\, \ket{\D,-\htop-{\ctop\over3}}^Q}{-q}{G} ,\nonumber \\
{\cal A}\, \kcc{l,\, \ket{-l,\,\htop}^G}{q}{G,Q} \to \kcc{l,\,
\ket{-l,-\htop-{\ctop\over3}}^Q}{-q}{G,Q}\ , \ \ 
\ \ {\cal A} \,\kcc{l,\, \ket{-l,\,\htop}^G}{q}{ } \to \kcc{l,
\, \ket{-l,-\htop-{\ctop\over3}}^Q}{-q}{ },\nonumber\\ 
{\cal A}\, \kcc{l,\, \ket{0,\,\htop}}{q}{G} \to \kcc{l,\,
\ket{0,-\htop-{\ctop\over3}}}{-q}{Q}\ , \ \ \ \ \ \ \ \
\ \ {\cal A}\, \kcc{l,\, \ket{0,\,\htop}}{q}{Q} \to \kcc{l,\,
\ket{0,-\htop-{\ctop\over3}}}{-q}{G} ,
\label{AADh}
\end{eqnarray}

\noi
and their inverses. 

\vskip 0.17in
\noi 
{\it The universal even-odd spectral flow $\, \hat\cU_{\pm 1}$}

As was explained in refs. \cite{BJI3}, \cite{B1} the action of
the even-odd spectral flow operators $\hat\cU_{\pm 1}$ on the
Topological algebra
is identical to the action of
$\cA$, except
for the fact that $\hat\cU_{\pm 1}$  connect
the two sets of topological generators corresponding to the 
two different twists of the NS generators,
given by \req{twa} and \req{twb}, \ie\

\BE\new\BA{rclcrcl}
\hat\cU_1 \, \cL_m^{(2)} \, \hat\cU_1^{-1}&=& \cL_m^{(1)} - m\cH_m^{(1)}\,,\\
\hat\cU_1 \, \cH_m^{(2)} \, \hat\cU_1^{-1}&=&-\cH_m^{(1)} - {\ctop\over3}
 \delta_{m,0}\,,\\
\hat\cU_1 \, \cQ_m^{(2)} \, \hat\cU_1^{-1}&=&\cG_m^{(1)}\,,\\
\hat\cU_1 {\ } \cG_m^{(2)} \, \hat\cU_1^{-1}&=&\cQ_m^{(1)}\,.\
\EA \label{sp21} \EE

\noi
with $\hat\cU_1^{-1} = \hat\cU_{-1}$. One can see this also using
the composition rules of the spectral flows \cite{B1}. One finds
$\, \hat\cU_1=\cA\hat\cA_0$, $\, \hat\cU_{-1}=\hat\cA_0\cA$, where $\hat\cA_0$
is the operator that interchanges the labels
$(1)\leftrightarrow (2)$ corresponding to the two sets of
topological generators.

\vskip 0.17in
\noi
{\it The even spectral flow $\, \cU_{\pm1}$}

The topological even spectral flow $\cU_{\th}$ is not universal for any
value of $\th$. In particular it
does not map chiral primary states back to chiral primary 
states. However, for $\th=1$ it maps 
h.w. states annihilated by \Gn\ into h.w. states annihilated by \Qn , 
and the other way around for $\th=-1$. As a result $\cU_1$ transforms
the singular vectors
of types $\kc_{l,\kp^G}^{(q)G}$ and $\kc_{l,\kp^G}^{(q)G,Q}$ into 
singular vectors of types $\kc_{l,\kp^Q}^{(q)Q}$
and $\kc_{l,\kp^Q}^{(q)G,Q}$ (both), while
mapping all other types of singular vectors
to various kinds of states which are not singular 
vectors.
The topological even spectral flow\footnote{This spectral flow was
written for the first time in \cite{SeTi}, although only in \cite{B1}
it has been analyzed properly.} is given by \cite{SeTi} \cite{B1}

\BE\new\BA{rclcrcl}
\cU_\th \, \cL_m \, \cU_\th^{-1}&=& \cL_m
 +\th \cH_m + {\ctop\over 6} (\th + \th^2) \delta_{m,0}\,,\\
\cU_\th \, \cH_m \, \cU_\th^{-1}&=&\cH_m + {\ctop\over3} \th \delta_{m,0}\,,\\
\cU_\th \, \, \cG_m \, \cU_\th^{-1}&=&\cG_{m+\th}\,,\\
\cU_\th \, \cQ_m \, \cU_\th^{-1}&=&\cQ_{m-\th}\,,\
\label{spflw} \EA\EE

\noi
and satisfies $\cU_{\th}^{-1} = \cU_{(-\th)}$. It transforms the
$(\cL_0,\cH_0)$ eigenvalues $(\D,\htop)$ of the primary states as
$(\D-\th \htop+{\ctop\over6}(\th^2-\th), \htop-{\ctop\over3}\th)$, and the 
level of the secondary states as $l \rightarrow l-\th q$, letting invariant
the relative charge $q$.
Under $\cU_1$ a \Gn-closed, or chiral, singular
vector at level $l$ with relative charge $q$, built on a
primary of type $\ket{\D,\htop}^G$, is
transformed into a \Qn-closed singular vector at level
$l-q$ with relative charge $q$, built on a primary
of type $\ket{\D-\htop, \htop-{\ctop\over3}}^Q$. This \Qn-closed
singular vector may ``become" chiral if
$\D-\htop=q-l$, and is certainly chiral if the \Gn-closed singular vector
was annihilated by $\cG_{-1}$. Therefore there are four possibilities:

\def\xggw  {\mbox{$\kc_{l,\, \ket{\D,\, \htop}^G}^{(q)G} $}}
\def\xqqw {\mbox{$\kc_{l-q,\,\ket{-l-\htop,\,\htop-{\ctop\over3}}^Q}^{(q)Q} $}}
\def\xgqgw  {\mbox{$\kc_{l,\, \ket{-l,\, \htop}^G}^{(q)G,Q} $}}
\def\yqqw {\mbox{$\kc_{l-q,\,\ket{\D-\htop,\,\htop-{\ctop\over3}}^Q}^{(q)Q} $}}

  \begin{equation}   \begin{array}{rcl} 
\xggw  \stackrel{\cU_1}{\mbox{------}\!\!\!\rightarrow} \yqqw \ , \qquad
\xgqgw  \stackrel{\cU_1}{\mbox{------}\!\!\!\rightarrow} \xqqw \ ,
  \end{array} \nonumber \end{equation}

\def\xggw  {\mbox{$\kc_{l,\, \ket{\D,\, \htop}^G}^{(q)G} $}}
\def\xqqw {\mbox{$\kc_{l-q,\,\ket{\D-\htop,\,\htop-{\ctop\over3}}^Q}^{(q)Q} $}}
\def\xgqgw  {\mbox{$\kc_{l,\, \ket{\D,\, \htop}^G}^{(q)G,Q} $}}
\def\ogqgw  {\mbox{$\kc_{l,\, \ket{-l,\, 0}^G}^{(0)G,Q} $}}
\def\xgqqw 
{\mbox{$\kc_{l-q,\,\ket{q-l,\,\htop-{\ctop\over3}}^Q}^{(q)G,Q}$}}
\def\ogqqw 
{\mbox{$\kc_{l,\,\ket{-l,\,-{\ctop\over3}}^Q}^{(0)G,Q}$}}

  \begin{equation}   \begin{array}{rcl} 
\xggw  \stackrel{\cU_1}{\mbox{------}\!\!\!\rightarrow} \xgqqw \ , \qquad
\ogqgw  \stackrel{\cU_1}{\mbox{------}\!\!\!\rightarrow} \ogqqw \ .
  \end{array} \label{diaw1} \end{equation}

\noi

We see that chiral singular vectors are
transformed, in most cases, into 
non-chiral singular vectors (although annihilated by $\cQ_{-1}$, 
as the reader can easily verify). The mapping to another chiral singular
vector only occurs under very restricted conditions:\footnote{There 
are no chiral singular vectors of  
types $\kc_{l,\,\ket{\D,\,\htop}^G}^{(1)G,Q}$ and
$\kc_{l,\,\ket{\D,\,\htop}^Q}^{(-1)G,Q}$.}
$q=0$, $\htop=0$, and 
the requirement that the singular vector is annihilated by $\cG_{-1}$. The
resulting chiral singular vector, in turn, is annihilated by $\cQ_{-1}$.

These are the only types of singular vectors transformed
under $\cU_1$ into singular vectors, and the other way around using the 
inverse $\cU_{-1}$.
The chiral primaries $\ket{0,\htop}^{G,Q}$ are
transformed under $\cU_1$ (or $\cU_{-1}$) into non-chiral
primaries annihilated by $\cQ_{-1}$ (or $\cG_{-1}$). Primaries with
such constraints, which generate incomplete Verma modules, 
are beyond the scope of this paper.

Observe that $\cU_{\pm1}$ distinguish between the different types of
singular vectors drastically. Regarding the pairs of
equivalent singular vectors for $\D\neq0$,
$\kc_{l,\,\ket{\D,\,\htop}^G}^{(q)G}$ and
$\kc_{l,\,\ket{\D,\,\htop-1}^Q}^{(q+1)G}$ on the one side, and
$\kc_{l,\,\ket{\D,\,\htop}^Q}^{(q)Q}$ and
$\kc_{l,\,\ket{\D,\,\htop+1}^G}^{(q-1)Q}$ on the other side, $\cU_1$ and
$\cU_{-1}$, respectively, transform only one member of the pair into
a singular vector. The situation is more involved for the pairs of
equivalent singular vectors which are chiral. Namely, $\cU_1$ transforms
the chiral singular vector $\kc_{l,\,\ket{-l,\,\htop}^G}^{(0)G,Q}$ into
a singular vector of type
$\kc_{l,\,\ket{-l-\htop,\,\htop-{\ctop\over3}}^Q}^{(0)Q}$ 
while $\cU_{-1}$ transforms the equivalent singular vector
$\kc_{l,\,\ket{-l,\,\htop-1}^Q}^{(1)G,Q}$ into a singular vector of type
$\kc_{l+1,\,\ket{-l+\htop-1+{\ctop\over3},
\,\htop-1+{\ctop\over3}}^G}^{(1)G}$,
and similarly with the pair of equivalent chiral singular vectors
$\kc_{l,\,\ket{-l,\,\htop}^Q}^{(0)G,Q}$ and
$\kc_{l,\,\ket{-l,\,\htop+1}^G}^{(-1)G,Q}$.

\vskip 0.17in
\noi
{\it Other spectral flow transformations}

The topological odd spectral flow $\cA_{\th}$ \cite{B1} transforms
certain types of singular vectors back into singular vectors for $\th=0$
and $\th=2$, in addition to $\th=1$ which corresponds to the
universal mapping: $\cA_1=\cA$. However, the action of $\cA_0$ and $\cA_2$
is already included in the composition of $\cA$ with $\, \cU_{\pm1}$.
Namely, $\cA_0=\cA{\ }\cU_1$ and $\cA_2=\cU_1\cA$.

Similarly, the
even-odd spectral flow transformations $\hat\cU_0$  and $\hat\cU_{\pm2}$
\cite{B1} map some types of topological singular vectors to singular 
vectors, but their action is identical to the action of $\cA_0$ and 
$\cA_2$, although interchaging the labels $(1)\leftrightarrow(2)$ of the
two sets of topological generators: $\, \hat\cU_0=\cA_0\hat\cA_0$,
$\, \hat\cU_2=\cA_2\hat\cA_0$ and $\, \hat\cU_{-2}=\hat\cA_0\cA_2$,
where $\hat\cA_0$ is the label-exchange operator.

Finally, the odd-even spectral flow transformations
$\hat\cA_{\pm1}$ \cite{B1} act
like $\cU_{\pm1}$ but interchanging the labels $(1)\leftrightarrow(2)$
since $\hat\cA_{\pm1}=\cU_{\pm1}\hat\cA_0$.

\section{Families of Topological Singular Vectors in Chiral Verma
Modules}\lvm

\subsection{Family Structure}\lvm

Now let us apply the results of sections 2 and 3 to analyze the family
structure of the topological singular vectors in chiral Verma modules,
\ie\ built on chiral primaries $\ket{0,\htop}^{G,Q}$.
We found that in chiral Verma modules the cascade effect allows
only four types of topological singular vectors: $\kcc{}{0}{G}$,
$\kcc{}{0}{Q}$, $\kcc{}{1}{G}$ and $\kcc{}{-1}Q$. These four types
of singular vectors have been constructed at low levels (see
subsection 4.4). They are
connected to each other by the action of \Qn, \Gn\ and ${\cal A}$, 
at the same level $l$, in the way shown by the diagram:

\def\xgo  {\mbox{$\kc_{l,\, \htop}^{(0)G} $}}
\def\xqo  {\mbox{$\kc_{l,\, -\htop-{\ctop\over3}}^{(0)Q} $}}
\def\xqm  {\mbox{$\kc_{l,\, \htop}^{(-1)Q} $}}
\def\xgp  {\mbox{$\kc_{l,\, -\htop-{\ctop\over3}}^{(1)G} $}}

  \begin{equation}
  \begin{array}{rcl} \xgo &
  \stackrel{\Qz}{\mbox{------}\!\!\!\longrightarrow}
  & \xqm \\[3 mm]
   \cA\,\updownarrow\ && \ \updownarrow\, \cA
  \\[3 mm]  \xqo \! & \stackrel{\Gz}
  {\mbox{------}\!\!\!\longrightarrow} & \! \xgp  \end{array}
\label{dia} \end{equation}

\vskip .2in

Hence the topological singular vectors built on chiral
primaries $\ket{0,\htop}^{G,Q}$
come in families of four: one of each kind at the same level.
Two of them, one charged and one uncharged, belong to the chiral Verma 
module $V(\ket{0,\htop}^{G,Q})$, whereas the other pair belong to a
different chiral Verma module
$V(\ket{0,-\htop-{\ctop\over 3}}^{G,Q})$. This implies that
it is sufficient
to compute only one of these four singular vectors from scratch, the
other three being generated by the action of \Gn, \Qn\ and $\cA$. 
For $\htop=-{\ctop\over6}$ the two chiral Verma modules related by
$\cA$ coincide.
Therefore, if there are singular vectors for this value of $\htop$ (see next 
subsection), they must come four by four {\it in the same} chiral Verma
module: one of each kind at the same level.

As was explained in section 3, the action of $\,\hat\cU_{\pm1}$
\req{sp21}, which is the composition of $\cA$ with $\hat\cA_0$, 
is identical to the action of $\cA$ \req{autom} except that it
distinguishes (and interchanges) the topological generators of the
two twisted theories, \ie\ it exchanges the labels (1) and (2).
Thus one can substitute
 $\cA$ by $\,\hat\cU_1$ or $\,\hat\cU_{-1}$ in diagram \req{dia}.
This amounts to differentiate the upper part and the lower part
of the diagram as corresponding to topological singular vectors
and generators labelled by (1) or by (2), respectively. For example
\vskip .13in

\def\xgo1  {\mbox{$\ket{\chi^{(1)}}_{l,\, \htop}^{(0)G} $}}
\def\xqo2  {\mbox{$\ket{\chi^{(2)}}_{l,\, -\htop-{\ctop\over3}}^{(0)Q} $}}
\def\xqm1  {\mbox{$\ket{\chi^{(1)}}_{l,\, \htop}^{(-1)Q} $}}
\def\xgp2  {\mbox{$\ket{\chi^{(2)}}_{l,\, -\htop-{\ctop\over3}}^{(1)G} $}}

  \begin{equation}
  \begin{array}{rcl} \xgo1 &
  \stackrel{\Qz^{(1)}}{\mbox{------}\!\!\!\longrightarrow}
  & \xqm1 \\[3 mm]
   \hat\cU_{1}\,\uparrow\ && \ \uparrow\, \hat\cU_1
  \\[3 mm]  \xqo2 \! & \stackrel{\Gz^{(2)}}
  {\mbox{------}\!\!\!\longrightarrow} & \! \xgp2  \end{array}
\label{dia12} \end{equation}

\vskip .2in
\noi
(one can reverse the arrows
$\hat\cU_1$ using $\hat\cU_{-1}$). Therefore, the topological
singular vectors in chiral Verma modules come actually in families of
four plus four vectors, four
of them labelled by (1) and the other four identical vectors labelled
by (2), the two sets being connected through the
action of $\hat\cU_{\pm1}$. From the point of view of the Topological
algebra \req{topalgebra} the two sets of singular vectors and
generators look identical, so that one can 
consider a unique set of four singular vectors, as we did before.
However, the topological generators (1) and (2) are different with
respect to the generators of the NS algebra.
As a consequence diagrams \req{dia} and
\req{dia12} give different results under the (un)twistings $T_{W1}$
\req{twa} and $T_{W2}$ \req{twb}.

Regarding the untwisting of these 
topological singular vectors, the \Gn-closed singular vectors
in chiral Verma modules are transformed into singular vectors of the
NS algebra, with the same U(1) charge and built on
antichiral primaries under $T_{W1}$
\req{twa}, and with the reverse U(1) charge and built on chiral
primaries under $T_{W2}$ \req{twb} (whereas
the \Qn-closed singular vectors are transformed into states which are not 
NS singular vectors). As a consequence, the fact that the 
four types of singular vectors in diagram \req{dia} are the only existing 
ones in chiral Verma modules, implies \cite{BJI5} 
that the singular vectors of the NS algebra 
built on chiral primaries have only relative charges $q=0$ or $q=-1$, 
while those built on antichiral primaries have $q=0$ or $q=1$.

Finally, let us point out a result for the NS algebra 
which is easier to obtain from the Topological algebra. It is the
fact that there are no chiral NS singular vectors in chiral Verma
modules neither antichiral NS singular vectors in antichiral Verma
modules. To see this one only needs to ``untwist" the fact that
there are no chiral topological singular vectors in chiral 
topological Verma modules.

\subsection{Spectrum of h}\lvm

Let us discuss now the spectrum of U(1) charges $\htop$ corresponding to
the topological chiral primaries $\ket{0,\htop}^{G,Q}$ which contain
singular vectors in their
Verma modules. This spectrum follows directly from the
spectrum corresponding to the
singular vectors of the NS algebra on antichiral Verma modules.
Namely, the \Gn-closed topological singular vectors on chiral 
topological primaries become NS singular vectors on antichiral 
primaries with the same U(1) charge, under $T_{W1}$ \req{twa}. 

Hence the spectrum of U(1) charges $\htop$ corresponding to the 
topological chiral Verma modules which contain \Gn-closed singular 
vectors is identical to the spectrum of U(1) charges $\htop$ 
corresponding to the antichiral Verma modules of the NS algebra
which contain singular vectors.
This spectrum has been conjectured
in \cite{BJI5} by analyzing the roots of the determinant
formula for the NS algebra in a rather non-trivial way 
(because the
determinant formula does not apply directly to chiral or antichiral
Verma modules, but only to complete Verma modules).

The results are as follows. A topological
chiral primary with U(1) charge given by

\BE \htop_{r,s}^{(0)} = {3-\ctop \over 6}(r+1)-{s \over 2} \ , \ \ \ \ \ 
r \in {\bf Z}^+ , \ \ s \in 2{\bf Z}^+ \label{hrs0} \EE

\noi
has one singular vector of type $\kc^{(0)G}$, and one singular vector of 
type $\kc^{(-1)Q}$, at level $rs\over2$ in its Verma module (and possibly 
more singular vectors at higher levels) . Similarly, a topological
chiral primary with U(1) charge given by

\BE \htop_{r,s}^{(1)} = {\ctop-3 \over 6}(r-1)+{s \over 2} -1 \ , \ \ \ \ \ 
r \in {\bf Z}^+ , \ \ s \in 2{\bf Z}^+ \label{hrs1} \EE

\noi
has one singular vector of type $\kc^{(1)G}$, and one singular vector of 
type $\kc^{(0)Q}$, at level $rs\over2$ in its
Verma module\footnote{The spectrum \req{hrs1} for the uncharged \Qn-closed
topological singular vectors $\kc^{(0)Q}$ was also written down in \cite{Sem} 
just by fitting the known data (until level 4), without any
derivation or further analysis.} (and possibly more singular vectors at
higher levels).

These expressions have been checked
until level 4 by explicit construction of the singular vectors and by
computing the chiral determinant formulae \cite{BJI5}.
Observe that between both expressions there exists the spectral
flow relation $\htop_{rs}^{(1)} = -\htop_{rs}^{(0)}-{\ctop\over3}{\ }$.
Therefore the two expressions coincide for the special case
$\htop_{rs}^{(1)} = \htop_{rs}^{(0)} = -{\ctop\over6}{\ }$.
The solutions to this give discrete
values of $\ctop<3$. Namely

\BE \ctop = {3(r-s+1)\over r}\,, \qquad r\in\oZ^+, \ \ s\in 2\oZ^+
  \label{crs} \EE

\noi
For example, for $s=2$ the solutions corresponding to
$\,r=1,\,2,\,3\,$ are $\,\ctop=0,\,3/2,\,2\,$ respectively.
Therefore for the discrete values \req{crs} the
topological chiral Verma modules
with $\htop=-{\ctop\over6}$ have four singular vectors,
 one of each type, at the same level $l={rs\over2}$.

\subsection{Subsingular vectors}\lvm

Subsingular vectors of the N=2 Superconformal algebras have been
recently discovered in ref. \cite{BJI5}. For the case of the 
Topological algebra we have found that most of the singular
vectors given by the spectrum \req{hrs1}, are subsingular
vectors in the complete Verma modules $V(\ket{0,\htop}^G)$ and
most of the singular vectors given by the spectrum \req{hrs0},
are subsingular vectors in the complete Verma modules $V(\ket{0,\htop}^Q)$.
The argument goes as follows (all the statements about spectra in complete 
Verma modules can be checked in subsection 5.2).
 
The spectrum $\htop_{r,s}^{(1)}$ \req{hrs1}, corresponding to
singular vectors
of types $\kc_l^{(1)G}$ and $\kc_l^{(0)Q}$ in chiral Verma modules, is
the ``fusion" of
two different spectra, $\htop_k$ and $\hat\htop_{r,s}$, given by

\BE  \htop_k= {\ctop-3 \over 6}(k-1) \ , \ \ \ \ \qquad 
\hat\htop_{r,s} = {\ctop-3 \over 6}(r-1)+{s\over2} \ . \label{hhrs} \EE

\noi
The spectrum $\htop_k$ corresponds to $\,\htop_{r,s}^{(1)}\,$ for $s=2$:
$\htop_k=\htop_{k,2}^{(1)}\,$, with the level of the singular vector
given by $l=k$. 
It coincides with the spectrum of singular vectors
of types $\kc_{l,\ket{0,\htop}^G}^{(1)G}$
and $\kc_{l,\ket{0,\htop}^G}^{(0)Q}$ in the complete Verma modules
$V(\ket{0,\htop}^G)$. The spectrum $\hat\htop_{r,s}$, on
the other hand, gives $\htop_{r,s}^{(1)}$ for $s>2$ in the way:
$\hat\htop_{r,s-2}=\htop_{r,s>2}^{(1)}$, where the level of the singular
vector is given by $l=rs/2$.
It turns out that $\hat\htop_{r,s}$
corresponds to half the spectrum of singular vectors of types
$\kc_{l,\ket{0,\htop}^G}^{(0)G}$ and $\kc_{l,\ket{0,\htop}^G}^{(-1)Q}$
in the complete Verma modules $V(\ket{0,\htop}^G)$ (the other half
being given by \req{hrs0}), with the level given by $l=rs/2$. Therefore
$\,\hat\htop_{r,s-2}=\htop_{r,s>2}^{(1)}\,$ corresponds either to singular
vectors of types $\kc_l^{(1)G}$ and $\kc_l^{(0)Q}$ in chiral Verma modules
$V(\ket{0,\htop}^{G,Q})$, at level $l=rs/2$, or to singular vectors of
types $\kc_{l',\ket{0,\htop}^G}^{(0)G}$ and
$\kc_{l',\ket{0,\htop}^G}^{(-1)Q}$ in complete Verma modules
$V(\ket{0,\htop}^G)$ at level $l'=r(s-2)/2$.

It happens, 
at least at levels 2 and 3, that the
latter singular vectors vanish, while the former ones appear from
secondaries which become singular, once one imposes the chirality
condition $\cQ_0\ket{0,\htop}^G=0$ on the primary $\ket{0,\htop}^G$, turning
it into the chiral primary $\ket{0,\htop}^{G,Q}$. But
$\cQ_0\ket{0,\htop}^G$ is a singular vector so that the chirality
condition is equivalent to take the quotient of the complete Verma
module $V(\ket{0,\htop}^G)$ by this singular vector.
Those secondaries are
therefore subsingular vectors \cite{FFsub}, since they are singular
only in the chiral Verma module, that is, in the 
quotient of the Verma module $V(\ket{0,\htop}^G)$ by the 
submodule generated by the singular vector $\cQ_0\ket{0,\htop}^G$.

Hence the singular vectors of types $\kc_l^{(1)G}$ and 
$\kc_l^{(0)Q}$ in chiral Verma modules are
subsingular vectors in the complete Verma modules $V(\ket{0,\htop}^G)$, 
for the values $\htop_{r,s>2}^{(1)}$. The subsingular vectors 
are null and, however, are located outside the (incomplete) Verma
modules built on top of the singular vectors. 

Similarly, the spectrum $\htop_{r,s}^{(0)}$ \req{hrs0}, corresponding to
singular vectors of types  $\kc_l^{(0)G}$ 
and $\kc_l^{(-1)Q}$ in chiral Verma modules, is
the fusion of two different spectra: $\htop_{r,2}^{(0)}$ which corresponds
to singular vectors of types $\kc_{l,\ket{0,\htop}^Q}^{(0)G}$ and
$\kc_{l,\ket{0,\htop}^Q}^{(-1)Q}$, and $\htop_{r,s>2}^{(0)}$ which, together
with $\htop_{r,s}^{(1)}$, corresponds to singular vectors of types
$\kc_{l,\ket{0,\htop}^Q}^{(0)Q}$ and $\kc_{l,\ket{0,\htop}^Q}^{(1)G}$, 
all these vectors in the complete Verma modules $V(\ket{0,\htop}^Q)$.
Therefore, the singular vectors of types $\kc_l^{(0)G}$
and $\kc_l^{(-1)Q}$ in chiral Verma modules are subsingular vectors in
the complete Verma modules $V(\ket{0,\htop}^Q)$ for the values
$\htop_{r,s>2}^{(0)}$; that is, they are not singular vectors in
$V(\ket{0,\htop}^Q)$ but they are singular vectors in the quotient of
$V(\ket{0,\htop}^Q)$ by the submodule generated by the singular vector
$\cG_0\ket{0,\htop}^Q$.

These results we have checked at levels 2 and 3 and we conjecture that
they hold at any level. In the next subsection we show the families of
singular vectors in chiral Verma modules until level 3 and we identify the
subsingular vectors contained in them.

\subsection{Families at Levels 1, 2 and 3}\lvm

Let us write down the complete families of topological 
singular vectors in 
chiral Verma modules until level 3. Some of these vectors have been
published before: $\kc_2^{(0)Q}$ and $\kc_3^{(0)Q}$ were written in
\cite{BeSe2} and \cite{BeSe3} respectively, while the complete family
at level 2 was given in \cite{BJI3}. The fact that most of these
singular vectors are subsingular vectors in the
complete Verma modules $V(\ket{0,\htop}^G)$ or $V(\ket{0,\htop}^Q)$ was
not realized, however.

The families of topological singular vectors in chiral Verma modules are
the following.

At level 1:

\BE        
 \ket\chi_1^{(0)G} = (\cL_{-1}+\cH_{-1})\ket{0,-{\ctop\over3}}^{G,Q},
  \qquad    \ket\chi_1^{(0)Q} = \cL_{-1}\ket{0,0}^{G,Q},  \label{unch1}\EE
\BE  \ket\chi_1^{(-1)Q} = \cQ_{-1} \ket{0,-{\ctop\over3}}^{G,Q}, \qquad 
      \ket\chi_1^{(1)G} = \cG_{-1}\ket{0,0}^{G,Q} . \label{char1}\EE

At level 2:

        \BE
        \ket\chi_2^{(0)G}=(\theta\cL_{-2}+\alpha\cL_{-1}^2+
        \Gamma\cH_{-1}\cL_{-1}+\beta\cH_{-1}^2+\gamma\cH_{-2}+
        \delta\cQ_{-1}\cG_{-1})\ket{0,\htop}^{G,Q}
        \EE

        \begin{eqnarray}
        \htop=\ccases{1-\ctop\over 2}{-{\ctop+3\over 3}}\!,\qquad\alpha=
        \ccases{6\over\ctop-3}{\ctop-3\over 6}\!,\qquad\theta=\ccases{9-\ctop
        \over 3-\ctop}{0}\!,\qquad 
        \Gamma=\ccases{18\over\ctop-3}{\ctop\over
        3}\!\nonumber\\
        \beta=\ccases{12\over\ctop-3}{\ctop+3\over 6}\!,\qquad
        \gamma=\ccases{2}{\ctop+3\over 6}\!,\qquad
        \delta=\ccases{3\over\ctop
        -3}{1\over 2}\!
        \end{eqnarray}
        
\vskip .17in

        \BE
        \ket\chi_2^{(0)Q}=(\cL_{-2}+\alpha\cL_{-1}^2+\Gamma\cH_{-1}\cL_{-1}
        +{1\over 2}\Gamma
        \cQ_{-1}\cG_{-1})\ket{0,\htop}^{G,Q}\EE

        \BE
        \htop=\ccases{\ctop-3\over 6}{1}\!,\qquad\alpha=\ccases{6\over\ctop-3}
        {\ctop-3\over 6}\!,\qquad\Gamma=\ccases{6\over 3-\ctop}{-1}\!
        \EE

\vskip .17in
        
        \BE
        \ket\chi_2^{(1)G}=(\cG_{-2}+\alpha\cL_{-1}\cG_{-1}
        +\Gamma\cH_{-1}\cG_{-1})\ket{0,\htop}^{G,Q}\EE

        \BE
        \htop=\ccases{\ctop-3\over 6}{1}\!,\qquad\alpha=\ccases{6\over\ctop
        -3}{\ctop-3\over 6}\!,\qquad\Gamma=\ccases{6\over 3-\ctop}{-1}\!
        \EE

\vskip .17in

        \BE
        \ket\chi_2^{(-1)Q}=(\cQ_{-2}+\alpha\cL_{-1}\cQ_{-1}
        +\beta\cH_{-1}\cQ_{-1})\ket{0,\htop}^{G,Q}\EE

        \BE
        \htop=\ccases{1-\ctop\over2}{-{\ctop+3\over3}}\!,\qquad\alpha=
        \ccases{6\over\ctop-3}{\ctop-3\over 6}\!,\qquad \beta=\ccases
        {12\over\ctop-3}{\ctop+3\over6}\!
        \EE

At level 3:        

        \begin{eqnarray*}
        \ket\chi_3^{(0)G}=(\alpha\cL_{-1}^3+\theta\cL_{-2}\cL_{-1}+
        \beta\cH_{-3}+\gamma\cH_{-2}
        \cL_{-1}+\delta\cL_{-3}+\\
        \epsilon\cH_{-1}\cL_{-2}+\mu\cH_{-1}^2
        \cL_{-1}+\nu\cH_{-1}\cL_{-1}^2+\kappa\cH_{-1}\cH_{-2}+
        \rho\cH_{-1}^3+\\
        a\cQ_{-2}\cG_{-1}+e\cQ_{-1}\cG_{-2}+f\cL_{-1}\cQ_{-1}\cG_{-1}
        +g\cH_{-1}\cQ_{-1}\cG_{-1})\ket{0,\htop}^{G,Q}\end{eqnarray*}

        \begin{eqnarray}
        \htop=\ccases{3-2\ctop\over 3}{-{\ctop+6\over 3}}\!,\qquad\alpha=
        \ccases{3\over 3-\ctop}{3-\ctop\over 12}\!,\qquad\theta=\ccases{15-2
        \ctop\over\ctop-3}{-{1\over 2}}\!,\qquad
        \beta=\ccases{3-\ctop}
        {\ctop^2+12\ctop+27\over 6(3-\ctop)}\!\nonumber\\
        \delta=\ccases{(\ctop-6)
        ^2\over 3(3-\ctop)}{-{1\over 2}}\!,\qquad\gamma=\ccases{5\ctop-12
        \over 3-\ctop}{\ctop^2+6\ctop-3\over 4(3-\ctop)}\!,\qquad
        \epsilon=
        \ccases{33-4\ctop\over\ctop-3}{-{1\over 2}}\!,\qquad\nu=\ccases{18
        \over 3-\ctop}{-{\ctop+3\over 4}}\!\nonumber\\
        \mu=\ccases{33\over 3-\ctop
        }{\ctop^2+6\ctop-3\over 4(3-\ctop)}\!,\qquad
        a=\ccases{-1}{6\over 3-
        \ctop}\!,\qquad e=\ccases{{1\over 2}{15-\ctop\over\ctop-3}}{1\over 2}
        \!,\qquad\kappa=\ccases{-9}{\ctop^2+12\ctop+27\over 4(3-\ctop)}\!
        \nonumber\\
        \rho=\ccases{18\over 3-\ctop}{\ctop^2+12\ctop+27\over 12(3-
        \ctop)}\!,\qquad f=\ccases{9\over 2(3-\ctop)}{-{3\over 4}}\!,\qquad 
        g=
        \ccases{21\over 2(3-\ctop)}{3\ctop+15\over 4(3-\ctop)}\!
        \end{eqnarray}

        \begin{eqnarray*}
        \ket\chi_3^{(0)Q}=(\alpha\cL_{-1}^3-2\cL_{-2}\cL_{-1}+
        \gamma\cH_{-2}\cL_{-1}+\delta\cL_{-3}+2e\cH_{-1}\cL_{-2}+
        g\cH_{-1}^2\cL_{-1}\\
        +2f\cH_{-1}\cL_{-1}^2+
        a\cQ_{-2}\cG_{-1}+e\cQ_{-1}\cG_{-2}+f\cL_{-1}\cQ_{-1}\cG_{-1}+g
        \cH_{-1}\cQ_{-1}\cG_{-1})\ket{0,\htop}^{G,Q}\end{eqnarray*}

        \begin{eqnarray}
        \htop=\ccases{\ctop-3\over 3}{2}\!,\qquad\alpha=\ccases{3\over 3-
        \ctop}{3-\ctop\over 12}\!,\qquad\gamma=\ccases{\ctop\over\ctop-3}{{1
        \over 2}{\ctop+9\over\ctop-3}}\!,\qquad
        \delta=\ccases{-{\ctop\over 3}}
        {9+\ctop\over 3-\ctop}\!\nonumber\\
        a=\ccases{{1\over 2}{15-\ctop\over 3-
        \ctop}}{-{1\over 2}}\!,\qquad e=\ccases{1}{6\over\ctop-3}\!
        ,\qquad
        f=\ccases{{1\over 2}{9\over\ctop-3}}{3\over 4}\!,\qquad
        g=\ccases{6\over 3-\ctop}{6\over 3-\ctop}\!\end{eqnarray}

        \begin{eqnarray*}
        \ket\chi_3^{(1)G}=(\alpha\cL_{-1}^2\cG_{-1}+\beta\cL_{-1}\cG_{-2}+
        \epsilon\cL_{-2}
        \cG_{-1}+\gamma\cH_{-2}\cG_{-1}+\theta\cG_{-3}+\\
        e\cH_{-1}\cG_{-2}+f
        \cH_{-1}\cL_{-1}\cG_{-1}
        +g\cH_{-1}^2\cG_{-1})\ket{0,\htop}^{G,Q}\end{eqnarray*}

        \begin{eqnarray}
        \htop=\ccases{\ctop-3\over 3}{2}\!,\qquad\alpha=\ccases{3\over 2(3-
        \ctop)}{3-\ctop\over 24}\!,\qquad\beta=\ccases{\ctop\over 2(3-\ctop)}
        {-{3\over 4}}\!\nonumber\\
        \epsilon=\ccases{\ctop-6 \over 2(3-\ctop)}{-{1
        \over 4}}\!,\qquad\gamma=\ccases{\ctop\over 2(\ctop-3)}{\ctop+9\over
        4(\ctop-3)}\!,\qquad
        \theta=\ccases{{\ctop\over 6}{\ctop-6\over 3-\ctop}}{27-
        \ctop\over 4(3-\ctop)}\!\nonumber\\
        e=\ccases{1}{6\over\ctop-3}\!,\qquad
        f=\ccases{9\over 2(\ctop-3)}{3\over 4}\!,\qquad
        g=\ccases{3\over 3-\ctop}{3\over 3-\ctop}\!
        \end{eqnarray}

        \begin{eqnarray*}
        \ket\chi_3^{(-1)Q}=(\alpha\cL_{-1}^2\cQ_{-1}+\beta\cL_{-1}\cQ_{-2}
        +\epsilon\cL_{-2}\cQ_{-1}+\gamma\cH_{-2}\cQ_{-1}\\
        +\theta\cQ_{-3}+e\cH_{-1}\cQ_{-2}+
        f\cH_{-1}\cL_{-1}\cQ_{-1}
        +g\cH_{-1}^2\cQ_{-1})\ket{0,\htop}^{G,Q}\end{eqnarray*}

        \begin{eqnarray}
        \htop=\ccases{3-2\ctop\over 3}{-{\ctop+6\over 3}}\!,\qquad
        \alpha=\ccases{9\over 3-\ctop}{3-\ctop\over 4}\!,\qquad
        \beta=\ccases{3\ctop\over 3-\ctop}{-{9\over 2}}\!\nonumber\\
        \epsilon=\ccases{18-3\ctop\over\ctop-3}{-{3\over 2}}\!,\qquad
        \gamma=\ccases{-9}{\ctop^2+12\ctop+27\over 4(3-\ctop)}\!,\qquad
        \theta=\ccases{\ctop^2-6\ctop\over 3-\ctop}
        {81-3\ctop\over 2(3-\ctop)}\!\nonumber\\
        e=\ccases{9\ctop-18\over 3-\ctop}{9\ctop+45\over 2(3-\ctop)}\!
        ,\qquad
        f=\ccases{45\over 3-\ctop}{-{\ctop+6\over 2}}\!,\qquad
        g=\ccases{54\over 3-\ctop}{\ctop^2+12\ctop+27\over 4(3-\ctop)}\!
        \end{eqnarray}

\vskip .2in

At level 1 there are no subsingular vectors because $(r,s)=(1,2)$, \ie\
the only possible value of $s$ is 2. At level 2 the singular vectors
$\kc_2^{(1)G}$ and $\kc_2^{(0)Q}$, for $\htop_{1,4}^{(1)}=1$, are
subsingular in the complete Verma module $V(\ket{0,1}^G)$. That is,
$\ket{\hat\chi}_{2,\ket{0,1}^G}^{(1)G}$ and
$\ket{\hat\chi}_{2,\ket{0,1}^G}^{(0)\hat{Q}}$ given by

\BE
\ket{\hat\chi}_{2,\ket{0,1}^G}^{(1)G}=(\cG_{-2}+
{\ctop-3\over6}\cL_{-1}\cG_{-1}-\cH_{-1}\cG_{-1})\ket{0,1}^G
\EE

\noi
and

\BE
\ket{\hat\chi}_{2,\ket{0,1}^G}^{(0)\hat{Q}}=(\cL_{-2}+
{\ctop-3\over6}\cL_{-1}^2-\cH_{-1}\cL_{-1}-
{1\over2}\cQ_{-1}\cG_{-1})\ket{0,1}^G
\EE

\begin{figure}[!ht]
\begin{center}
\leavevmode
\epsfxsize= 12cm
\epsffile{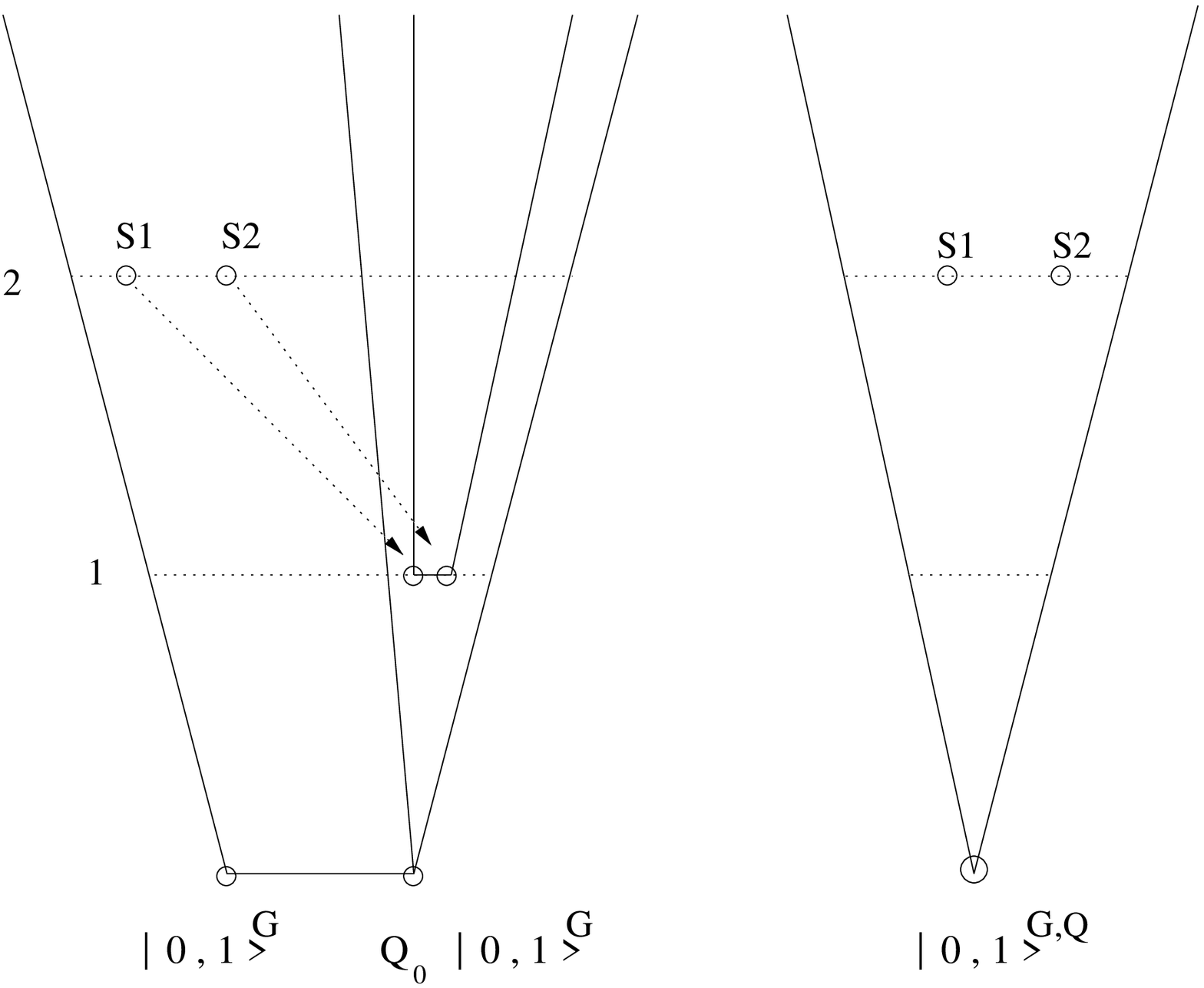}
\caption{$S1$ and $S2$ represent the subsingular vectors in the complete
  Verma module $V(|0,1>^{G})$, on the left, which become the singular
  vectors $\kc^{(1)G}_{2}$ and $\kc^{(0)Q}_{2}$ in the chiral
  Verma module $V(|0,1>^{G,Q})$, on the right. $V(|0,1>^{G,Q})$ is the
  quotient of $V(|0,1>^{G})$ by the submodule generated by the
  level-zero singular vector $\cQ_{0}|0,1>^{G}$. The subsingular
  vectors sit outside the (incomplete) Verma modules built on the
  singular vectors at levels zero and 1, to which they descend.}
\label{fig:subsing}
\end{center}
\end{figure}

\noi are subsingular vectors (the hat denotes that they are not h.w.
vectors while $\hat{Q}$ denotes that the vector is not $\cQ_0$-closed
anymore).  The positive mode $\cQ_1$ brings
$\ket{\hat\chi}_{2,\ket{0,1}^G}^{(1)G}$ ``down" to the singular vector
$\ket{\chi}_{1,\ket{0,1}^G}^{(0)G}=\cG_{-1}\cQ_0\ket{0,1}^G$ (for
$\ctop\neq9$ since $\cQ_1\ket{\hat\chi}_{2,\ket{0,1}^G}^{(1)G}=
{(9-\ctop)\over6}\cG_{-1}\cQ_0\ket{0,1}^G$), which is a secondary
singular vector built on the level-zero singular vector
$\ket{\chi}_{0,\ket{0,1}^G}^{(-1)G,Q}= \cQ_0\ket{0,1}^G$.  Similarly,
the positive modes $\cL_1$ and $\cH_1$ bring
$\ket{\hat\chi}_{2,\ket{0,1}^G}^{(0)\hat{Q}}$ down to the singular
vector $\ket{\chi}_{1,\ket{0,1}^G}^{(0)G}$. From this one it is not
possible to reach the subsingular vectors acting with the negative
modes, however. That is, the subsingular vectors sit outside the
(incomplete) Verma modules built on top of the singular vectors
$\ket{\chi}_{1,\ket{0,1}^G}^{(0)G}$ and
$\ket{\chi}_{0,\ket{0,1}^G}^{(-1)G,Q}$.  This is shown in
Figure~\ref{fig:subsing}.

Similarly, the singular vectors $\kc_2^{(0)G}$ and $\kc_2^{(-1)Q}$,
for $\htop_{1,4}^{(0)}=-{\ctop+3\over3}$, are subsingular in the
complete Verma module $V(\ket{0,-{\ctop+3\over3}}^Q)$. Both of them
descend to the singular vector
$\kc_{1,\ket{0,-{\ctop+3\over3}}^Q}^{(0)Q}=
\cQ_{-1}\cG_0\ket{0,-{\ctop+3\over3}}^Q$, for $\ctop\neq9$, which is a
secondary singular vector built on the level-zero singular vector
$\kc_{0,\ket{0,-{\ctop+3\over3}}^Q}^{(1)G,Q}=
\cG_0\ket{0,-{\ctop+3\over3}}^Q$.

At level 3 the singular vectors $\kc_3^{(1)G}$ and
$\kc_3^{(0)Q}$, for $\htop_{1,6}^{(1)}=2$, are subsingular in the
complete Verma module $V(\ket{0,2}^G)$  whereas the singular vectors
$\kc_3^{(0)G}$ and $\kc_3^{(-1)Q}$, for $\htop_{1,6}^{(0)}=-{\ctop+6\over3}$,
are subsingular in the complete Verma module $V(\ket{0,-{\ctop+6\over3}}^Q)$.
In these cases the subsingular vectors do not descend to any 
secondary singular vectors but to the level-zero singular vectors
 $\ket{\chi}_{0,\ket{0,2}^G}^{(-1)G,Q}=\cQ_0\ket{0,2}^G$ and
$\kc_{0,\ket{0,-{\ctop+6\over3}}^Q}^{(1)G,Q}=
\cG_0\ket{0,-{\ctop+6\over3}}^Q$, for $\ctop\neq9$ as before. 

\vskip .3in

\section{Families of Topological Singular Vectors in Complete Verma
  Modules}\lvm

\subsection{Family Structure}\lvm

Let us apply the results of sections 2 and 3 to derive the family 
structure of the singular vectors in complete Verma modules.
One can expect a much richer structure than in the case
of chiral Verma modules since, on the one hand, there are many more types
of singular vectors, twenty-nine versus four, and, on the other,
there is an additional spectral flow mapping, $\cU_{\pm1}$, which
can extend the four-member subfamilies of singular vectors given by
the actions of $\,\cG_0$, $\cQ_0$ and $\cA\,$ (two-member subfamilies rather
if the singular vectors are chiral)\footnote{We remind the reader
that chiral singular vectors $\kc_l^{(q)G,Q}$, \ie\, annihilated by
both $\cG_0$ and $\cQ_0$, do not exist in chiral Verma modules, they exist
in generic Verma modules $V(\ket{\D,\htop}^G)$ and
$V(\ket{\D,\htop}^Q)$ with $l=-\D$, and at level zero also in no-label
Verma modules $V(\ket{0,\htop})$:
$\kc_{0,\ket{0,\htop}}^{(0)G,Q}=\cQ_0\cG_0\ket{0,\htop}$.}. The 
label-exchange operator $\hat\cA_0\,$ can be composed with 
$\cA\,$ and $\,\cU_{\pm1}$ to produce the same transformations as 
these ones but connecting the two sets of topological generators and
states labeled by (1) and (2). In the diagrams presented below we will 
not consider $\hat\cA_0\,$, for simplicity, although this possibility 
must be taken into account, as we did in the case of chiral Verma 
modules in last section (diagram \req{dia12}).

Let us start with the generic Verma modules
$V(\ket{\D,\htop}^G)$ and $V(\ket{\D,\htop}^Q)$. The types of
singular vectors allowed by the cascade effect
are twenty, as shown in tables \req{tabl2} and \req{tabl3}:
twelve types of generic singular vectors (six \Gn-closed and
six \Qn-closed), plus four types of chiral singular vectors, plus four
types of no-label singular vectors. The twelve types of generic
singular vectors and the four types of chiral singular vectors
can be mapped to singular vectors of the
NS algebra, as we will see. Therefore one can write down general
construction formulae for them, using the construction formulae for 
the NS singular vectors \cite{Doerr1} \cite{Doerr2}. The four types 
of no-label singular vectors, however, are not related to NS singular
vectors. These twenty types of singular vectors exist already at level 1
(see Appendix B).

\vskip .17in
\noi
{\it First kind of generic families of topological singular vectors}

Let us analyze how the generic and chiral types of singular vectors are
organized into families. The key fact is that using $\cU_{\pm1}$ one can
extend a topological subfamily of four (or two) singular
vectors given by the actions of \Gn, \Qn\ and $\cA$, resulting in, at least,
two subfamilies
located in four different Verma modules, two each. These two
subfamilies will be denoted as ``the skeleton-family".

Let us start with an uncharged singular vector of type
 $\kc^{(0)G}_{l,{\kp^G}}{\ }$ in the Verma module
$V(\ket{\D,\htop}^G)$. As diagram \req{diab0} shows,
the members of its skeleton-family are, in the most general case,
singular vectors of the types $\kc^{(0)G}_{l,{\kp^G}}{\ }$ and
$\kc^{(-1)Q}_{l,{\kp^G}}{\ }$, in both the Verma modules 
$V(\ket{\D,\htop}^G)$ and $V(\ket{\D-\htop,-\htop}^G)$, and singular 
vectors of the types $\kc^{(0)Q}_{l,{\kp^Q}}{\ }$ 
and $\kc^{(1)G}_{l,{\kp^Q}}{\ }$, in both the Verma modules 
$V(\ket{\D-\htop,\htop-\ctop/3}^Q)$ and $V(\ket{\D,-\htop-\ctop/3}^Q)$. 

In the special case $\,\D=-l\,$ the conformal weight of the singular 
vectors in the lower subdiagram is zero, so that the corresponding 
arrows $\Qz$, $\Gz$ cannot be reversed, producing
secondary chiral singular vectors
$\kc_{l,\,\ket{\D,\,\htop}^G}^{(-1)G,Q}$ and
$\kc_{l,\,\ket{\D,-\htop-\ctop/3}^Q}^{(1)G,Q}$ on the right-hand side,
at level zero with respect to the singular vectors on the left-hand side.

In the special case $\D-\htop=-l$ the conformal weight of the 
singular vectors in the upper subdiagram is zero so that 
one of the following two possibilities must happen:

a) The singular vectors denoted by
$\kc_{l,\,\ket{\D-\htop,-\htop}^G}^{(0)G}$ and
$\kc_{l,\,\ket{\D-\htop,\,\htop-\ctop/3}^Q}^{(0)Q}$ 
turn out to be chiral, \ie\ of types 
$\kc_{l,\,\ket{\D-\htop,-\htop}^G}^{(0)G,Q}$ and
$\kc_{l,\,\ket{\D-\htop,\,\htop-\ctop/3}^Q}^{(0)G,Q}$ 
instead, so that the
singular vectors $\kc_{l,\,\ket{\D-\htop,-\htop}^G}^{(-1)Q}$ and
$\kc_{l,\,\ket{\D-\htop,\,\htop-\ctop/3}^Q}^{(1)G}$ are absent.

b) The singular vectors 
$\kc_{l,\,\ket{\D-\htop,-\htop}^G}^{(0)G}$ and
$\kc_{l,\,\ket{\D-\htop,\,\htop-\ctop/3}^Q}^{(0)Q}$ 
are not chiral, therefore the corresponding arrows \Qn\ ,  \Gn\ cannot
be reversed producing secondary chiral singular vectors 
$\kc_{l,\,\ket{\D-\htop,-\htop}^G}^{(-1)G,Q}$ and
$\kc_{l,\,\ket{\D-\htop,\,\htop-\ctop/3}^Q}^{(1)G,Q}$ at level zero 
with respect to the singular vectors on the left-hand side.

\vskip .2in

\def\btggo  {\mbox{$\kc_{l,\, \ket{\D,\,\htop}^G}^{(0)G} $}}
\def\btqqo  {\mbox{$\kc_{l,\, \ket{\D,-\htop-{\ctop\over3}}^Q}^{(0)Q} $}}
\def\btqqm  {\mbox{$\kc_{l,\, \ket{\D,\,\htop}^G}^{(-1)Q} $}}
\def\btggp  {\mbox{$\kc_{l,\, \ket{\D,-\htop-{\ctop\over3}}^Q}^{(1)G} $}}
\def\bbggo  {\mbox{$\kc_{l,\, \ket{\D-\htop,-\htop}^G}^{(0)G} $}}
\def\bbqqo 
 {\mbox{$\kc_{l,\, \ket{\D-\htop,\,\htop-{\ctop\over3}}^Q}^{(0)Q} $}}
\def\bbqqm  {\mbox{$\kc_{l,\, \ket{\D-\htop,-\htop}^G}^{(-1)Q} $}}
\def\bbggp 
 {\mbox{$\kc_{l,\, \ket{\D-\htop,\,\htop-{\ctop\over3}}^Q}^{(1)G} $}}

  \begin{equation}
  \begin{array}{rcl}
 \cU_1\,\uparrow\ & \\[3 mm]
  \bbggo &
  \stackrel{\Qz}{\mbox{------}\!\!\!\longrightarrow}
  & \bbqqm \\[3 mm]
   \cA\,\updownarrow\ && \ \updownarrow\, \cA
  \\[3 mm]  \bbqqo \! & \stackrel{\Gz}
  {\mbox{------}\!\!\!\longrightarrow} & \! \bbggp
  \\[5 mm]  \cU_1\,\uparrow\ &  \\[4mm]
   \btggo &
  \stackrel{\Qz}{\mbox{------}\!\!\!\longrightarrow}
  & \btqqm \\[3 mm]
   \cA\,\updownarrow\ && \ \updownarrow\, \cA
  \\[3 mm]   \btqqo \! & \stackrel{\Gz}
  {\mbox{------}\!\!\!\longrightarrow} & \! \btggp
  \end{array} \label{diab0} \end{equation}

\vskip .2in

The upper row and the lower row of diagram \req{diab0} are
connected by $\cU_1$, as indicated by the arrow on top,
since $\cU_1 \cA \, \cU_1 \cA = \oI$.
It has therefore the topology of a circle. There are no 
arrows $\cU_{\pm1}$ coming in or out of the singular vectors on the
right-hand side (unless they turn out to be chiral) 
because these types do not transform into singular vectors under
$\cU_{\pm1}$. In the most general case, however, it
will be possible to go beyond this limitation and attach more subfamilies
to the skeleton-family, as we will see.

If we start with an uncharged chiral singular vector in the Verma module
$V(\ket{\D,\htop}^G)$, with $\D=-l$, then the skeleton-family is as
in diagram \req{diab0} but with the lower subdiagram reduced to the couple
of chiral singular vectors $\kc_{l,\,\ket{-l,\,\htop}^G}^{(0)G,Q}$ and
$\kc_{l,\,\ket{-l,-\htop-\ctop/3}^Q}^{(0)G,Q}$. In addition, if $\htop=0$
the singular vectors in the upper subdiagram have zero conformal weight 
also, so that this subdiagram contains chiral singular vectors as well. 

The first kind of generic families 
contains therefore the four generic types of singular vectors 
$\kc^{(0)G}_{{\kp^G}}{\ }$, $\kc^{(0)Q}_{{\kp^Q}}{\ }$,
$\kc^{(1)G}_{{\kp^Q}}{\ }$ and $\kc^{(-1)Q}_{{\kp^G}}\,$,
and the four chiral types of singular vectors
as particular cases, all of them at the same level $l$. The untwisting 
of the singular vectors of types $\kc^{(0)G}_{{\kp^G}}{\ }$ and
$\kc^{(0)G,Q}_{{\kp^G}}{\ }$ produces uncharged singular vectors
of the NS algebra. As a result, all the singular vectors of the first
kind of generic families can be mapped to uncharged NS singular vectors.

\vskip .17in
\noi
{\it Second kind of generic families of topological singular vectors}

Now let us take a charged singular vector of type
 $\kc^{(1)G}_{l,{\kp^G}}{\ }$ in the Verma module $V(\ket{\D,\htop}^G)$.
 As diagram \req{diab1} shows,
the other members of its skeleton-family are, in the most general case,
singular vectors of the types:
$\kc^{(0)Q}_{l,{\kp^G}}{\ }$ in the same Verma module,
$\kc^{(1)Q}_{l-1,{\kp^Q}}{\ }$ and $\kc^{(2)G}_{l-1,{\kp^Q}}{\ }$ in the
Verma module $V(\ket{\D-\htop,\htop-\ctop/3}^Q)$,
$\kc^{(-1)Q}_{l,{\kp^Q}}{\ }$ and $\kc^{(0)G}_{l,{\kp^Q}}{\ }$ in the
Verma module $V(\ket{\D,-\htop-\ctop/3}^Q)$, and
$\kc^{(-1)G}_{l-1,{\kp^G}}{\ }$ and $\kc^{(-2)Q}_{l-1,{\kp^G}}{\ }$ in the
Verma module $V(\ket{\D-\htop,-\htop}^G)$. 
For $l=1$ the upper subdiagram reduces to a chiral couple 
at level zero (there are no level-zero singular vectors with $|q|=2$).

\vskip .2in

\def\btggob  {\mbox{$\kc_{l,\, \ket{\D,\,\htop}^G}^{(1)G} $}}
\def\btqqob  {\mbox{$\kc_{l,\, \ket{\D,-\htop-{\ctop\over3}}^Q}^{(-1)Q} $}}
\def\btqqmb  {\mbox{$\kc_{l,\, \ket{\D,\,\htop}^G}^{(0)Q} $}}
\def\btggpb  {\mbox{$\kc_{l,\, \ket{\D,-\htop-{\ctop\over3}}^Q}^{(0)G} $}}
\def\bbggob  {\mbox{$\kc_{l-1,\, \ket{\D-\htop,-\htop}^G}^{(-1)G} $}}
\def\bbqqob 
 {\mbox{$\kc_{l-1,\, \ket{\D-\htop,\,\htop-{\ctop\over3}}^Q}^{(1)Q} $}}
\def\bbqqmb 
 {\mbox{$\kc_{l-1,\, \ket{\D-\htop,-\htop}^G}^{(-2)Q} $}}
\def\bbggpb 
 {\mbox{$\kc_{l-1,\, \ket{\D-\htop,\,\htop-{\ctop\over3}}^Q}^{(2)G} $}}

  \begin{equation} \begin{array}{rcl}
   \cU_1\,\uparrow\ & \\[3 mm]
  \bbggob &
  \stackrel{\Qz}{\mbox{------}\!\!\!\longrightarrow}
  & \bbqqmb \\[3 mm]
   \cA\,\updownarrow\ && \ \updownarrow\, \cA
  \\[3 mm]  \bbqqob \! & \stackrel{\Gz}
  {\mbox{------}\!\!\!\longrightarrow} & \! \bbggpb
  \\[5 mm] \cU_1\,\uparrow\ & \\[4 mm]
  \btggob &
  \stackrel{\Qz}{\mbox{------}\!\!\!\longrightarrow}
  & \btqqmb \\[3 mm]
   \cA\,\updownarrow\ && \ \updownarrow\, \cA
  \\[3 mm]  \btqqob \! & \stackrel{\Gz}
  {\mbox{------}\!\!\!\longrightarrow} & \! \btggpb
 \end{array} \label{diab1} \end{equation}

\vskip .2in

We can repeat the same analysis as for the
previous skeleton-family, for the particular cases $\D=-l$ and 
$\D-\htop=1-l$. Now we have to take into account, however, that the
cascade effect forbids chiral
singular vectors 
of types $\kc^{(1)G,Q}_{{\kp^G}}{\ }$,
$\kc^{(-1)G,Q}_{{\kp^Q}}{\ }$, $\kc^{(2)G,Q}_{{\kp^G}}{\ }$ and
$\kc^{(-2)G,Q}_{{\kp^Q}}{\ }$ which otherwise 
could appear associated to the skeleton-family \req{diab1}.

The second kind of generic families contains therefore the remaining
eight generic types of singular vectors:
$\kc^{(0)G}_{{\kp^Q}}{\ }$, $\kc^{(0)Q}_{{\kp^G}}{\ }$,
$\kc^{(1)G}_{{\kp^G}}{\ }$ and $\kc^{(-1)Q}_{{\kp^Q}}{\ }$, at level $l$,
plus $\kc^{(1)Q}_{{\kp^Q}}{\ }$, $\kc^{(-1)G}_{{\kp^G}}{\ }$,
$\kc^{(2)G}_{{\kp^Q}}{\ }$ and $\kc^{(-2)Q}_{{\kp^G}}{\ }$ at level 
$l-1$, plus the four chiral types of singular vectors 
as particular cases. The untwisting of 
the singular vectors of types $\kc^{(1)G}_{{\kp^G}}{\ }$ and
$\kc^{(-1)G,Q}_{{\kp^G}}{\ }$ produces charged singular vectors
of the NS algebra. As a result, all the singular vectors of the second
kind of generic families can be mapped to charged NS singular vectors.

Observe that the four types of chiral singular vectors can appear as
particular cases, both in the first kind and in the second kind of
generic families. They produce therefore the intersection of these
two kinds of families, which in the absence of chiral singular
vectors are completely disconnected from each other.

\vskip .17in
\noi
{\it Enlargements of the generic families}

When the conformal weight of the primary state
$\kp^G$ or $\kp^Q$ is different
from zero the highest weight of the Verma module is degenerate, 
so that one has the choice to express the primaries of
type $\kp^Q$ as $\kp^Q= \cQ_0\,\kp^G$
(\ie\ $\ket{\D,\htop}^Q = \cQ_0\,\ket{\D,\htop+1}^G$), or the other
way around using \Gn\ .
As a result, for $\D\neq0$ the singular vectors of types
$\kc^{(q)G}_{\ket{\D,\,\htop}^Q}$ and $\kc^{(q)Q}_{\ket{\D,\,\htop}^G}$
can be expressed as singular vectors of types
$\kc^{(q-1)G}_{\ket{\D,\,\htop+1}^G}$ and 
$\kc^{(q+1)Q}_{\ket{\D,\,\htop-1}^Q}$,
respectively (observe that $\htop \to \htop \pm1$ and $q \to q\mp1$),
and the uncharged chiral singular vectors (for which $\D\neq0$) can
be expressed always as charged chiral singular vectors.

This brings about important consequences because the spectral flow mappings
$\cU_{\pm1}$ distinguish drastically between a given type of singular
vector and the equivalent type.
On the one hand, there are arrows $\cU_{\pm1}$ coming in or out of
the singular vectors of types $\kc_{\kp^G}^{(q)G}$ and
$\kc_{\kp^Q}^{(q)Q}$ while there are none for the 
singular vectors of types $\kc_{\kp^Q}^{(q)G}$ and $\kc_{\kp^G}^{(q)Q}$.
On the other hand, there are different arrows $\cU_{\pm1}$ coming in or 
out of the uncharged chiral singular vectors and of their charged partners.

Therefore if the conformal weights of the four Verma modules 
involved in diagrams \req{diab0} and \req{diab1} are different
from zero, \ie\ $\D \not= 0$ and $\D \not=\htop$, then one can express
the corresponding singular vectors in the way shown in diagrams 
\req{diab2} and \req{diab3}. In this case the first kind
of skeleton-families can be viewed as those consisting of
the uncharged singular vectors of the types
$\kc^{(0)G}_{{\kp^G}}{\ }$ and $\kc^{(0)Q}_{{\kp^Q}}\,$, including the chiral
ones $\kc^{(0)G,Q}_{{\kp^G}}{\ }$ and $\kc^{(0)G,Q}_{{\kp^Q}}\,$
as particular cases,
while the second kind of skeleton-families can be viewed as
those consisting of the
charged singular vectors of the types
$\kc^{(1)Q}_{{\kp^Q}}{\ }$, $\kc^{(1)G}_{{\kp^G}}{\ }$,
$\kc^{(-1)G}_{{\kp^G}}{\ }$ and $\kc^{(-1)Q}_{{\kp^Q}}\,$, including the
chiral ones $\kc^{(1)G,Q}_{{\kp^Q}}{\ }$ and $\kc^{(-1)G,Q}_{{\kp^G}}{\ }$
as particular cases.

Moreover, now one can attach arrows $\cU_{\pm1}$ to each of these
singular vectors, and complete subfamilies with them, since
they belong to the appropriate types.
To be precise, using $\cU_{\pm1}$ one can attach a
subfamily of singular vectors, \ie\ either a four-member ``box" 
or a chiral couple, to every pair of
singular vectors of types $\kc^{(q)G}_ {\kp^G}$  and
$\kc^{(-q)Q}_ {\kp^Q}$ connected by $\cA$. These arrows $\cU_{\pm1}$,
that we draw horizontally on the right of diagrams
\req{diab2} and \req{diab3}, do not return to the left of the
diagrams, consequently, unlike in the case of the vertical upper-most
arrows which return to the bottom of the diagrams.

Observe that the chiral singular vectors produce the intersection of the
two kinds of skeleton-families 
also because the uncharged chiral singular vectors
are equivalent to charged chiral singular vectors.

\vskip .17in
\noi
{\it Sequences of attachements}

Starting with the subfamily which contain the \Gn-closed 
singular vector $\kc^{(q)G}_{l,\ket{\D,\htop}^G}$ 
one can distinguish two different sequences
of attachements: the sequence starting
as $\cU_1{\ }\ket{\D,\htop}^G \to \ket{\D-\htop,\htop-\ctop/3}^Q$,
which will be denoted as sequence A,
and the sequence starting as
$\cU_1{\ }\ket{\D,-\htop-\ctop/3+1}^G \to
\ket{\D+\htop+\ctop/3-1,-\htop-2\ctop/3+1}^Q$, which will be
denoted as sequence B. 

Let us analyze in some detail sequences A and B,
with the help of diagrams \req{diab0}, \req{diab1},
\req{diab2} and \req{diab3}, focusing on the
singular vectors of type $\kc^{(q)G}_{\kp^G}$. Sequence A starts

\def\btgqo  {\mbox{$\kc_{l,\, \ket{\D,\,\htop}^G}^{(0)G} $}}
\def\btqgo  {\mbox{$\kc_{l,\, \ket{\D,-\htop-{\ctop\over3}}^Q}^{(0)Q} $}}
\def\btqgm  {\mbox{$\kc_{l,\, \ket{\D,\,\htop-1}^Q}^{(0)Q} $}}
\def\btgqp  {\mbox{$\kc_{l,\, \ket{\D,-\htop-{\ctop\over3}+1}^G}^{(0)G} $}}
\def\bbgqo  {\mbox{$\kc_{l,\, \ket{\D-\htop,-\htop}^G}^{(0)G} $}}
\def\bbqgo 
 {\mbox{$\kc_{l,\, \ket{\D-\htop,\,\htop-{\ctop\over3}}^Q}^{(0)Q} $}}
\def\bbqgm  {\mbox{$\kc_{l,\, \ket{\D-\htop,-\htop-1}^Q}^{(0)Q} $}}
\def\bbgqp 
 {\mbox{$\kc_{l,\, \ket{\D-\htop,\,\htop-{\ctop\over3}+1}^G}^{(0)G} $}}

  \begin{equation} \begin{array}{rcl}
   \cU_1\,\uparrow\ & \\[3 mm]
 \bbgqo &
  \stackrel{\Qz}{\mbox{------}\!\!\!\longrightarrow}
  & \bbqgm \qquad  \stackrel{\cU_{-1}} {\longrightarrow} \\[3 mm]
  {\rm for} \D \not=\htop \qquad \ \ \ \ \qquad \qquad
   \cA\,\updownarrow\ && \ \updownarrow\, \cA
  \\[3 mm]  \bbqgo \! & \stackrel{\Gz}
  {\mbox{------}\!\!\!\longrightarrow} & \! \bbgqp \qquad \stackrel{\cU_{1}}
  {\longrightarrow} \\[5 mm] \cU_1\,\uparrow\ & \\[4 mm]
  \btgqo &
  \stackrel{\Qz}{\mbox{------}\!\!\!\longrightarrow}
  & \btqgm \qquad \stackrel{\cU_{-1}} {\longrightarrow} \\[3 mm]
  {\rm for} \D \not=0 \qquad \ \ \ \ \qquad \qquad
   \cA\,\updownarrow\ && \ \updownarrow\, \cA
  \\[3 mm]  \btqgo \! & \stackrel{\Gz}
  {\mbox{------}\!\!\!\longrightarrow} & \! \btgqp
   \qquad \stackrel{\cU_{1}} {\longrightarrow}
 \end{array} \label{diab2} \end{equation}

\vskip .2in

\noi
with the mapping ${\ }\cU_1 \, \kc^{(q)G}_{l,\,\ket{\D,\,\htop}^G} \to
\kc^{(q)Q}_{l-q,\,\ket{\D-\htop,\,\htop-\ctop/3}^Q}{\ } $
(a chiral singular vector $\, \kc^{(q)G,Q}_{l,\,\ket{\D,\,\htop}^G}$ 
can also be taken instead).
Then using $\cA,\, \Gz$ and \Qn\ one obtains a
second subfamily attached to the former
by $\cU_1$ (more precisely $\cU_{\pm1}$ since we
can invert the arrow).
Next, one looks for further singular vectors
of type $\kc^{(q)G}_{\kp^G}$, in the second subfamily,
 to which attach new arrows $\cU_1$.
The upper-most singular vector of this type does not lead to a new
subfamily, but to the previous subfamily, because
${\ }\cU_1\cA \, \cU_1\cA = \oI{\ }$, as pointed 
out before. For this reason, if the second
subfamily consists of a chiral couple this sequence is finished. If the
second subfamily is a four-member ``box", like in the diagrams
above, there are three different possibilities depending on whether 
the conformal weights $\D-\htop$ and $\D-\htop+l-q$, of the primary states
and the singular vectors, respectively, are zero or different from zero. 
If $\D-\htop \not= 0$ and  $\D-\htop \not= q-l$, then the 
singular vector of type

\def\btgqob  {\mbox{$\kc_{l,\, \ket{\D,\,\htop}^G}^{(1)G} $}}
\def\btqgob  {\mbox{$\kc_{l,\, \ket{\D,-\htop-{\ctop\over3}}^Q}^{(-1)Q} $}}
\def\btqgmb  {\mbox{$\kc_{l,\, \ket{\D,\,\htop-1}^Q}^{(1)Q} $}}
\def\btgqpb  {\mbox{$\kc_{l,\, \ket{\D,-\htop-{\ctop\over3}+1}^G}^{(-1)G} $}}
\def\bbgqob  {\mbox{$\kc_{l-1,\, \ket{\D-\htop,-\htop}^G}^{(-1)G} $}}
\def\bbqgob  
{\mbox{$\kc_{l-1,\, \ket{\D-\htop,\,\htop-{\ctop\over3}}^Q}^{(1)Q} $}}
\def\bbqgmb  {\mbox{$\kc_{l-1,\, \ket{\D-\htop,-\htop-1}^Q}^{(-1)Q} $}}
\def\bbgqpb
{\mbox{$\kc_{l-1,\, \ket{\D-\htop,\,\htop-{\ctop\over3}+1}^G}^{(1)G} $}}

   \begin{equation} \begin{array}{rcl}
   \cU_1\, \uparrow\ & \\[3 mm]
  \bbgqob &
  \stackrel{\Qz}{\mbox{------}\!\!\!\longrightarrow}
  & \bbqgmb \qquad \stackrel{\cU_{-1}} {\longrightarrow}\\[3 mm]
  {\rm for} \D \not=\htop \qquad \ \ \ \ \qquad  \qquad
   \cA\,\updownarrow\ && \ \updownarrow\, \cA
  \\[3 mm]  \bbqgob \! & \stackrel{\Gz}
  {\mbox{------}\!\!\!\longrightarrow} & \! \bbgqpb \qquad \stackrel{\cU_1}
  {\longrightarrow}  \\[5 mm] \cU_1\,\uparrow\ & \\[4 mm]
   \btgqob &
  \stackrel{\Qz}{\mbox{------}\!\!\!\longrightarrow}
  & \btqgmb \qquad \stackrel{\cU_{-1}} {\longrightarrow}\\[3 mm]
  {\rm for} \D \not=0 \qquad \ \ \ \ \qquad  \qquad
   \cA\,\updownarrow\ && \ \updownarrow\, \cA
  \\[3 mm]  \btqgob \! & \stackrel{\Gz}
  {\mbox{------}\!\!\!\longrightarrow} & \! \btgqpb \qquad
  \stackrel{\cU_{1}} {\longrightarrow}
  \end{array} \label{diab3} \end{equation}

\vskip .2in

\noi
$\kc^{(q+1)G}_{\kp^Q}$
can be expressed as a singular vector of type $\kc^{(q)G}_{\kp^G}$,
as in diagrams \req{diab2} and \req{diab3}. Namely,
${\ } \kc^{(q+1)G}_{l-q,\,\ket{\D-\htop,\,\htop-\ctop/3}^Q}{\ } =
\kc^{(q)G}_{l-q,\,\ket{\D-\htop,\,\htop-\ctop/3+1}^G}{\ }$.
Attaching an arrow $\cU_1$ to this singular vector
one obtains a new subfamily different from the previous one,
although involving the same types of singular vectors which are already
in the skeleton-family. If $\D-\htop =0$ and $\D-\htop \not= q-l$, then the
singular vector of type $\kc^{(q+1)G}_{\kp^Q}$ cannot be 
expressed as a singular vector of type $\kc^{(q)G}_{\kp^G}$, and the 
sequence is finished since one cannot attach any more arrows $\cU_{\pm1}$.
Finally, if $\D-\htop = q-l$ (and the singular vectors on the left-hand 
side are not chiral, as we assume) then \Qn\ and \Gn\ produce secondary 
chiral singular vectors on the right-hand side,
to which one can attach new arrows $\cU_{\pm1}$. 

Repeating the same reasoning $n$ times one finds that sequence A continues 
or stops after $n$ steps depending on whether the conformal weights
$\hat\D$ and/or $\hat\D +l-nq$ are zero or different from zero, where

\BE \hat\D = \D-n\htop-n(n-1)({3-\ctop\over6}) \ \,, \qquad n\in \oZ^+
\label{condDh} \EE

\noi
Namely, if $\hat\D\not=0$ and $\hat\D\not=nq-l$, the sequence of 
singular vectors goes on attaching an arrow $\cU_1$ to the \Gn-closed
singular vector ${\ }\kc^{(q)G}_{l-nq,\,\ket{\hat\D,\,\hat\htop}^G}{\ }$ 
at level $\,l-nq$, with $\,\hat\htop=\htop+n({3-\ctop\over3})$. 
If $\hat\D=0$ and $\hat\D\not=nq-l$ the sequence is finished, whereas 
if $\hat\D=nq-l$ the sequence may finish depending on whether or not
the \Gn-closed singular vector ``becomes" chiral, \ie\ of type
${\ }\kc^{(q)G,Q}_{l-nq,\,\ket{\hat\D,\,\hat\htop}^G}{\ }$.
Starting with charge $q=1$ singular
 vectors $\kc^{(1)G}_{l,\ket{\D,\htop}^G}$ this
sequence can finish before one of these conditions is met, since 
$l-nq$ must be positive. Starting with charge $q=-1$ or with
uncharged singular vectors, however, this sequence of mappings and
attachements of four-member subfamilies only ends if $\hat\D=0$ and
$\hat\D\not=nq-l$ or if $\hat\D=nq-l$ and the corresponding
\Gn-closed singular vector turns out to be chiral. In addition
this sequence has no loops, although individually a given Verma module can
appear more than once, in different subfamilies of the sequence (see
Appendix C). {\it Therefore, depending on the values of 
$\D,{\ }\htop$, and $\ctop$, sequence A can give rise to an infinitely 
large family of singular vectors}. 

Sequence B starts with the mapping 
${\ }\Gz\,\cA{\ }\kc^{(q)G}_{l,\,\ket{\D,\,\htop}^G} \to
\kc^{(-q)G}_{l,\,\ket{\D,-\htop-\ctop/3+1}^G}{\ }$, 
with $\D\not= 0$, followed by the mapping 
${\ }\cU_1 {\ }\kc^{(-q)G}_{l,\,\ket{\D,-\htop-\ctop/3+1}^G} \to
\kc^{(-q)Q}_{l+q,\,\ket{\D+\htop+\ctop/3-1,
-\htop-2\ctop/3+1}^Q}{\ } $
(diagrams \req{diab2} and \req{diab3}). Following similar
reasoning as for sequence A one finds that sequence B continues after 
$n$ steps if the condition $\hat\D\not=0$, $\hat\D\not=-(nq+l)$, with
\BE \hat\D = \D+n\htop-n(n+1)({3-\ctop\over6}) \ \,, \qquad n\in\oZ^++\{0\}
\label{condDhb} \EE

\noi
is satisfied. Then one can attach an arrow
$\cU_1$ to the singular vector 
${\ }\kc^{(-q)G}_{l+nq,\,\ket{\hat\D,\,\hat\htop}^G}{\ }$ 
at level $\,l+nq$, with $\,\hat\htop=-\htop+(n+1)({3-\ctop\over3})$. 
Sequence B stops if $\hat\D=0$, $\hat\D\not=-(nq+l)$, whereas it may stop,
although not necessarily, if $\hat\D=-(nq+l)$.
Starting with charge $q=-1$ singular
vectors $\kc^{(-1)G}_{l,\ket{\D,\htop}^G}$ this
sequence can end before one of these conditions is satisfied, because
$\,l+nq\,$ must be positive. Starting with charge $q=1$ or with 
uncharged singular vectors this sequence only ends if $\hat\D=0$,  
$\hat\D\not=-(nq+l)$, or if $\hat\D=-(nq+l)$ and the corresponding 
\Gn-closed singular vector turns out to be chiral, \ie\ of type
${\ }\kc^{(-q)G,Q}_{l+nq,\,\ket{\hat\D,\,\hat\htop}^G}{\ }$. 
{\it Therefore, depending on the values of 
$\D,{\ }\htop$, and $\ctop$, sequence B can also give rise to an 
infinitely large family of singular vectors}. 

 However, unlike in the previous case, this sequence can come back
to a given Verma module, \ie\ it has loops (although these loops
do not involve necessarily the same singular vectors, as we will see). 
Namely, for 
$\htop=(n+1)({3-\ctop\over6}){\ }$ one finds $\hat\D=\D$ and
$\hat\htop=\htop$ after $n$ steps. 

Notice that sequences A and B are complementary
in that sequence A has the bound $l-nq \geq 0$, which only applies to
charge $q=1$ singular vectors, whereas sequence B has the bound
$l+nq \geq 0$, which only applies to charge $q=-1$ singular vectors.

\vskip .17in
\noi
{\it Families of no-label singular vectors in generic Verma modules}

The no-label types of singular vectors in generic Verma modules are
decoupled from
the other types in the sense that $\cA$ transforms the no-label 
singular vectors into each other, and $\cU_{\pm1}$ transforms
these singular vectors into states that are not singular. The action of
$\cG_0$ and $\cQ_0$, however, produces $\cG_0$-closed and
$\cQ_0$-closed secondary singular vectors, respectively. Therefore we
can distinguish two kinds of skeleton-families involving no-label
singular vectors: those containing
$\kc_{\kp^G}^{(0)}$ and $\kc_{\kp^Q}^{(0)}$ and those containing 
$\kc_{\kp^G}^{(-1)}$ and $\kc_{\kp^Q}^{(1)}$. However
the singular vectors of types $\kc_{\kp^Q}^{(1)}$ and $\kc_{\kp^G}^{(-1)}$
are always equivalent to singular vectors of types $\kc_{\kp^G}^{(0)}$ and
$\kc_{\kp^Q}^{(0)}$, 
respectively, so that there is only one skeleton-family left. 

\vskip .17in
\noi
{\it Families of singular vectors in no-label Verma modules}

The singular vectors in no-label Verma modules
$V(\ket{0,\htop})$ are organized into four-member families, like in the 
case of chiral Verma modules. However, in no-label Verma modules
there are two different kinds of four-member families instead of one,
as shown in diagram \req{diab4}. These two kinds of families exist
already at level 1
(see Appendix B). In addition, at level zero the families
containing the uncharged singular vectors reduce to two uncharged chiral
singular vectors whereas the other kind of families does not exist.

\vskip .17in

\def\lggo  {\mbox{$\kc_{l,\,\ket{0,\,\htop}}^{(0)G} $}}
\def\lqqo  {\mbox{$\kc_{l,\,\ket{0,-\htop-{\ctop\over3}}}^{(0)Q} $}}
\def\lqqm  {\mbox{$\kc_{l,\,\ket{0,\,\htop}}^{(-1)Q} $}}
\def\lggp  {\mbox{$\kc_{l,\,\ket{0,-\htop-{\ctop\over3}}}^{(1)G} $}}
\def\lgqo  {\mbox{$\kc_{l,\,\ket{0,\,\htop}}^{(2)G} $}}
\def\lqgo  {\mbox{$\kc_{l,\,\ket{0,-\htop-{\ctop\over3}}}^{(-2)Q} $}}
\def\lqgm  {\mbox{$\kc_{l,\,\ket{0,\,\htop}}^{(1)Q} $}}
\def\lgqp  {\mbox{$\kc_{l,\,\ket{0,-\htop-{\ctop\over3}}}^{(-1)G} $}}

  \begin{equation} \begin{array}{rcl}
  \begin{array}{rcl} \lggo &
  \stackrel{\Qz}{\mbox{------}\!\!\!\longrightarrow}
  & \lqqm \\[3 mm]
   \cA\,\updownarrow\ && \ \updownarrow\, \cA
  \\[3 mm]  \lqqo \! & \stackrel{\Gz}
  {\mbox{------}\!\!\!\longrightarrow} & \! \lggp \end{array}
              &  \quad {\rm and}\quad &
 \begin{array}{rcl} \lgqo &
  \stackrel{\Qz}{\mbox{------}\!\!\!\longrightarrow}
  & \lqgm \\[3 mm]
   \cA\,\updownarrow\ && \ \updownarrow\, \cA
  \\[3 mm]  \lqgo \! & \stackrel{\Gz}
  {\mbox{------}\!\!\!\longrightarrow} & \! \lgqp  \end{array} 
 \end{array} \label{diab4} \end{equation}

\subsection{Spectrum of $\D$ and h}\lvm

Now let us consider the spectra of conformal weights $\D$ and U(1) 
charges $\htop$ corresponding to the complete Verma modules
which contain singular vectors.

\vskip .17in
\noi
{\it Generic singular vectors}

For the case of generic singular vectors
the spectra of $\D$ and $\htop$ can be obtained
easily from the spectra corresponding to the
NS algebra, given by the zeroes of the determinant formula
\cite{BFK}, \cite{Nam}, \cite{KaMa3}.
The argument goes as follows. From the determinant formula one deduces
that the Verma module of the NS algebra
 $V_{NS}(\Delta',\htop')$
 has (at least) one uncharged singular vector, at level $l'={rs\over2}$,
if $(\Delta',\htop')$ lies on the
quadratic vanishing surface $f_{r,s} (\Delta',\htop')=0$, where

\BE
f_{r,s}(\Delta',\htop')=-2t\Delta'-\htop'^2-{1\over4}t^2+{1\over4}(s-tr)^2
{\ \ \ }, {\ \ \ } r\in\oZ^+, {\ } s\in 2\oZ^+, \label{frs}
\EE
\noi
with $t=(3-\ctop)/3$.
If $(\Delta',\htop')$ lies on the
vanishing plane $g_k (\Delta',\htop')=0$ instead, where

\BE
g_k(\Delta',\htop')=2\Delta'-2k\htop'-t(k^2-{1\over4}){\ \ \ },{\ \ \ }
k\in\oZ+1/2,
\label{gk}
\EE

\noi
then $V_{NS}(\Delta',\htop')$ has (at least) one charged
singular vector, at level $l'=|k|$,
with relative charge $q'=sgn(k)$.
On the other hand, by inspecting the
twists $T_{W1}$ \req{twa} and $T_{W2}$ \req{twb}, one finds that they
transform the primary states and singular vectors of the
NS algebra, with $(L_0, H_0)$ eigenvalues $(\D',\htop')$,
into \Gn-closed primary states and \Gn-closed singular vectors
of the Topological algebra with $(\cL_0, \cH_0)$ eigenvalues
$(\D,\htop)$, where $\D=\D'\pm {\htop'\over2}$ and $\htop=\pm \htop'$. 
Consequently, the level and the relative U(1) charge
of the topological singular vectors are given by $l=l'\pm {q'\over2}$ and
$q=\pm q'$ respectively.

It is convenient to rewrite $f_{r,s}(\Delta',\htop')$ and
$g_k(\Delta',\htop')$ in terms of the topological $(\cL_0, \cH_0)$
eigenvalues, using $(\D',\htop') = (\D-{\htop\over2}, \pm \htop)$,
resulting in

\BE
f_{r,s}(\D,\htop)=-2t\D+t\htop-\htop^2-{1\over4}t^2+{1\over4}(s-tr)^2
{\ \ \ }, {\ \ \ } r\in\oZ^+, {\ } s\in 2\oZ^+ \label{tfrs} \EE

\noi
and

\BE
g_k(\D,\htop)=2\D-\htop\mp 2k\htop-t(k^2-{1\over4}){\ \ \ },{\ \ \ }
k\in\oZ+1/2
\label{tgk} \EE

As a consequence if $f_{r,s}(\Delta,\htop)=0$
the topological Verma module $V(\ket{\Delta,\htop}^G)$
has (at least) one uncharged \Gn-closed singular vector 
$\ket\chi_{\ket\phi^G}^{(0)G}$, at level $l={rs\over2}$.
If  $g_k(\D,\htop)=0$ instead, then the topological Verma module 
$V(\ket{\Delta,\htop}^G)$ has (at least) one charged \Gn-closed 
singular vector $\ket\chi_{\ket\phi^G}^{(q)G}$, at level
$l=|k|\pm {q'\over2}$ with relative charge $q=\pm q'$, where $q'=sgn(k)$.
In addition, the singular vectors of type $\ket\chi_{\ket\phi^G}^{(-1)Q}$
are in the same
Verma modules and at the same levels as the singular vectors of type
$\ket\chi_{\ket\phi^G}^{(0)G}$, and the same happens
with the singular vectors
of types $\ket\chi_{\ket\phi^G}^{(0)Q}$
and $\ket\chi_{\ket\phi^G}^{(1)G}$, on the one side,
and with the singular vectors
of  types $\ket\chi_{\ket\phi^G}^{(-2)Q}$
and $\ket\chi_{\ket\phi^G}^{(-1)G}$, on the other.

As to the singular vectors
built on \Qn-closed topological primaries, one gets the corresponding
spectra straightforwardly just by applying the spectral flow
automorphism $\cA$ \req{autom} which interchanges $G \leftrightarrow Q$,
$q \leftrightarrow -q$ and $\htop \leftrightarrow -\htop-{\ctop\over3}$.
For example, the singular vectors of  
types $\ket\chi_{\ket\phi^Q}^{(0)Q}$  and $\ket\chi_{\ket\phi^Q}^{(1)G}$
are at the same level in the Verma module $V(\ket{\D,\hat\htop}^Q)$ 
satisfying $f_{r,s}(\D,-\hat\htop-{\ctop\over3})=0$ in \req{tfrs},
and similarly the remaining cases.

\vskip .17in
\noi
{\it Chiral singular vectors}

Chiral singular vectors can be viewed
as particular cases of $\cG_0$-closed singular vectors which 
become chiral when $\D+l=0$, but they can also be viewed
as particular cases of $\cQ_0$-closed singular vectors which 
become chiral when $\D+l=0$ . As a consequence, 
the spectra of $\D$ and $\htop$ corresponding to chiral singular
vectors of type $\kc_{l,\,\ket{\D,\,\htop}^G}^{(0)G,Q}\,$ must lie
on the intersection of the spectra corresponding to singular vectors
of types $\kc_{l,\,\ket{\D,\,\htop}^G}^{(0)G}\,$ and
$\kc_{l,\,\ket{\D,\,\htop}^G}^{(0)Q}\,$, with $\D=-l$.
These are given by the solutions to the quadratic
vanishing surface $f_{r,s}(-l,\htop)=0$, and the solutions to
the vanishing plane $g_{l-1/2}(-l,\htop)=0$, respectively
(the type $\kc_{l,\,\ket{\phi}^G}^{(0)Q}\,$ has the same
spectra as the type $\kc_{l,\,\ket{\phi}^G}^{(1)G}\,$).
The intersection of $f_{r,s}(-l,\htop)=0$ \req{tfrs} and
$g_{l-1/2}(-l,\htop)=0$ \req{tgk} gives 

\BE \htop={t\over2}(1-l)-1, \qquad (s-tr)^2=(2-tl)^2
\label{ch0} \EE
\noi
with $l={rs\over2}>0$. The obvious general solution for any $t$ is
$\, s=2, \, r=l\,$, and there is also the particular solution
$\, r=2/t, \, s=lt\, $, which only makes sense for the values of $t$
such that $r \in \oZ^+\,$ and $s\in 2\oZ^+$. At level zero there are no 
uncharged chiral singular vectors (they exist only on no-label primaries). 

 Similarly, the spectra corresponding to chiral
singular vectors of type $\kc_{l,\,\ket{\D,\,\htop}^G}^{(-1)G,Q}$ 
must lie on the intersection of the spectra corresponding to singular 
vectors of types $\kc_{l,\,\ket{\D,\,\htop}^G}^{(-1)G}$ 
and $\kc_{l,\,\ket{\D,\,\htop}^G}^{(-1)Q}\,$, with $\D=-l$, given by the
solutions to $g_{-(l+1/2)}(-l,\htop)=0$, and the solutions to 
$f_{r,s}(-l,\htop)=0$, respectively
(the type $\kc_{l,\,\ket{\phi}^G}^{(-1)Q}\,$ has the same
spectra as the type $\kc_{l,\,\ket{\phi}^G}^{(0)G}\,$).
The intersection of $f_{r,s}(-l,\htop)=0$ and
$g_{-(l+1/2)}(-l,\htop)=0$  gives 

\BE \htop={t\over2}(1+l)+1, \qquad (s-tr)^2=(2-tl)^2
\label{ch1} \EE

\noi
with $l={rs\over2}>0$. As before, the general solution is $\,s=2, \, r=l\,$,
for any $t$, and there is also the particular solution 
$\,r=2/t, \, s=lt\,$. For $l=0$ there are charged chiral singular vectors 
$\, \kc_{0,\,\ket{0,\,\htop}^G}^{(-1)G,Q}= \Qz \ket{0,\,\htop}^G \,$
for every value of $\htop$. 

The untwisting of equations
\req{ch0} and \req{ch1} give the same equations back, with $T_{W2}$ reversing 
the sign of $\htop$. Therefore  $\htop$ and  $(-\htop)$ in 
\req{ch0} give the spectrum corresponding to {\it uncharged antichiral
and chiral singular vectors of the NS algebra} at level $l$,
 respectively, while $\htop$ and  $(-\htop)$ in
\req{ch1} give the spectrum corresponding to {\it charge $q=-1$ 
antichiral and charge $q=1$ chiral singular vectors of the NS algebra} at
level $l+\half$, respectively. As far as we know these results for the 
chiral and antichiral singular vectors of the NS algebra were unknown.  

\vskip .17in
\noi
{\it Other types of singular vectors}

Regarding singular vectors in no-label Verma modules and 
no-label singular vectors in generic Verma modules, 
we ignore the corresponding spectra of $\D$ and/or $\htop$. 
The existing singular vectors at level 1, shown in Appendix B, 
suggest however that 
the spectrum of $\htop$ for no-label Verma modules 
$V(\ket{0,\htop})$ is the ``sum" of the spectra corresponding 
to the Verma modules $V(\ket{0,\htop}^G)$ and $V(\ket{0,\htop}^Q)$,
and also suggest that the no-label singular vectors are very scarce
(at level 1 they only exist for $\ctop=-3$).

\section{Internal Mappings and D\"orrzapf Pairs}\lvm

An interesting issue is to analyze under which conditions the 
chains of mappings
involving $\cA$ and $\cU_{\pm1}$ act {\it inside} a Verma module.
By inspecting diagrams \req{diab0}
and \req{diab1} one sees only two possibilities: the
Verma modules $V(\ket{\D,\htop}^G)$ and $V(\ket{\D-\htop,-\htop}^G)$,
 connected by $\cA \, \cU_1$,
coincide for $\htop=0$, and so do the Verma modules
$V(\ket{\D,-\htop-\ctop/3}^Q)$ and
 $V(\ket{\D-\htop,\htop-\ctop/3}^Q)$, related by $\cU_1 \cA$.
In diagrams \req{diab2} and \req{diab3} there are
more possibilities. For example, the Verma modules $V(\ket{\D,\htop}^G)$
and $V(\ket{\D,-\htop-\ctop/3+1}^G)$ coincide for
$\htop={3-\ctop\over6}$ while the Verma modules
$V(\ket{\D-\htop,-\htop}^G)$ 
 and $V(\ket{\D-\htop,\,\htop-\ctop/3+1}^G)$
coincide for $\htop={\ctop-3\over6}$.

The case $\htop= {3-\ctop\over6}={t\over2}$ is the
simplest example of a loop associated to sequence B, corresponding to
$n=0$. In general, for $\htop=(n+1)t/2$ 
${\ } (n=0, 1, 2 ....) {\ }$ one finds $\hat{\D}=\D$ and
$\hat{\htop}=\htop$ after $n$ attachements of four-member subfamilies. For
this value of $\htop$ and $\,\D_{r,s}={1\over8t}[(s-tr)^2-n^2t^2]\,$, 
satisfying $f_{r,s}(\D,\htop)=0$ in eq. \req{tfrs}, the loop of mappings 
connect in the same Verma module and at the same level
two singular vectors of type $\kc_{l,\kp^G}^{(0)G}$ and two singular
vectors of type $\kc_{l,\kp^G}^{(-1)Q}$. For $\htop=(n+1)t/2$
eqns. \req{ch0} and  \req{ch1} give $n+l=-{2\over t}$ and $n-l={2\over t}$
respectively. Therefore for these values the mappings connect 
a couple of chiral singular vectors $\kc_{l,\kp^G}^{(0)G,Q}$ 
and $\kc_{l,\kp^G}^{(-1)G,Q}$ instead. Finally, for $\htop=(n+1)t/2$ 
and $\,\D_k={nt\over4}(1\pm2k)+{t\over2}(k\pm{1\over2})^2\,$, satisfying 
$g_k(\D,\htop)=0$ in eq. \req{tgk}, the mappings connect four singular 
vectors of
different types in the same Verma module: $\kc_{l,\kp^G}^{(1)G}$ and
$\kc_{l,\kp^G}^{(0)Q}$ at level $l$, and $\kc_{l+n,\kp^G}^{(-1)G}$ and
$\kc_{l+n,\kp^G}^{(-2)Q}$ at level $l+n$.

The question naturally arises whether the two uncharged singular
vectors $\kc_{l,\kp^G}^{(0)G}$ and the two charged singular vectors
$\kc_{l,\kp^G}^{(-1)Q }\,$, which
are together in the same Verma module at the same level, coincide
(up to constants)
or they are linearly independent. In the NS algebra,
the possibility that two uncharged singular vectors,
at the same level, in the same Verma module,  may be linearly
independent, was discovered by D\"orrzapf in ref. \cite{Doerr2}. Under
the topological twistings $T_{W1}$ \req{twa} and $T_{W2}$ \req{twb}
this possibility extends also to the topological singular vectors of
type $\kc_{\kp^G}^{(0)G}$, and to all the types 
related {\it necessarily} to them via the
mappings that we have analyzed; \ie\ the three types of generic
singular vectors involved, together with $\kc_{\kp^G}^{(0)G}$,
in the first kind of generic families, 
depicted in diagram \req{diab0}: $\kc_{\kp^G}^{(-1)Q}$,
$\kc_{\kp^Q}^{(0)Q}$ and $\kc_{\kp^Q}^{(1)G}$. This possibility does
not extend, however, to the eight types of generic singular vectors 
involved in the second kind of generic families, depicted in diagram
\req{diab1} because they are related 
{\it necessarily} to the charged singular vectors of the NS
algebra, and D\"orrzapf proved that for these singular vectors there are
no such two dimensional spaces. As to the chiral types of singular 
vectors, they do not admit two dimensional spaces \cite{DB1}, in spite
of the fact that they may appear in generic families of the first kind
as particular cases. However, chiral singular vectors may appear as 
partners of \Gn-closed or \Qn-closed singular vectors in D\"orrzapf 
pairs since they are just particular cases of \Gn-closed 
and \Qn-closed singular vectors (see Appendix C).

Observe that the loops of sequence B, mapping singular
vectors into singular vectors inside the same Verma module, 
connect singular vectors of the same type, at the
same level, only for the types which admit two dimensional spaces.  
This fact strongly suggests that these loops of
mappings may transform, under appropriate conditions, 
 a given singular vector of any of the types
$\kc_{\kp^G}^{(0)G}$, $\kc_{\kp^G}^{(-1)Q}$, $\kc_{\kp^Q}^{(0)Q}$, or
$\kc_{\kp^Q}^{(1)G}$, into a linearly independent singular vector of
exactly the same type, at the same level.

The D\"orrzapf conditions for the appearance of two linearly
independent uncharged NS singular vectors at the same level, in the same
Verma module, consist of the simultaneous vanishing of two curves,
$\epsilon_{r,s}^+(t,\htop)=0$ and $\epsilon_{r,s}^-(t,\htop)=0$, given by

\BE
\epsilon_{r,s}^{\pm}(t,\htop)= \prod_{m=1}^r
(\pm{s-rt\over2t}+{\htop\over t}\mp{1\over2}\pm m)\,\  ,\ r\in\oZ^+,
\,\, s\in2\oZ^+
\label{Dcond}
\EE

\noi
where $t={3-\ctop\over3}$ and the level of the singular vector is $l=rs/2$.
The topological twists $T_{W1}$ \req{twa} and $T_{W2}$ \req{twb} let 
these conditions invariant, extending the existence of the two dimensional 
space of singular vectors to the topological singular vectors of types
$\kc_{\kp^G}^{(0)G}$, $\kc_{\kp^G}^{(-1)Q}$, $\kc_{\kp^Q}^{(1)G}$ and
$\kc_{\kp^Q}^{(0)Q} \,$, as we pointed out.

Let us analyze the conditions $\epsilon_{r,s}^+(t,\htop)=0$ and
$\epsilon_{r,s}^-(t,\htop)=0$ for $\htop=(n+1)t/2$, which correspond to
the loops in sequence B. One gets the expressions:

\BE
\epsilon_{r,s}^+(t,(n+1)t/2)= \prod_{m=1}^r
({s-rt\over2t}+{n\over2}+m)\nonumber
\EE

\BE
\epsilon_{r,s}^-(t,(n+1)t/2)= \prod_{m=1}^r
({rt-s\over2t}+{n\over2}+1-m)
\EE

\noi
Inspecting these, one realizes that the vanishing of the term
corresponding to $m=m_1$ in the first expression results in the vanishing
of the term corresponding to $m=m_2$ in the second expression provided
$m_2=m_1+n+1$. This implies that $r\geq n+2$ for the singular vector,
but the equality $r=n+2$, $m_1=1$ has no solution for the vanishing
of the first term in $\epsilon_{r,s}^+$. Therefore $r>n+2$, implying
$l>n+2$ since $l\geq r$.

We see that one has to look at level 3, at least, to check if the loops
in sequence B connect ``D\"orrzapf pairs" of singular vectors, since $l>2$
already for $n=0$ ($\htop=t/2$). We have checked that this is indeed the
case for the singular vectors which correspond to the solutions to 
$\epsilon_{r,s}^+(t,t/2)=\epsilon_{r,s}^-(t,t/2)=0$ at level 3
($r=3$, $s=2$), which are $t=2$ ($\ctop=-3$) and $t=-2$ ($c=9$). In
Appendix C we have written down the pairs of singular vectors of types
$\kc_{\kp^G}^{(0)G}$ and $\kc_{\kp^Q}^{(0)Q}$ (equivalent to pairs of singular
vectors of types $\kc_{\kp^Q}^{(1)G}$ and $\kc_{\kp^G}^{(-1)Q}$,
respectively), for the case $\htop=t/2=-1$, $\ctop=9$, together with the
complete family to which they belong. It turns out that this family contains
 two more D\"orrzapf pairs, which are not connected by any loops of sequence B,
and with the particularity that one member in each pair is chiral. 

When the singular vectors of types $\kc_{\kp^G}^{(0)G}$, 
$\kc_{\kp^G}^{(-1)Q}$, $\kc_{\kp^Q}^{(0)Q}$ and $\kc_{\kp^Q}^{(1)G}$ 
do not satisfy the D\"orrzapf conditions \req{Dcond}, then they  
must be mapped back to themselves (up to constants) by the loops of sequence
B, and by any other mappings acting inside a given Verma module.
In Appendix B we show an example of a singular vector of type 
$\kc_{\kp^G}^{(0)G}$ at level 1, with $\htop=t/2$, which comes back to
itself under the first loop of sequence B.

To finish, we conjecture that the two partners in every 
D\"orrzapf pair of singular vectors of types
$\kc_{l, \kp^G}^{(0)G}\,$, $\kc_{l,\kp^G}^{(-1)Q}\,$, 
$\kc_{\kp^Q}^{(1)G}\,$ and $\kc_{\kp^Q}^{(0)Q}\,$, are 
connected by the mappings made out of $\cA$, $\cU_{\pm1}$, 
\Gn\ and \Qn\ , and therefore they belong to the same family.

\section{Conclusions and Final Remarks}\lvm

We have analyzed several issues concerning the singular vectors of
the Topological algebra, considering chiral
Verma modules as well as complete Verma modules. 

\vskip .17in

{\it First}, we have investigated which are the different types 
of topological singular
vectors which may exist, taking into account the relative U(1)  charge
$q$ and the BRST-invariance properties of the singular vectors
themselves and of the primary states
 on which they are built (this is of utmost importance).
We have identified an algebraic mechanism, ``the cascade effect", 
which provides a necessary, although not sufficient, condition for 
the existence of a given type of singular vector. 
In chiral Verma modules $V(\ket{0,\htop}^{G,Q})$, \ie\ built on 
chiral primaries, there are only
four types of singular vectors allowed by the cascade effect, all four 
with $|q| \leq 1$,
as shown in table \req{tabl1}. By explicit construction we 
know that the four types exist at low levels, even at level 1. 
In complete Verma modules, built on topological primaries without
additional constraints other than the ones imposed by the algebra,
one finds a total of twenty-nine different types
of singular vectors allowed by the cascade effect, all of them with
$|q|\leq2$, as shown in tables \req{tabl2}, \req{tabl3} and \req{tabl4}.
Twenty of these types (twelve generic, four chiral and four no-label)
correspond to singular vectors in generic Verma modules,
$V(\ket{\D,\htop}^G)$ and $V(\ket{\D,\htop}^Q)$, 
and the remaining nine types correspond to singular vectors in
no-label Verma modules $V(\ket{0,\htop})$.
Twenty-eight of these types can be constructed already 
at level 1 and one type only exists at level zero. In addition, the  
twelve generic types and the four chiral types of singular vectors in
generic Verma modules can be mapped to the singular vectors of the NS
algebra and therefore they must necessarily exist.

\vskip .17in

{\it Second}, we have analyzed a set of mappings between the topological
singular vectors: the spectral flow mappings $\cA$ and $\cU_{\pm1}$,
and the action of the zero modes \Gn\ and \Qn. While \Gn\ and \Qn\ act 
inside a given Verma module, $\cA$ and $\cU_{\pm1}$
interpolate between different Verma modules. The universal spectral flow
automorphism $\cA$ \req{autom} transforms all kinds of 
singular vectors back into singular vectors, and chiral
states into chiral states. The even spectral flow transformation
$\cU_1$ \req{spflw} is very selective, on the contrary, mapping 
singular vectors of types $\kc_{\kp^G}^{(q)G}$ and $\kc_{\kp^G}^{(q)G,Q}$
into singular vectors of types $\kc_{\kp^Q}^{(q)Q}$ and
$\kc_{\kp^Q}^{(q)G,Q}$ (both), and the other way around for $\cU_{-1}$,
while mapping all other types of singular vectors
into various kinds of states which are not singular vectors.
One can consider several other spectral flow mappings, but they are
equivalent to the compositions of $\cA$, $\cU_{\pm1}$ and $\hat\cA_0$
(the label-exchange operator which interchanges the labels 
$(1) \leftrightarrow (2)$ of the two sets of topological generators 
resulting from the two different topological twists).

\vskip .17in

{\it Third}, using \Gn , \Qn\ and the 
spectral flow mappings $\cA$ and $\cU_{\pm1}$, we have derived the
family structure of the topological singular vectors, every member of a
family being connected to any other member by a chain of transformations.
The singular vectors in chiral Verma modules 
come in families of four with a unique pattern.
The families of singular vectors in complete Verma modules follow an
infinite number of different patterns, which we have grouped roughly in
five main kinds. Two main kinds consist of generic singular vectors, 
and may contain chiral singular vectors as particular cases, producing 
the intersection of the two main kinds of families. 
Another main kind of family 
contains no-label singular vectors in generic Verma modules, and  
the remaining two main kinds 
of families correspond to singular vectors in no-label Verma modules. 

\vskip .17in

{\it Fourth}, we have written down the complete families of topological 
singular vectors in chiral Verma modules until level 3, and we have 
shown that, starting at level 2,  
half at least of these singular vectors 
are subsingular in the complete Verma modules. Namely,
we have found that the singular vectors of types
$\kc^{(1)G}$ and $\kc^{(0)Q}$ corresponding to the spectrum
$\htop_{r,s>2}^{(1)}$ in eq. \req{hrs1}, are subsingular vectors in the
generic Verma modules $V(\ket{0,\htop}^G)$, whereas the singular
vectors of types $\kc^{(0)G}$ and $\kc^{(-1)Q}$
corresponding to the spectrum $\htop_{r,s>2}^{(0)}$ in
eq. \req{hrs0}, are subsingular vectors in the generic Verma modules
$V(\ket{0,\htop}^Q)$. We conjecture that this result is true at any level.

\vskip .17in

{\it Fifth}, we have derived the spectra of conformal weights $\D$ and 
U(1) charges $\htop$
corresponding to the generic Verma modules which contain generic and
chiral topological singular vectors. From the spectra corresponding to
the chiral singular vectors we deduced straightforwardly the spectra
corresponding to chiral and antichiral singular vectors of the
NS algebra, which were unknown. Regarding the
no-label singular vectors in generic Verma modules and the singular
vectors in no-label Verma modules we ignore the spectra. The
results at level 1 suggest that the no-label singular vectors are
very scarce (they only exist for $\ctop=-3$), and also
suggest that the spectrum of $\htop$ for no-label Verma modules
$V(\ket{0,\,\htop})$ is the ``sum" of the spectra corresponding to
the generic Verma modules with zero conformal weight
$V(\ket{0,\,\htop}^G)$ and $V(\ket{0,\,\htop}^Q)$.

\vskip .17in

{\it Sixth}, we have analyzed some conditions under which 
the chains of mappings involving $\cA$ and $\cU_{\pm1}$
act inside a Verma module.
In generic Verma modules there are many possibilities. 
In particular we have found a sequence of
mappings (sequence B) which has loops, \ie\ which comes back to the same
Verma modules after $n$ steps ($n=0,1,2...$).
For singular vectors of the types $\kc_{\kp^G}^{(0)G}$,
$\kc_{\kp^G}^{(-1)Q}$, $\kc_{\kp^Q}^{(0)Q}$ and $\kc_{\kp^Q}^{(1)G}$
these loops connect pairs of singular vectors of the same type at the
same level. With this motivation we have analyzed the D\"orrzapf
conditions \req{Dcond} (originally written for the NS algebra)
for the appearance of two linearly independent singular vectors of the  
same type, at the same level and in the same Verma module,
for the case of the loops of mappings of sequence B, 
finding solutions for every $n$, starting at level 3.
We have constructed singular vectors at level
3 which satisfy these conditions, and belong to the four types
mentioned before, and we have checked that the loop of
mappings for $n=0$ transform these singular vectors
into singular vectors of the same type, at the same levels, 
but linearly independent from the initial ones. 
These facts and several more examples 
suggest that the loops of mappings made out of  
$\cA$, $\cU_{\pm1}$, \Gn\ and \Qn\ , in particular the ones
of sequence B, produce pairs of linearly independent
singular vectors of types $\kc_{\kp^G}^{(0)G}$,$\kc_{\kp^G}^{(-1)Q}$,
$\kc_{\kp^Q}^{(0)Q}$ and $\kc_{\kp^Q}^{(1)G}$ provided the D\"orrzapf
conditions are satisfied; otherwise the
pairs of singular vectors coincide (up to constants). We conjecture
that, in all cases, the two partners of a given D\"orrzapf pair are
connected by these mappings, \ie\ they belong to the same family.

\vskip .17in

{\it Finally}, in Appendix A we have given a detailed account of the 
``cascade effect", in
Appendix B we have written down the complete set of topological singular
vectors at level 1 for complete Verma modules (twenty-eight 
different types), and in Appendix C we have shown a specially interesting 
thirty-eight-member family of generic and chiral singular vectors at
levels 3, 4, 5, and 6. This family contains four ``D\"orrzapf pairs"
at level 3, with two partners, corresponding to two different pairs, 
being chiral singular vectors. 

\vskip .17in

A final comment is that the mappings we have analyzed simplify enormously
the construction of topological singular vectors because
for a given family it is only necessary to compute one of the singular
vectors from scratch. In addition, the
singular vectors of the second main kind of families are connected
at different levels, and therefore it is possible to construct singular
vectors at high levels starting from singular vectors at lower levels.
An important observation is that any
topological singular vector annihilated by \Gn\ (\ie\ \Gn-closed or chiral)
built on a primary annihilated also by \Gn, is transformed into a singular
vector of the NS algebra under the untwistings, and the
other way around under the twistings. As a consequence
the generic singular vectors of types
 $\kc_{\kp^G}^{(q)G}$, \ie\  \Gn-closed built on \Gn-closed primaries,
can be constructed using the corresponding formulae for constructing
singular vectors of
the NS algebra \cite{Doerr1}, \cite{Doerr2}, and performing the 
topological twists $T_{W1}$ \req{twa} and/or $T_{W2}$ \req{twb} afterwards. 
The other types of generic singular vectors and the chiral singular
vectors can be obtained from those
ones via the mappings that we have analyzed.
For non-generic singular vectors in complete Verma modules and for
singular vectors in chiral Verma modules, however, there are no
construction formulae\footnote{Some attemps have appeared in the
literature to write down construction formulae for some types of 
singular vectors of the Topological algebra, but we find that work 
very premature and confusing.}.

In a forthcoming paper \cite{BJI8} we will analyze thoroughly the 
implications that the results presented here have on the singular 
vectors of the NS algebra and on the singular vectors of the R algebra. 

The rigorous proof that the cascade effect provides a necessary condition 
for the existence of a given type of singular vector
will be presented in a next publication \cite{DB1} together with the
analysis of the dimensionalities of the spaces associated to every
type of singular vector and some related issues.

\vskip 1cm

\centerline{\bf Acknowledgements}

We are indebted to M. D\"orrzapf and A. Kent for illuminating discussions.
We also thank J.W. van Holten, E.B. Kiritsis, S.~Mukhi, 
A.N.~Schellekens and A.~Tiemblo for useful comments. Finally, we 
are very grateful to the (second) referee of Nuclear Physics B 
for reading this work with so much interest and for many 
important suggestions to improve its quality.

\setcounter{equation}{0}
\def\theequation{A.\arabic{equation}}

\subsection*{Appendix A}\lvm

The ``cascade effect" is an algebraic mechanism which consists of the
vanishing in cascade of the coefficients of the ``would-be" singular vector
when imposing the highest weight conditions, alone or in combination
with the conditions for BRST or anti-BRST invariance. The vanishing in
cascade of the coefficients starts in all cases either with the h.w.
conditions $\cQ_1 \kc=0$ or $\cG_1 \kc=0$, or with the BRST or anti-BRST
invariance conditions $\cQ_0 \kc=0$ or $\cG_0 \kc=0$. Once it starts,
the cascade effect goes on until the end getting rid of all the
coefficients of the would-be singular vectors. The proof of this
statement will be presented in a forthcoming publication \cite{DB1}.

For topological singular vectors in chiral Verma modules this 
effect occurs for all the would-be
singular vectors with relative charges $|q|>1$ and for those with 
$|q|=1$ of the types $\kc^{(1)Q}$ and $\kc^{(-1)G}$.
For topological singular vectors in complete Verma modules
the cascade effect occurs for all the
would-be singular vectors with $|q|>2$, for most with $|q|=2$ 
and for some with $|q|=1$. 

The types of topological singular vectors for which
the cascade effect does not take place are the ones shown in 
tables \req{tabl1}, \req{tabl2}, \req{tabl3} and \req{tabl4}, a total
of four different types in chiral Verma modules and twenty-nine 
different types in complete Verma modules, from which one type exists only
at level zero ($\kc_{0,\ket{0,\htop}}^{(0)G,Q}=\cG_0\cQ_0\ket{0,\htop}$).
The cascade effect not taking place for these types of singular vectors is
a necessary, although not sufficient, condition for their existence.
However, we know by explicit construction that all these types
of singular vectors exist at low levels, even at level 1 (except the
type that only exists at level zero).
In what follows we will analyze the cascade effect.

\vskip .17in
\noi
{\it Cascade effect in chiral Verma modules}

Let us consider topological singular vectors built on chiral primaries
$\ket{0,\htop}^{G,Q}$.
To see how the cascade effect
 prevents the existence of such singular vectors with
relative charge $|q| >1$ let us take a
general level-$l$ secondary state with $q=2$. 
For convenience we will organize the different terms
 into shells, the first shell consisting of the terms
 with maximum number of bosonic modes, the second shell
containing the terms with maximum number of bosonic modes minus
one, and so on.
Now by imposing the h.w. condition $\cQ_1 \kc = 0 $ one deduces easily
that the first shell
of terms must vanish. The argument goes as follows.
 For $q=2$ the first shell consists of all possible
terms with $l-3$ bosonic modes. Those terms have the structure

\BE
C_{11}^{pr} \ \ {\cal H}_{-1}^{p}
 \ \ {\cal L}_{-1}^{r} \ \ \cG_{-2}\
 \ \cG_{-1}, \ \ \ {p+r} = l-3
\EE

\noi
where $C_{11}^{pr}$ are the coefficients to be determined.
The action of $\cQ_1$ on $\cG_{-2}$ produces another bosonic mode,
since $\{ \cQ_1, \cG_{-2}\} = 2 {\cal L}_{-1} - 2 {\cal H}_{-1}$.
Consequently, the terms with $p=0,\,r=l-3$ , giving rise to
$\cL_{-1}^{l-2} \cG_{-1}$, and $r=0,\,p=l-3$, giving rise to
$\cH_{-1}^{l-2} \cG_{-1}$
cannot be compensated  by any other terms
and must vanish (in other words, there are no other terms which will produce
$\cL_{-1}^{l-2} \cG_{-1}$ or  $\cH_{-1}^{l-2} \cG_{-1}$ under the action of
$\cQ_1$). Applying the same reasoning again 
we see that the terms with $p=1,\,r=l-4$ and
$r=1,\,p=l-4$ must also vanish, due to the vanishing
 of the terms with $p=0$
and $r=0$. In turn this produces the vanishing of the terms with $p=2$ and 
$r=2$, and so on. Therefore, {\it the first shell of terms disappears because 
of a cascade effect only by imposing the h.w. condition} $\cQ_1 \kc = 0$.

The vanishing of the first shell induces, in turn, the
vanishing of the second shell, consisting of the terms
with $l-4$ bosonic modes, coming in three different structures:
$ \cH_{-2} \ \cH_{-1}^{p} \ \cL_{-1}^{r} \ \cG_{-2}
 \ \cG_{-1}{\ }{\rm and}{\ }\ \cH_{-1}^p \ \cL_{-2} \ \cL_{-1}^r \ \cG_{-2}
 \ \cG_{-1}, {\ }{\rm for}{\ } \ \ {p+r} = l-5 ,$ and 
${\ } {\cal H}_{-1}^{p} \ {\cal L}_{-1}^{r} \ \cG_{-3}
 \ \cG_{-1}$ for $ {p+r} = l-4 {\ }$.
By imposing $\cH_1 \kc = 0$ and again $\cQ_1 \kc = 0$, one
finds three inequivalent blocks containing the terms which
compensate each other in such a way that all the coefficients must 
vanish. The vanishing of the second shell of terms induces, in turn, 
the vanishing of the third shell,
with $l-5$ bosonic modes, and so on, as the reader can verify.
Hence there are no topological singular vectors
with relative charge $|q|=2$ in chiral Verma modules (the
same analysis can be repeated for the case $q=-2$, giving the same
results, just by interchanging $\cQ \leftrightarrow \cG$
everywhere).

The same mechanism takes place for the would-be singular vectors
with $q>2$ (and $q<-2$ as a consequence) 
because the first shell of terms with maximum
number of bosonic modes has the unique structure 
$ \ \ {\cal H}_{-1}^{p} \ \ {\cal L}_{-1}^{r} ........\ \cG_{-3}
 \ \cG_{-2}\ \cG_{-1},{\ } $ in chiral Verma modules,
where the number of $\cG$ modes is equal to $q$.
The action of $\cQ_1$ on these terms produces a new bosonic mode
when hitting a $\cG$-mode other than $\cG_{-1}$ . As a result
the terms with $p=0$
and $r=0$ cannot be compensated by any other terms
and must vanish, giving rise to the vanishing of the complete
shell of terms in a cascade effect, as in the case $q=2$. 

\vskip .12in
In the case $|q|=1$, for secondary states of types
 $\kc^{(1)G}$ and $\kc^{(-1)Q}$ there is no cascade effect
getting rid of the terms. However, there is such effect
for the would-be singular vectors of types $\kc^{(1)Q}$ and 
$\kc^{(-1)G}$. For $q=1$ the action of $\cQ_1$ on $\cG_{-1}$,
in the first shell of terms $\cH_{-1}^p \ \cL_{-1}^r \ \cG_{-1}$,
does not create a new bosonic mode but
a number: $\cH_{-1}^p \ \cL_{-1}^r\ \cL_0 \rightarrow
\Delta \cH_{-1}^p \ \cL_{-1}^r$ and
 $\cH_{-1}^p \ \cL_{-1}^r\ \cH_0 \rightarrow \htop \cH_{-1}^p \ \cL_{-1}^r$,
where $\htop$ and $\Delta$ are the U(1) charge
and the conformal weight of the topological
primary state on which the secondary is built
(for topological chiral primaries $\Delta =0$).
 However, if one imposes BRST-invariance on the $q=1$ would-be
singular vector, \ie\ $\cQ_0 \kc = 0 $, the vanishing of the terms
starts again since \Qn\ does indeed create a bosonic mode
($\cL_{-1}$) when acting on $\cG_{-1}$. The first shell of terms, 
with $l-1$ bosonic modes vanishes in this way. Therefore 
there is a cascade effect getting rid of the terms of the would-be
singular vectors of the types $\kc^{(1)Q}$ and $\kc^{(-1)G}$
(for the latter by interchanging $\cQ\leftrightarrow\cG$ 
everywhere in the previous expressions).

Let us consider finally the case $q=0$. Since the first shell
$\cH_{-1}^p \ \cL_{-1}^r$ contains only bosonic modes, it is not 
possible to create any other bosonic mode by acting with the positive 
modes or with \Qn\ or \Gn\ . Therefore the cascade effect does not
take place in the case $q=0$.

\vskip .17in
\noi
{\it Cascade effect in complete Verma modules}

When the topological primary is \Qn-closed $\ket{\D,\htop}^Q$
or no-label $\ket{\D,\htop}$, the \Gn-modes also
contribute to build the Verma module, unlike in the chiral case. As a
result, for $q>0$ the first shell of terms with maximum number of bosonic
modes contains one \Gn-mode. 
Similarly, when the topological primary is \Gn-closed $\ket{\D,\htop}^G$ 
or no-label, the \Qn-modes contribute to build the Verma module, 
and for $q<0$ the first shell contains one \Qn-mode.

The resulting structures for the first shells of terms, depending on
the type of primary state,
are therefore the following. For $q>2$ one has
$\cH_{-1}^{p} \ \cL_{-1}^{r} \ .......\cG_{-3}\ \cG_{-2}\ \cG_{-1}\ 
\ket{\D,\htop}^G {\ }$, 
${\ } \cH_{-1}^{p} \ \cL_{-1}^{r} \ .......\cG_{-2}\ \cG_{-1}\
\cG_0\ \ket{\D,\htop}^Q $ and ${\ }\cH_{-1}^{p} \ \cL_{-1}^{r} \ .......
\cG_{-2}\ \cG_{-1}\ \cG_0\ \ket{0,\htop}$.
The cascade effect occurs in all three cases simply by imposing
$\cQ_1 \kc=0$. For $q<-2$ one finds the same result by
interchanging $\cQ \leftrightarrow \cG$ everywhere.
Hence the cascade effect always takes place for the would-be
topological singular vectors with
$|q|>2$. 

 For $q=2$ the first shells are
$\cH_{-1}^{p} \ \cL_{-1}^{r} \ \cG_{-2}\ \cG_{-1}\ \ket{\D,\htop}^G$,
${\ }\cH_{-1}^{p} \ \cL_{-1}^{r} \ \cG_{-1}\ \cG_0\ \ket{\D,\htop}^Q{\ }$ and
${\ }\cH_{-1}^{p} \ \cL_{-1}^{r} \ \cG_{-1}\ \cG_0\ \ket{0,\htop}$.
The cascade effect starts in the first case by imposing
$\cQ_1 \kc=0$ whereas it starts in the second and third cases only
by imposing $\cQ_0 \kc=0$. Therefore
for $q=2$ the cascade effect takes place for
the would-be singular vectors of the types
$\kc^{(2)G}_{\kp^G}{\ }$, $\kc^{(2)G,Q}_{\kp^G}{\ }$,
$\kc^{(2)Q}_{\kp^G}{\ }$, $\kc^{(2)}_{\kp^G}{\ }$, $\kc^{(2)Q}_{\kp^Q}{\ }$,
$\kc^{(2)G,Q}_{\kp^Q}{\ }$ 
and $\kc^{(2)Q}_{\ket{0,\htop}}{\ }$. One finds similar results for 
$q=-2$, interchanging $\cQ \leftrightarrow \cG$ in all these types.

For $q=1$ the first shells of terms are
$\cH_{-1}^{p} \ \cL_{-1}^{r} \ \cG_{-1}\ \ket{\D,\htop}^G{\ }$,
${\ }\cH_{-1}^{p} \ \cL_{-1}^{r} \ \cG_0\ \ket{\D,\htop}^Q{\ }$ and
$\cH_{-1}^{p} \ \cL_{-1}^{r} \ \cG_0\ \ket{0,\htop}$.
The cascade effect occurs only in the first case, by imposing $\cQ_0 \kc=0$;
that is, only for the would-be singular vectors of the types
$\kc^{(1)Q}_{\kp^G}{\ }$ and $\kc^{(1)G,Q}_{\kp^G}{\ }$.
One finds similar results for $q=-1$, interchanging $\cQ \leftrightarrow \cG$.

For $q=0$ the cascade effect does not take place, obviously,
like in the case of chiral Verma modules.

The cascade effect also prevents, indirectly, the existence of no-label
singular vectors of types $\kc^{(2)}_{\kp^Q}{\ }$, 
$\kc^{(-2)}_{\kp^G}{\ }$, $\kc^{(1)}_{\kp^G}{\ }$ and 
$\kc^{(-1)}_{\kp^Q}{\ }$. The reason is that acting with \Gn\ or \Qn\
on these singular vectors one obtains \Gn-closed or \Qn-closed singular
vectors forbidden by the cascade effect. For example, 
$\Gz\kc^{(1)}_{\kp^G}{\ }=\kc^{(2)G}_{\kp^G}{\ }$, but the latter
type does not exist, so that $\kc^{(1)}_{\kp^G}{\ }$ must be \Gn-closed,
\ie\ of type $\kc^{(1)G}_{\kp^G}{\ }$ instead\footnote{We thank
M. D\"orrzapf for pointing out this to us.}.

\setcounter{equation}{0}
\def\theequation{B.\arabic{equation}}

\subsection*{Appendix B}\lvm

Let us write down the complete set of topological singular vectors at 
level 1 in complete Verma modules.

\vskip .15in
\noi
{\it Generic singular vectors}

First let us consider the four types of singular
vectors associated to the skeleton-family \req{diab0}; that is, 
the generic singular vectors of the first main kind of families.
The uncharged \Gn-closed topological singular vectors at level 1, built on
\Gn-closed primaries, are given by

\BE
\kc_{1,\ket{\D,\,\htop}^G}^{(0)G}=((\htop-1)\cL_{-1}
+(\htop-1-2\D)\cH_{-1}+\cG_{-1}\cQ_0)\ket{\D,\,\htop}^G , \label{sv1}
\EE

\noi
where $\D$ and $\htop$ are related by the quadratic vanishing
surface $f_{1,2}(\D,\,\htop)=0$, eq. \req{tfrs}, as
$2\D t=-\htop^2+t\htop+1-t$, with $t=(3-\ctop)/3$. 
By using $\cA$ \req{autom} one obtains $\kc_{1,\ket{\D,\,\htop'}^Q}^{(0)Q} =
\cA \, \kc_{1,\ket{\D,\,\htop}^G}^{(0)G}$, with $\htop'=-\htop-{\ctop\over3}$.
That is,

\BE
\kc_{1,\ket{\D,\,\htop'}^Q}^{(0)Q}=((-\htop'-{\ctop\over3}-1)\cL_{-1}
+2\D\cH_{-1}+\cQ_{-1}\cG_0)\ket{\D,\,\htop'}^Q , \label{sv0qq}
\EE

\noi
with $\D$ and $\htop'$ related by $2\D t=-\htop'^2+t\htop'-2\htop'$. Now
acting with \Qn\ and \Gn\ on $\kc_{1,\ket{\D,\,\htop}^G}^{(0)G}$ and
$\kc_{1,\ket{\D,\,\htop'}^Q}^{(0)Q}$, respectively, one obtains

\BE
\kc_{1,\ket{\D,\,\htop}^G}^{(-1)Q}=({3-\ctop\over3(\htop-1)}\cL_{-1}\Qz 
+\cH_{-1}\Qz+\cQ_{-1})\ket{\D,\,\htop}^G , \label{svm1qg}
\EE

\BE
\kc_{1,\ket{\D,\,\htop'}^Q}^{(1)G}=({\ctop-3\over3\htop'+\ctop+3}\cL_{-1}\Gz
-{6+3\htop'\over3\htop'+\ctop+3}\cH_{-1}\Gz+\cG_{-1})\ket{\D,\,\htop'}^Q . 
 \label{sv1gq}
\EE

Therefore for given values of $\D$ and $\htop$, and setting 
 $\htop'=-\htop-{\ctop\over3}$, one gets a four-member subfamily attached
to another similar four-member subfamily with $\hat\D=\D-\htop$ 
and $\hat\htop=-\htop$, as one can see in diagram \req{diab0}. Whether
or not there are further attachements depends on the values of
$\D$, $\htop$ and $\ctop$.

Let us consider now the eight types of singular vectors associated 
to the skeleton-family \req{diab1}; that is, the generic singular
vectors of the second main kind of families. The singular vectors of
types $\kc_{1,\ket{\D,\,\htop}^G}^{(1)G}$ and
$\kc_{1,\ket{\D,\,\htop'}^Q}^{(-1)Q}=\cA\,\kc_{1,\ket{\D,\,\htop}^G}^{(1)G}$
are given by

\BE \kc_{1,\ket{\D,\,\htop}^G}^{(1)G} = \cG_{-1}\,\ket{\D,\,\htop}^G,
\qquad \kc_{1,\ket{\D,\,\htop'}^Q}^{(-1)Q} = \cQ_{-1}\,\ket{\D,\,\htop'}^Q\,,
\EE

\noi
with $\htop'=-\htop-{\ctop\over3}$ and $\D=\htop$, corresponding to the
vanishing plane $g_{1/2}(\D,\htop)=0$, eq. \req{tgk}. Acting with \Qn\ and
\Gn\ on these vectors one completes the four-member subdiagram associated to
 $\kc_{1,\ket{\D,\,\htop}^G}^{(1)G}$, obtaining

\BE \kc_{1,\ket{\D,\,\htop}^G}^{(0)Q} = \Qz\cG_{-1}\,\ket{\D,\,\htop}^G,
\qquad \kc_{1,\ket{\D,\,\htop'}^Q}^{(0)G}=\Gz\cQ_{-1}\,\ket{\D,\,\htop'}^Q\,.
\EE

\noi
This subdiagram is attached to a chiral couple at level zero, and more
attachements may be possible (following sequence B) depending on the 
values of $\htop$ and $\ctop$.

The remaining four types of generic singular vectors at level 1, which
also provide a four-member subdiagram, attached to a four-member
subdiagram at level 2, are:

\BE \kc_{1,\ket{\D,\,\htop'}^Q}^{(2)G} = \cG_{-1}\Gz\,\ket{\D,\,\htop'}^Q ,
\qquad \kc_{1,\ket{\D,\,\htop}^G}^{(-2)Q} = \cQ_{-1}\Qz\,\ket{\D,\,\htop}^G ,
 \EE
\BE \kc_{1,\ket{\D,\,\htop'}^Q}^{(1)Q} = 
(\cL_{-1}\Gz - \D \cG_{-1})\ket{\D,\,\htop'}^Q , \EE 
\BE \kc_{1,\ket{\D,\,\htop}^G}^{(-1)G} =
( \cL_{-1}\Qz + \cH_{-1} \Qz - \D \cQ_{-1})\ket{\D,\,\htop}^G ,\EE

\noi
where $\htop'=-\htop-{\ctop\over3}$ and $\htop$ and $\D$ are related by the
vanishing plane $g_{-3/2}(\D,\htop)=0$ as $\D+\htop-1+\ctop/3=0$. As
before, it may be further attachements depending on the values of
$\D$, $\htop$ and $\ctop$.

As is explained in section 2, all the singular vectors built on primaries
of type $\ket{\D,\htop}^G$ are equivalent to singular vectors built
on primaries of type $\ket{\D,\htop}^Q$ (with different assignements of
the U(1) charges $q$ and $\htop$), provided $\D \not= 0$, as the reader
can easily verify.

\vskip .17in
\noi
{\it Chiral singular vectors}

Chiral singular vectors only exist in the types
$\kc_{\kp^G}^{(0)G,Q}\,$, $\kc_{\kp^Q}^{(0)G,Q}\,$, $\kc_{\kp^G}^{(-1)G,Q}\,$  
 and $\kc_{\kp^Q}^{(1)G,Q}\,$, and at level 1 they require $\D=-1$.
The uncharged chiral singular vectors are given by

\BE
\kc_{1,\ket{-1,-1}^G}^{(0)G,Q}=(-2\cL_{-1}
+\cG_{-1}\cQ_0)\ket{-1,-1}^G , \label{sv0gqg}
\EE

\BE
\kc_{1,\ket{-1,\,{3-\ctop\over3}}^Q}^{(0)G,Q}=(-2\cL_{-1}
-2\cH_{-1}+\cQ_{-1}\cG_0)\ket{-1,{3-\ctop\over3}}^Q , \label{sv0gqq}
\EE

\noi
and the charged chiral singular vectors are given by

\BE
\kc_{1,\,\ket{-1,\,{6-\ctop\over3}}^G}^{(-1)G,Q}=(\cL_{-1}\Qz 
+\cH_{-1}\Qz+\cQ_{-1})\ket{-1,\,{6-\ctop\over3}}^G , \label{svm1gqg}
\EE

\BE
\kc_{1,\ket{-1,-2}^Q}^{(1)G,Q}=(\cL_{-1}\Gz
+\cG_{-1})\ket{-1,-2}^Q . \label{sv1gqq}
\EE

Observe that the chiral singular vectors are the intersection of the
corresponding \Gn-closed and \Qn-closed singular vectors in each case.
Observe also that $\kc_{1,\ket{-1,-1}^G}^{(0)G,Q}$ 
and $\kc_{1,\ket{-1,-2}^Q}^{(1)G,Q}$ on the one side, 
and $\kc_{1,\ket{-1,\,{3-\ctop\over3}}^Q}^{(0)G,Q}$ and
$\kc_{1,\ket{-1,{6-\ctop\over3}}^G}^{(-1)G,Q}$ on the other side,
are equivalent just by expressing $\ket{-1,-1}^G= \Gz \ket{-1,-2}^Q$ and
$\ket{-1,{3-\ctop\over3}}^Q= \Qz  \ket{-1,\,{6-\ctop\over3}}^G$.

\vskip .17in
\noi
{\it No-label singular vectors}

The no-label singular
vectors, \ie\ which cannot be expressed as linear combinations of
$\cG_0$-closed and $\cQ_0$-closed singular vectors,
exist only in the types $\kc_{\kp^Q}^{(1)}\,$
and $\kc_{\kp^G}^{(-1)}\,$, and the equivalent types
$\kc_{\kp^G}^{(0)}\,$, and $\kc_{\kp^Q}^{(0)}\,$. At level 1 they 
require $\D=-1$. One finds solutions only for $\ctop=-3$. They are:

\BE
\kc_{1,\ket{-1,-2}^Q}^{(1)}=(\cL_{-1}\Gz
-\cH_{-1}\Gz)\ket{-1,-2}^Q , \label{sv1nl}
\EE

\BE
\kc_{1,\,\ket{-1,\,3}^G}^{(-1)}=(\cL_{-1}\Qz 
+2\cH_{-1}\Qz)\ket{-1,\,3}^G , \label{svm1nl}
\EE

\BE
\kc_{1,\ket{-1,-1}^G}^{(0)}=(\cL_{-1} -\cH_{-1})\ket{-1,-1}^G ,
\label{sv0gnl}\EE

\BE
\kc_{1,\ket{-1,\,2}^Q}^{(0)}=(\cL_{-1}
+2\cH_{-1})\ket{-1,\,2}^Q . \label{sv0qnl}
\EE

The conformal weight of these singular vectors is zero.
Thus acting with \Gn\ and \Qn\ one obtains \Gn-closed and \Qn-closed
secondary singular vectors, respectively.

\vskip .17in
\noi
{\it Singular vectors in no-label Verma modules}

Now let us consider no-label Verma modules built on
no-label primaries $\ket{0,\htop}$. The corresponding singular vectors
are grouped in two different kinds of families, 
as diagram \req{diab4} shows. At level 1 one finds one solution 
for the families on the right-hand side of the diagram:

\BE
\kc_{1,\ket{0,-1}}^{(2)G}=\cG_{-1}\cG_0\,\ket{0,-1} ,\qquad
\kc_{1,\ket{0,{3-\ctop\over3}}}^{(-2)Q}=
\cQ_{-1}\cQ_0\,\ket{0,{3-\ctop\over3}} ,
\EE

\BE
\kc_{1,\ket{0,-1}}^{(1)Q}=\Qz\cG_{-1}\cG_0\,\ket{0,-1} ,\qquad 
\kc_{1,\ket{0,{3-\ctop\over3}}}^{(-1)G}=
\Gz\cQ_{-1}\cQ_0\,\ket{0,{3-\ctop\over3}} ,
\EE

\noi
and two solutions for the families on the left-hand side of the diagram
(which intersect for $\ctop=-3$):

\BE
\kc_{1,\ket{0,0}}^{(1)G}=\cG_{-1}\cG_0\cQ_0\,\ket{0,0} ,\qquad
\kc_{1,\ket{0,-{\ctop\over3}}}^{(-1)Q}=
\cQ_{-1}\cQ_0\cG_0\,\ket{0,-{\ctop\over3}} , \EE
\BE
\kc_{1,\ket{0,0}}^{(0)Q}=\cL_{-1}\cG_0\cQ_0\,\ket{0,0} ,\qquad  
\kc_{1,\ket{0,-{\ctop\over3}}}^{(0)G}=(\cL_{-1}\cG_0\cQ_0+\cH_{-1}\cG_0\cQ_0)
\ket{0,-{\ctop\over3}} , \EE

\noi
and

\BE
\kc_{1,\ket{0,-{\ctop+3\over3}}}^{(1)G}=({\ctop+3\over3}\cL_{-1}\cG_0+
{\ctop+3\over3}\cH_{-1}\cG_0+\cG_{-1}\cG_0\cQ_0)\ket{0,-{\ctop+3\over3}} , 
\EE
\BE \kc_{1,\ket{0,1}}^{(-1)Q}=({\ctop+3\over3}\cL_{-1}\cQ_0+
\cQ_{-1}\cQ_0\cG_0)\ket{0,1} , \EE
\BE
\kc_{1,\ket{0,-{\ctop+3\over3}}}^{(0)Q}=((\ctop-3)\cL_{-1}\cQ_0\cG_0+
(\ctop+3)\cH_{-1}\cQ_0\cG_0+(\ctop+3)\cQ_{-1}\cG_0)
\ket{0,-{\ctop+3\over3}} , \EE
\BE \kc_{1,\ket{0,1}}^{(0)G}=({3-\ctop\over6}\cL_{-1}\cG_0\cQ_0+
\cH_{-1}\cG_0\cQ_0-{\ctop+3\over6}\cG_{-1}\cQ_0)\ket{0,1} . \EE

The spectrum of $\htop$ that we have found at level 1
for no-label Verma modules $V(\ket{0,\htop})$ is the ``sum"
of the spectra of $\htop$ at level 1 corresponding to the 
generic Verma modules with zero conformal weight $V(\ket{0,\htop}^G)$ 
and $V(\ket{0,\htop}^Q)$, as the interested reader can verify.

\vskip .17in 
\noi
{\it Loop of sequence B}

To finish, let us show that the loop of sequence B for 
$\htop={3-\ctop\over6}=t/2$ transforms a singular vector of type
$\kc_{1,\,\ket{\D,\,\htop}^G}^{(0)G}$, with $\D  \not= 0$, 
back to itsef. This result is to be expected because at level 1 
the D\"orrzapf conditions \req{Dcond} cannot be satisfied. 

For $\D \not= 0$ the subfamily associated to 
$\kc_{1,\,\ket{\D,\,\htop}^G}^{(0)G}$, given by \req{sv1}, can be 
expressed as shown in diagram \req{diab2}.
In order to 
obtain $\kc_{1,\ket{\D,-\htop-\ctop/3+1}^G}^{(0)G}
=\cG_0\,\cA\kc_{1,\ket{\D,\,\htop}^G}^{(0)G}$ one has to 
apply $\cG_0 \, \cA$ first, then one has to express the
resulting $\cQ_0$-closed primary as
$\ket{\D,-\htop-\ctop/3}^Q=\cQ_0\,\ket{\D,-\htop-\ctop/3+1}^G =
\cQ_0\,\ket{\D,-\htop+t}^G $, resulting finally in the expression

\BE
\kc_{1,\ket{\D,-\htop+t}^G}^{(0)G}
=((\htop+1)2\D\cL_{-1}+4(\D^2+\D)\cH_{-1}
+(\htop-1-2\D)\cG_{-1}\cQ_0)\ket{\D,-\htop+t}^G . \label{sv1p}
\EE

Now setting $\htop=t/2$ in both vectors, \req{sv1} and 
\req{sv1p}, and taking into
account that $\D=(t-2)^2/8t$, given by $f_{1,2}(\D,t/2)=0$ in
\req{tfrs}, one arrives at the same expression:

\BE
\kc_{1,\ket{\D,t/2}^G}^{(0)G}
=({t-2\over2}\cL_{-1}+{t^2-4\over4t}\cH_{-1}
+\cG_{-1}\cQ_0)\ket{\D,t/2}^G
\EE

\noi
Therefore, the two uncharged singular vectors
$\kc_{1,\ket{\D,\,\htop}^G}^{(0)G}$ and
$\cG_0\,\cA\kc_{1,\ket{\D,\,\htop}^G}^{(0)G}$ 
coincide for $\htop={3-\ctop\over6}=t/2$, provided $\D \not= 0$.

To see that the mapping given by $\Gz\, \cA$ 
is far from being the identity,   
let us repeat the same procedure for a general \Gn-closed uncharged
secondary state at level 1 instead 
of a singular vector, \ie\ for the state

\BE
\ket\gamma_{1,\,\ket{\D,\,\htop}^G}^{(0)G}=(\alpha\cL_{-1}+\beta\cH_{-1}+
\cG_{-1}\cQ_0)\ket{\D,\,\htop}^G{\ },
\EE

\noi
with $\alpha-\beta=2\D$. One finds that $\cG_0\,\cA$ maps the 
secondary state back
to itself, for $\htop=t/2$, only if $\beta^2=4\D(\D+1)$. This
is in fact the case for the singular vector \req{sv1} for that particular
value of $\htop$.

\setcounter{equation}{0}
\def\theequation{C.\arabic{equation}}

\subsection*{Appendix C}\lvm

Here we show a specially interesting thirty-eight-member 
family of generic and chiral topological singular vectors at 
levels 3, 4, 5, and 6, for $\ctop=9$ ($t=-2$). It contains four 
D\"orrzapf pairs, at level 3, and three chiral couples which produce 
the intersection of the singular vectors associated to the
skeleton-family \req{diab0} (first kind of generic families)
with the singular vectors associated to the 
skeleton-family \req{diab1} (second kind of generic families).
Due to the length of this family we write down explicitely only  
the four D\"orrzapf pairs, displayed in
diagram \req{dia6}. The upper row and the 
lower row of the diagram are connected by $\cU_1$, 
as indicated by the arrow on top, since $\cU_1\cA{\ }\cU_1\cA={\bf I}$

\def\stgqo  {\mbox{$\kc_{3,\, \ket{-3,\,1}^G}^{(0)G,Q} $}}
\def\stqgo  {\mbox{$\kc_{3,\, \ket{-3,-4}^Q}^{(0)G,Q} $}}
\def\stqgm  {\mbox{$\kc_{3,\, \ket{-3,-4}^Q}^{(0)Q} $}}
\def\stgqp  {\mbox{$\kc_{3,\, \ket{-3,\,1}^G}^{(0)G} $}}
\def\sbgqo  {\mbox{$\kc_{3,\, \ket{-4,-1}^G}^{(0)G} $}}
\def\sbqgo 
 {\mbox{$\kc_{3,\, \ket{-4,-2}^Q}^{(0)Q} $}}
\def\sbqgm  {\mbox{$\ket{\hat\chi}_{3,\, \ket{-4,-2}^Q}^{(0)Q} $}}
\def\sbgqp 
 {\mbox{$\ket{\hat\chi}_{3,\, \ket{-4,-1}^G}^{(0)G} $}}

  \begin{equation} \begin{array}{rclclc}
   \cU_1\,\uparrow\ & \\[3 mm]
 \sbgqo &
  \stackrel{\Qz}{\mbox{------}\!\!\!\longrightarrow}
  & \sbqgm  &
  \stackrel{\cU_{-1}}{\mbox{------}\!\!\!\longrightarrow} &
   \stgqp & \stackrel{\Qz} {\longrightarrow} \\[3 mm]
   \cA\,\updownarrow\ && \ \updownarrow\, \cA &&  \updownarrow\,\cA\, 
  \\[3 mm]  \sbqgo \! & \stackrel{\Gz}
  {\mbox{------}\!\!\!\longrightarrow} & \! \sbgqp &
\stackrel{\cU_1}
  {\mbox{------}\!\!\!\longrightarrow} &  
\stqgm &  \stackrel{\Gz} {\longrightarrow} \\[5 mm] 
\cU_1\,\uparrow\ & \\[4 mm]
  \stgqo & \\[3 mm]
   \cA\,\updownarrow\ &
  \\[3 mm]  \stqgo \!  
 \end{array} \label{dia6} \end{equation}

\vskip .2in

The uncharged chiral couple of singular vectors  
$\kc_{3,\,\ket{-3,\,1}^G}^{(0)G,Q}$ and $\kc_{3,\,\ket{-3,-4}^Q}^{(0)G,Q}\,$,

\begin{eqnarray*}
\kc_{3,\,\ket{-3,\,1}^G}^{(0)G,Q}=\cQ_0\,(3\cL_{-2}\cG_{-1}+\cL_{-1}^2\cG_{-1}
-\cH_{-2}\cG_{-1}+2\cH_{-1}^2\cG_{-1}-3\cL_{-1}\cH_{-1}\cG_{-1}+\\
4\cL_{-1}\cG_{-2}-8\cH_{-1}\cG_{-2}+12\cG_{-3})
\ket{-3,1}^G ,
\end{eqnarray*}

\begin{eqnarray*}
\kc_{3,\,\ket{-3,-4}^Q}^{(0)G,Q}=\cG_0\,(3\cL_{-2}\cQ_{-1}+\cL_{-1}^2\cQ_{-1}
+6\cH_{-2}\cQ_{-1}+6\cH_{-1}^2\cQ_{-1}+5\cL_{-1}\cH_{-1}\cQ_{-1}+\\
4\cL_{-1}\cQ_{-2}+12\cH_{-1}\cQ_{-2}+12\cQ_{-3})
\ket{-3,-4}^Q \,,
\end{eqnarray*}

\noi
is connected by $\cU_{\pm1}$ to the pair of singular vectors
$\kc_{3,\,\ket{-4,-2}^Q}^{(0)Q}$ and $\kc_{3,\,\ket{-4,-1}^G}^{(0)G}\,$.
Acting further with \Gn\ and \Qn\ 
one obtains the four-member subfamily

\begin{eqnarray*}
\kc_{3,\,\ket{-4,-2}^Q}^{(0)Q}=(3\cL_{-2}\cQ_{-1}\cG_0+\cL_{-1}^2\cQ_{-1}\cG_0
+\cH_{-2}\cQ_{-1}\cG_0-\cL_{-1}\cH_{-1}\cQ_{-1}\cG_0+
4\cL_{-1}\cQ_{-1}\cG_{-1}+\\
-3\cL_{-1}\cQ_{-2}\cG_0 -4\cH_{-1}\cQ_{-1}\cG_{-1}+
\cH_{-1}\cQ_{-2}\cG_0-8\cQ_{-2}\cG_{-1}
+12\cQ_{-1}\cG_{-2}+\cQ_{-3}\cG_0)\ket{-4,-2}^Q ,
\end{eqnarray*}

\begin{eqnarray*}
\kc_{3,\,\ket{-4,-1}^G}^{(0)G}=(3\cL_{-2}\cG_{-1}\cQ_0+\cL_{-1}^2\cG_{-1}\cQ_0
+4\cH_{-2}\cG_{-1}\cQ_0+2\cH_{-1}^2\cG_{-1}\cQ_0+\\
 3\cL_{-1}\cH_{-1}\cG_{-1}\cQ_0+4\cL_{-1}\cG_{-1}\cQ_{-1}
+8\cH_{-1}\cG_{-1}\cQ_{-1}-3\cL_{-1}\cG_{-2}\cQ_0+\\
-4\cH_{-1}\cG_{-2}\cQ_0-8\cG_{-2}\cQ_{-1}+12\cG_{-1}\cQ_{-2}
+\cG_{-3}\cQ_0)\ket{-4,-1}^G ,
\end{eqnarray*}
 
\begin{eqnarray*}
\kc_{3,\,\ket{-4,-1}^G}^{(-1)Q}=
\ket{\hat\chi}_{3,\,\ket{-4,-2}^Q}^{(0)Q}=(-4\cL_{-3}-8\cL_{-1}\cL_{-2}
-2\cL_{-1}^3-8\cH_{-3}-8\cH_{-1}\cH_{-2}+\\
-8\cL_{-2}\cH_{-1}
-6\cL_{-1}^2\cH_{-1}-10\cL_{-1}\cH_{-2}-4\cL_{-1}\cH_{-1}^2
+\cL_{-1}^2\cQ_{-1}\cG_0+
2\cL_{-1}\cH_{-1}\cQ_{-1}\cG_0+\\
\cL_{-1}\cQ_{-1}\cG_{-1}+
4\cH_{-1}\cQ_{-1}\cG_{-1}+3\cL_{-1}\cQ_{-2}\cG_0+6\cQ_{-2}\cG_{-1}
-4\cQ_{-1}\cG_{-2})\ket{-4,-2}^Q ,
\end{eqnarray*}

\begin{eqnarray*}
\kc_{3,\,\ket{-4,-2}^Q}^{(1)G}=
\ket{\hat\chi}_{3,\,\ket{-4,-1}^G}^{(0)G}=(-4\cL_{-3}-8\cL_{-1}\cL_{-2}
-2\cL_{-1}^3+2\cH_{-1}\cH_{-2}+2\cL_{-1}\cH_{-1}^2+\\
\cL_{-1}^2\cG_{-1}\cQ_0
-\cH_{-2}\cG_{-1}\cQ_0-\cH_{-1}^2\cG_{-1}\cQ_0+\cL_{-1}\cG_{-1}\cQ_{-1}-
3\cH_{-1}\cG_{-1}\cQ_{-1}+\\
3\cL_{-1}\cG_{-2}\cQ_0+3\cH_{-1}\cG_{-2}\cQ_0
+6\cG_{-2}\cQ_{-1}-4\cG_{-1}\cQ_{-2})\ket{-4,-1}^G \,,
\end{eqnarray*}

\noi
where the two last singular vectors have been expressed as singular 
vectors of their equivalent types, using 
$\ket{-4,-1}^G=\cG_0\,\ket{-4,-2}^Q$ and
$\ket{-4,-2}^Q=\cQ_0\,\ket{-4,-1}^G$. We see that this four-member
subfamily consists of two D\"orrzapf pairs: one pair of linearly
independent singular vectors of type $\kc_{\kp^G}^{(0)G}$ (equivalently 
$\kc_{\kp^Q}^{(1)G}$) at the same level in the same Verma module ,
and one pair of linearly independent singular vectors of type
$\kc_{\kp^Q}^{(0)Q}$ (equivalently $\kc_{\kp^G}^{(-1)Q}$) at
the same level in the same Verma module. Observe that the Verma module 
$V(\ket{-4,-1}^G)$ has $\htop=-1={t\over2}$, therefore the singular 
vectors of the same type are transformed into each other by the simplest
loop of sequence B, \ie\ just by the mapping given by $\Gz\,\cA$.
This subfamily is attached further, acting with $\cU_{\pm1}$, 
to the pair of singular vectors $\kc_{3,\,\ket{-3,-4}^Q}^{(0)Q}$ and 
$\kc_{3,\,\ket{-3,\,1}^G}^{(0)G}\,$,
which in spite of having zero conformal weight $(\D+l=0)$ are not chiral,
although they are in the same Verma modules at the same level than the
chiral singular vectors $\kc_{3,\,\ket{-3,\,1}^G}^{(0)G,Q}$ and 
$\kc_{3,\,\ket{-3,-4}^Q}^{(0)G,Q}\,$ : 

\begin{eqnarray*}
\kc_{3,\,\ket{-3,-4}^Q}^{(0)Q}=(-4\cL_{-3}-8\cL_{-1}\cL_{-2}-2\cL_{-1}^3
-8\cL_{-2}\cH_{-1}-6\cL_{-1}^2\cH_{-1}-2\cL_{-1}\cH_{-2}+\\
-4\cL_{-1}\cH_{-1}^2+\cL_{-1}^2\cG_0\cQ_{-1}
-2\cH_{-2}\cG_0\cQ_{-1}+2\cL_{-1}\cH_{-1}\cG_0\cQ_{-1}
+\cL_{-1}\cG_0\cQ_{-2}+\\
-2\cH_{-1}\cG_0\cQ_{-2}+3\cL_{-1}\cG_{-1}\cQ_{-1}+6\cH_{-1}\cG_{-1}\cQ_{-1}
+6\cG_{-1}\cQ_{-2}-4\cG_0\cQ_{-3})\ket{-3,-4}^Q ,
\end{eqnarray*}

\begin{eqnarray*}
\kc_{3,\,\ket{-3,\,1}^G}^{(0)G}=(-4\cL_{-3}-8\cL_{-1}\cL_{-2}-2\cL_{-1}^3
-8\cH_{-3}+2\cH_{-1}\cH_{-2}-8\cL_{-1}\cH_{-2}+\\
2\cL_{-1}\cH_{-1}^2+\cL_{-1}^2\cQ_0\cG_{-1}
+\cH_{-2}\cQ_0\cG_{-1}-\cH_{-1}^2\cQ_0\cG_{-1}+3\cL_{-1}\cQ_{-1}\cG_{-1}+\\
-3\cH_{-1}\cQ_{-1}\cG_{-1}+\cL_{-1}\cQ_0\cG_{-2}+3\cH_{-1}\cQ_0\cG_{-2}
+6\cQ_{-1}\cG_{-2}-4\cQ_0\cG_{-3})\ket{-3,1}^G .
\end{eqnarray*}

Hence we have encountered two more D\"orrzapf
pairs with the particularity that in each pair one of the singular vectors
turns out to be chiral (chiral vectors are just particular cases of 
\Gn-closed vectors, as well as particular cases of \Qn-closed vectors).
We have checked that these singular vectors satisfy the D\"orrzapf
conditions \req{Dcond}. 
Observe that in this case the singular vectors of the same type are not
connected to each other by a loop of sequence B, but rather by the
mapping $\kc_{3,\,\ket{-3,\,1}^G}^{(0)G} =
\cA\,\cU_1\,\Gz\,\cU_1\,\kc_{3,\,\ket{-3,\,1}^G}^{(0)G,Q}$.

Acting now with \Gn\ and \Qn\ one obtains {\it secondary} 
chiral singular vectors $\kc_{3,\,\ket{-3,-4}^Q}^{(1)G,Q}$ and
$\kc_{3,\,\ket{-3,\,1}^G}^{(-1)G,Q}$ at level zero with respect to 
the singular vectors on which they are built. Therefore these arrows
\Gn , \Qn\ cannot be reversed unlike the previous ones. 
One can attach arrows $\cU_{\pm1}$ to these secondary singular vectors,
and with them the four-member subfamily:
$\kc_{4,\,\ket{-4,-2}^Q}^{(-1)Q}$, $\kc_{4,\,\ket{-4,-1}^G}^{(1)G}$,
$\kc_{4,\,\ket{-4,-2}^Q}^{(0)G,Q}$, $\kc_{4,\,\ket{-4,-1}^G}^{(0)G,Q}$,
at level 4, where the chiral singular vectors are, in turn, secondary
singular vectors of the non-chiral ones. Using $\cU_{\pm1}$ again
a new subfamily is attached:
$\kc_{4,\,\ket{-3,-4}^Q}^{(0)Q}$, $\kc_{4,\,\ket{-3,\,1}^G}^{(0)G}$,
$\kc_{4,\,\ket{-3,\,0}^Q}^{(0)Q}$, $\kc_{4,\,\ket{-3,-3}^G}^{(0)G}$.
Using $\cU_{\pm1}$ once more one attaches the subfamily:
$\kc_{4,\,\ket{0,-6}^Q}^{(0)Q}$, $\kc_{4,\,\ket{0,\,3}^G}^{(0)G}$,
$\kc_{4,\,\ket{0,-6}^Q}^{(1)G}$, $\kc_{4,\,\ket{0,\,3}^G}^{(-1)Q}$.
There are no further attachements to this subfamily since $\D=0$, so
that it is not possible to write singular vectors equivalent to
$\kc_{4,\,\ket{0,-6}^Q}^{(1)G}$ and $\kc_{4,\,\ket{0,\,3}^G}^{(-1)Q}$
where to attach more arrows $\cU_{\pm1}$.

The attachements of subfamilies is not finished, however, because the
uncharged chiral singular vectors are equivalent to charged chiral
singular vectors, and the spectral flow mappings $\cU_{\pm1}$
distinguish betwen a given chiral singular vector and the equivalent one.

\def\etggob  {\mbox{$\kc_{4,\, \ket{0,\,3}^G}^{(1)G} $}}
\def\etqqob  {\mbox{$\kc_{4,\, \ket{0,-6}^Q}^{(-1)Q} $}}
\def\etqqmb  {\mbox{$\kc_{4,\, \ket{0,\,3}^G}^{(0)Q} $}}
\def\etggpb  {\mbox{$\kc_{4,\, \ket{0,-6}^Q}^{(0)G} $}}
\def\ebggob  {\mbox{$\kc_{3,\, \ket{-3,-3}^G}^{(-1)G,Q} $}}
\def\ebqqob 
 {\mbox{$\kc_{3,\, \ket{-3,\,0}^Q}^{(1)G,Q} $}}

  \begin{equation} \begin{array}{rcl}
   \cU_1\,\uparrow\ & \\[3 mm]
  \ebggob &  \\[3 mm]
   \cA\,\updownarrow\ &   \\[3 mm]  \ebqqob \! & 
  \\[5 mm] \cU_1\,\uparrow\ & \\[4 mm]
  \etggob &
  \stackrel{\Qz}{\mbox{------}\!\!\!\longrightarrow}
  & \etqqmb \\[3 mm]
   \cA\,\updownarrow\ && \ \updownarrow\, \cA
  \\[3 mm]  \etqqob \! & \stackrel{\Gz}
  {\mbox{------}\!\!\!\longrightarrow} & \! \etggpb
 \end{array} \label{dia7} \end{equation}

\vskip .2in

Let us move to diagram \req{dia7}.
The uncharged chiral couple of singular vectors 
$\kc_{3,\,\ket{-3,\,1}^G}^{(0)G,Q}$ and 
$\kc_{3,\,\ket{-3,-4}^Q}^{(0)G,Q}\,$, in diagram \req{dia6}, 
is equivalent to the charged chiral couple
$\kc_{3,\,\ket{-3,\,0}^Q}^{(1)G,Q}$ and 
$\kc_{3,\,\ket{-3,-3}^G}^{(-1)G,Q}\,$ in diagram \req{dia7}.
These singular vectors are attached by $\cU_{\pm1}$ to the 
four-member subfamily: $\kc_{4,\,\ket{0,\,3}^G}^{(1)G}\,$, 
$\kc_{4,\,\ket{0,-6}^Q}^{(-1)Q}\,$, $\kc_{4,\,\ket{0,\,3}^G}^{(0)Q}\,$, 
$\kc_{4,\,\ket{0,-6}^Q}^{(0)G}\,$, at level 4. Since $\D=0$ for this 
subfamily there are no further attachements.  

Now let us consider the 
charged chiral couple $\kc_{3,\,\ket{-3,-4}^Q}^{(1)G,Q}$
and $\kc_{3,\,\ket{-3,\,1}^G}^{(-1)G,Q}$. 
It is equivalent to the 
uncharged chiral couple $\kc_{3,\,\ket{-3,-3}^G}^{(0)G,Q}$
and $\kc_{3,\,\ket{-3,\,0}^Q}^{(0)G,Q}$,
that can be attached using $\cU_{\pm1}$ to the subfamily:
$\kc_{3,\,\ket{0,-6}^Q}^{(0)Q}$, $\kc_{3,\,\ket{0,\,3}^G}^{(0)G}$,
$\kc_{3,\,\ket{0,-6}^Q}^{(1)G}$, $\kc_{3,\,\ket{0,\,3}^G}^{(-1)Q}$.
Since $\D=0$ for this subfamily there are no further attachements.

Finally let us consider the uncharged chiral couple
$\kc_{4,\,\ket{-4,-1}^G}^{(0)G,Q}$ and $\kc_{4,\,\ket{-4,-2}^Q}^{(0)G,Q}$.
It is equivalent to the 
charged chiral couple $\kc_{4,\,\ket{-4,-2}^Q}^{(1)G,Q}$ 
and $\kc_{4,\,\ket{-4,-1}^G}^{(-1)G,Q}$, that can be attached using 
$\cU_{\pm1}$ to the subfamily:
$\kc_{5,\,\ket{-3,-4}^Q}^{(-1)Q}$, $\kc_{5,\,\ket{-3,\,1}^G}^{(1)G}$,
$\kc_{5,\,\ket{-3,\,0}^Q}^{(1)Q}$, $\kc_{5,\,\ket{-3,-3}^G}^{(-1)G}$,
at level 5. This subfamily is in turn attached by $\cU_{\pm1}$ 
to the subfamily:
$\kc_{6,\,\ket{0,-6}^Q}^{(-1)Q}$, $\kc_{6,\,\ket{0,\,3}^G}^{(1)G}$,
$\kc_{6,\,\ket{0,-6}^Q}^{(0)G}$, $\kc_{6,\,\ket{0,\,3}^G}^{(0)Q}$,
at level 6. There are no further attachements since $\D=0$ for this
subfamily.

The complete family consists therefore 
of thirty-eight singular vectors at levels
3, 4, 5, and 6 distributed in five different Verma modules. Two of them
have $\D=0$ and consequently a single primary state:  
$V(\ket{0,\,3}^G)$ and $V(\ket{0,-6}^Q)$. The other three Verma 
modules have $\D\neq0$ and consequently two primary states. 
Namely, $V(\ket{-4,-1}^G)= V(\ket{-4,-2}^Q)\,$, 
$V(\ket{-3,\,1}^G) = V(\ket{-3,\,0}^Q)\,$, and   
$V(\ket{-3,-3}^G) = V(\ket{-3,-4}^Q)$.

\end{document}